\documentclass{bmcart}

\usepackage[utf8]{inputenc} 

\def\includegraphics{}

\startlocaldefs
\endlocaldefs

\usepackage{graphicx}
\usepackage{amsmath}
\usepackage[authoryear]{natbib}
\usepackage{float}

\usepackage{hyperref}

\usepackage{leftidx}

\usepackage{pdflscape}
\usepackage{rotating}

\definecolor{darkgreen}{rgb}{0,.5,0}
\definecolor{matlabred}{rgb}{0.8500    0.3250    0.0980}
\definecolor{matlabblue}{rgb}{0    0.4470    0.7410}
\definecolor{matlabmagenta}{rgb}{0.4940    0.1840    0.5560}
\definecolor{matlabgreen}{rgb}{0.4660    0.6740    0.1880}

\begin{document}

\begin{frontmatter}

\begin{fmbox}
\dochead{Research}


\title{The CHAOS-7 geomagnetic field model and \\ observed changes in the South Atlantic Anomaly}

\author[
   addressref={aff1},                   
   corref={aff1},                       
   email={cfinlay@space.dtu.dk}   
]{\inits{CC}\fnm{Christopher C.} \snm{Finlay}}
\author[
   addressref={aff1},
   email={ancklo@space.dtu.dk}
]{\inits{C}\fnm{Clemens} \snm{Kloss}}
\author[
   addressref={aff1},
   email={nio@space.dtu.dk}
]{\inits{N}\fnm{Nils} \snm{Olsen}}
\author[
   addressref={aff1},
   email={magdh@space.dtu.dk}
]{\inits{N}\fnm{Magnus D.} \snm{Hammer}}
\author[
   addressref={aff1},
   email={lastec@space.dtu.dk}
]{\inits{L}\fnm{Lars} \snm{T\o ffner-Clausen}}
\author[
   addressref={aff2},
   email={xx}
]{\inits{L}\fnm{Alexander} \snm{Grayver}}
\author[
   addressref={aff2},
   email={xx}
]{\inits{L}\fnm{Alexey} \snm{Kuvshinov}}


\address[id=aff1]{
  \orgname{Division of Geomagnetism, $\qquad$ DTU Space, Technical University of Denmark}, 
  \street{Centrifugevej 356},                     %
  \city{Kongens Lyngby},                              
  \cny{Denmark}                                    
}

\address[id=aff2]{
  \orgname{Institute of Geophysics, ETH Zurich} 
  \street{Sonneggstrasse 5},                     %
  \city{Zurich},                              
  \cny{Switzerland}                                    
}

\begin{artnotes}
\end{artnotes}

\end{fmbox}


\begin{abstractbox}

\begin{abstract} 
We present the CHAOS-7 model of the time-dependent near-Earth geomagnetic field between 1999 and 2020 based on magnetic field observations collected by the low-Earth orbit satellites {\it Swarm}, CryoSat-2, CHAMP, SAC-C and {\O}rsted, and on annual differences of monthly means of ground observatory measurements. The CHAOS-7 model consists of a time-dependent internal field up to spherical harmonic degree 20, a static internal field which merges to the LCS-1 lithospheric field model above degree 25, a model of the magnetospheric field and its induced counterpart, estimates of Euler angles describing the alignment of satellite vector magnetometers, and magnetometer calibration parameters for CryoSat-2. Only data from dark regions satisfying strict geomagnetic quiet-time criteria (including conditions on IMF $B_z$ and $B_y$ at all latitudes) were used in the field estimation.  Model parameters were estimated using an iteratively-reweighted regularized least-squares procedure; regularization of the time-dependent internal field was relaxed at high spherical harmonic degree compared with previous versions of the CHAOS model. We use CHAOS-7 to investigate recent changes in the geomagnetic field, studying the evolution of the South Atlantic weak field anomaly and rapid field changes in the Pacific region since 2014. At Earth's surface a secondary minimum of the South Atlantic Anomaly is now evident to the south west of Africa.  Green's functions relating the core-mantle boundary radial field to the surface intensity show this feature is connected with the movement and evolution of a reversed flux feature under South Africa.  The continuing growth in size and weakening of the main anomaly is linked to the westward motion and gathering of reversed flux under South America.  In the Pacific region at Earth's surface between 2015 and 2018 a sign change has occurred in the second time derivative (acceleration) of the radial component of the field. This acceleration change took the form of a localized, east-west oriented, dipole. It was clearly recorded on ground, for example at the magnetic observatory at Honolulu, and was seen in {\it Swarm} observations over an extended region in the central and western Pacific.  Downward continuing to the core-mantle boundary we find this event originated in field acceleration changes at low latitudes beneath the central and western Pacific in 2017. 
\end{abstract}




\end{abstractbox}
%

\end{frontmatter}

\section{Introduction}
The Earth's magnetic field is a fundamental part of our planetary environment and an integral component of many modern navigational systems, providing a natural and readily available source of orientation information.  To make use of the geomagnetic field for navigation one requires a good quality magnetometer to measure it, and a reference field model that relates the magnetic vector, at the location and time of the measurement, to the geographic directions.  The International Geomagnetic Reference Field (IGRF) is a prominent example of such a reference model, and a trusted source of information on the Earth's magnetic field for the wider scientific community including space physicists, high-energy particle physicists, exploration geologists, engineers and biologists.  This article reports on the parent model of candidates submitted by the Technical University of Denmark (DTU) for IGRF-13 in October 2019. 

This parent model, called CHAOS-7, is the latest in a series of time-dependent geomagnetic field models developed at DTU over the past 15 years (\citealp{Olsen:2006a,Olsen:2014a}; \citealp{Finlay:2016}).  CHAOS-7 spans a twenty-one year period from 1999 to 2020 for which both satellite and ground observatory data are available. For the past six years, satellite data have been delivered by the {\it Swarm} satellite trio, providing a particularly complete and homogeneous data coverage.  Here, we take this opportunity to report in detail on two particularly intriguing aspects of recent geomagnetic field change.  First we document changes since 2014 in the South Atlantic weak field anomaly which has important implications for the radiation dose experienced by satellites, and second we investigate patterns of rapid field change observed in the Pacific region over the past six years.    

The South Atlantic Anomaly (SAA) was originally detected by early low-earth satellite missions in the late 1950s as a region of enhanced flux of energetic charged particles \citep{Yoshida:1960,Vernov:1960,Ginzburg:1962}.   It has been well documented over the intervening years as a region of geospace where satellites systematically experience an enhanced radiation dose \cite[e.g.][]{Gledhill:1976, Heirtzler:2002}.  The depth to which charged particles in the radiation belts penetrate, known as their bounce point along a field magnetic field line \citep{Walt:2005}, depends on the intensity of the geomagnetic field. Anomalously weak magnetic field in the South Atlantic (compared with the field of a centered dipole) thus gives rise to enhanced charged particle flux in this region i.e. a radiation anomaly.  By tracking the evolution of the weak magnetic field region we are therefore able to track the development radiation anomaly. Here we document the evolution of the South Atlantic Anomaly as observed by the \textit{Swarm} satellites, both in the magnetic field and in single event electronic upsets monitored by onboard instruments.  We go on to investigate changes in the field intensity mapped down to Earth's surface using CHAOS-7. Ultimately processes in the Earth's core  determine the future evolution of the South Atlantic Anomaly; we use Green's functions relating the radial magnetic field at the core-mantle boundary to changes in the field intensity at satellite altitude to study the origin of these processes at the outer edge of the fluid outer core.  

A second focus point in this article is rapid secular variation observed since the launch of \textit{Swarm} in the Pacific region.  Traditionally the Pacific has been thought of as a quiet region for secular variation \citep{Vestine:1966}, but observatory records from Honolulu have indicated large amplitude changes in secular variation in the past decade (see the discussion in \cite{Finlay:2016} and the section "\nameref{sec:rapid_SA_Pac}"  below).  With 6 years of data now available from \textit{Swarm} we are able to track a change in sign of the second time derivative or secular acceleration in this region, and to study its spatial signature at the Earth's surface and at the core surface.

Compared to its predecessor, CHAOS-6 \citep{Finlay:2016}, CHAOS-7 uses a stricter criteria for selecting geomagnetically quiet times, it makes use of uncalibrated vector magnetic field data from the CryoSat-2 satellite between 2010 and 2014 by means of co-estimating magnetometer calibration parameters, and the temporal regularization is relaxed at higher spherical harmonic degrees.   We recall that compared to other modern geomagnetic field models \cite[e.g.][]{Lesur:2010,Maus:2005a, Sabaka:2015}, the CHAOS models are of intermediate complexity.  They involve co-estimation of alignment parameters, internal, magnetospheric and lithospheric fields. A variety of strategies  have been pursued by other recent field models, for example \cite{Alken:2020} did not co-estimate magnetospheric fields while \cite{Sabaka:2020} and \cite{Ropp:2020} co-estimate estimate ionospheric, magnetospheric and induced fields.  No attempt is made in CHAOS-7 to deterministically predict the future field evolution, rather our strategy is to provide regular updates, typically every 4 to 6 months, using the latest satellite and ground data. 

In the "\nameref{sec:Data}" section below we present details of the ground-based and satellite geomagnetic measurements used to derive the CHAOS-7 model.  In the "\nameref{sec:Model}" section we describe the CHAOS-7 model parameterization and its estimation, including details of the co-estimated calibration model used for CryoSat-2 data.  Special attention is given to how we treat fields resulting from currents induced in the electrically-conducting Earth by the time-varying magnetospheric fields.  In the section "\nameref{sec:Results}" we report diagnostics for the CHAOS-7 field model including data misfit statistics, comparisons with ground observatory time series, spatial power spectra and maps of the internal field. Validation tests against data not used to determine the field model are presented.  We also provide details on how the IGRF-13 candidate models were extracted.  In the "\nameref{sec:SAA}" section we report on the recent changes in the region of weakest field intensity and their origin at the core surface.  In section "\nameref{sec:rapid_SA_Pac}" we focus on rapid secular variation during the {\it Swarm} era and place this in historical context.  A summary and some final remarks are given in the "\nameref{sec:Conc}" section.

\section{Data}
\label{sec:Data}

CHAOS-7 is based on magnetometer data collected onboard the satellites {\O}rsted, CHAMP, SAC-C, CryoSat-2, and most importantly for the past six years, the three {\it Swarm} satellites, as well as an updated version of the revised monthly mean ground observatory secular variation data series \citep{Olsen:2014a}. 

\subsection{{\bf Satellite data}}
From the {\O}rsted mission we used vector data between March 1999 and December 2004 and scalar data between March 1999 and June 2013 for quasi-dipole (QD) latitudes \citep[see][for a definition]{Richmond:1995a} poleward of $\pm55^\circ$, or if attitude data were not available, each with 1 minute sampling.  In addition, along-track differences of {\O}rsted scalar data, separated by 15~seconds along track and with 1 minute sampling, were used as a source of scalar gradient information.  From the CHAMP mission we used vector data between August 2000 and September 2010 (converted to scalar data at quasi-dipole latitudes poleward of $\pm55^\circ$) with 1 minute sampling, as well as vector and scalar along-track differences (again based on measurements separated by 15~seconds along track), again with 1 minute sampling.  Vector and vector gradient data were used only when data from two star trackers were available. From the SAC-C mission, we used scalar data with 1 minute sampling between January 2001 and December 2004.   Along-track differences were not used from SAC-C. From the CryoSat-2 mission we used uncalibrated vector data from August 2010 to December 2014, from magnetometer FGM1, but with corrections applied for temperature variations, magnetotorquer currents, and other spacecraft effects \citep{Olsen:2020a}.  Data were averaged to 1 minute values using a robust linear fit in the magnetometer frame.

From the {\it Swarm} mission, we used the MAGX$\_$LR$\_$1B 1 Hz calibrated data product, baseline 0505/0506,  with an initial 1 minute sampling from the three satellites, Alpha, Bravo and Charlie from November 2013 to the end of August 2019.  In addition along-track gradient information was obtained from each satellite using differences between data 15~seconds apart, and East-West gradients were estimated from the lower pair Alpha and Charlie, using the 1 Hz data on Charlie with geocentric latitude closest to that of Alpha, sampled each minute, with the condition that the time difference was less than 50~sec. Field differences are less affected by correlated noise than the data themselves and their use has been shown to improve the quality of both core and lithospheric field models \citep{Olsen:2015}.  Finally, Swarm data were further down-sampled by a factor of three (i.e.\ effectively to a 3 minute sampling rate) to account for the fact that there were three satellites contributing data during this time interval. 

\begin{figure}[!ht]
\centerline{\includegraphics[angle=0, width=0.49\textwidth]{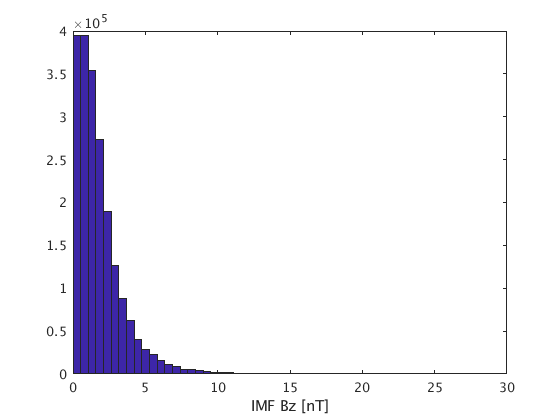}\includegraphics[angle=0, width=0.51\textwidth]{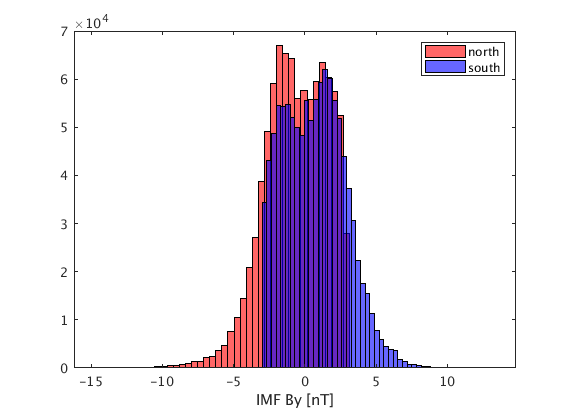}}
\centerline{\includegraphics[angle=0, width=0.6\textwidth]{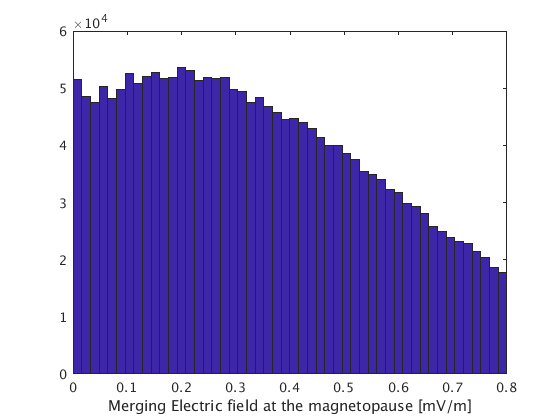}}

\caption{Histograms showing number of selected satellite data (combination of all the scalar, scalar gradient, vector and vector gradient data) distributed according to IMF $B_z$ (top left), IMF $B_y$ (top right) and the Merging Electric field at the Magnetopause $E_m$, as estimated by coupling function $0.33 v^{4/3} B_t^{2/3} \sin^{8/3} \left( |\Theta| / 2 \right)\ $~mV/m with $v$ the solar wind speed in km/s, $B_t=\sqrt{B_y^2 + B_z^2}$ the Interplanetary Magnetic Field magnitude in the $y-z$ plane in GSM coordinates in~nT, and  $\Theta = \arctan (B_y/B_z)$.  IMF and $Em$ values are averages of 1 minute values for 2 hours prior to the time of the observation. y-axis shows the number of observations per bin.
\label{fig:IMF}}
\end{figure}
The following data selection criteria were applied to all data sets in an effort to focus on the internal field of interest for IGRF.

\begin{itemize}
\item $Kp\leq 2^0$ ($3^0$ for gradients) and  $RC$-index  \citep{Olsen:2014a},  changing at most by 2~nT/hr (3~nT/hr for gradients)
\item Merging electric field at the magnetopause averaged over the previous 2 hrs, $E_m \leq 0.8$~mV/m
\item IMF $B_z$ at the magnetopause averaged over the previous 2 hrs is positive
\item IMF $B_y$ at the magnetopause averaged over the previous 2 hrs is less than +3~nT  when QD latitude is positive (northern QD hemisphere) i.e. $- \infty < B_y < 3 \mbox{nT}$, while when the QD latitude is negative (southern QD hemisphere) it is greater than -3nT i.e. $- 3 \mbox{nT} < B_y < \infty$ \citep{Friis-Christensen2017}
\item Only data from dark regions  (sun at least $10^{\circ}$ below  horizon), except for CryoSat-2 where calibration parameters and Euler angles are co-estimated using vector data from both dark and sunlit regions
\item Vector field data and vector field gradients used only equatorward of $\pm55^\circ$ QD latitude. 
\end{itemize}

\begin{figure}[!ht]
\includegraphics[angle=0, width=0.95\textwidth]{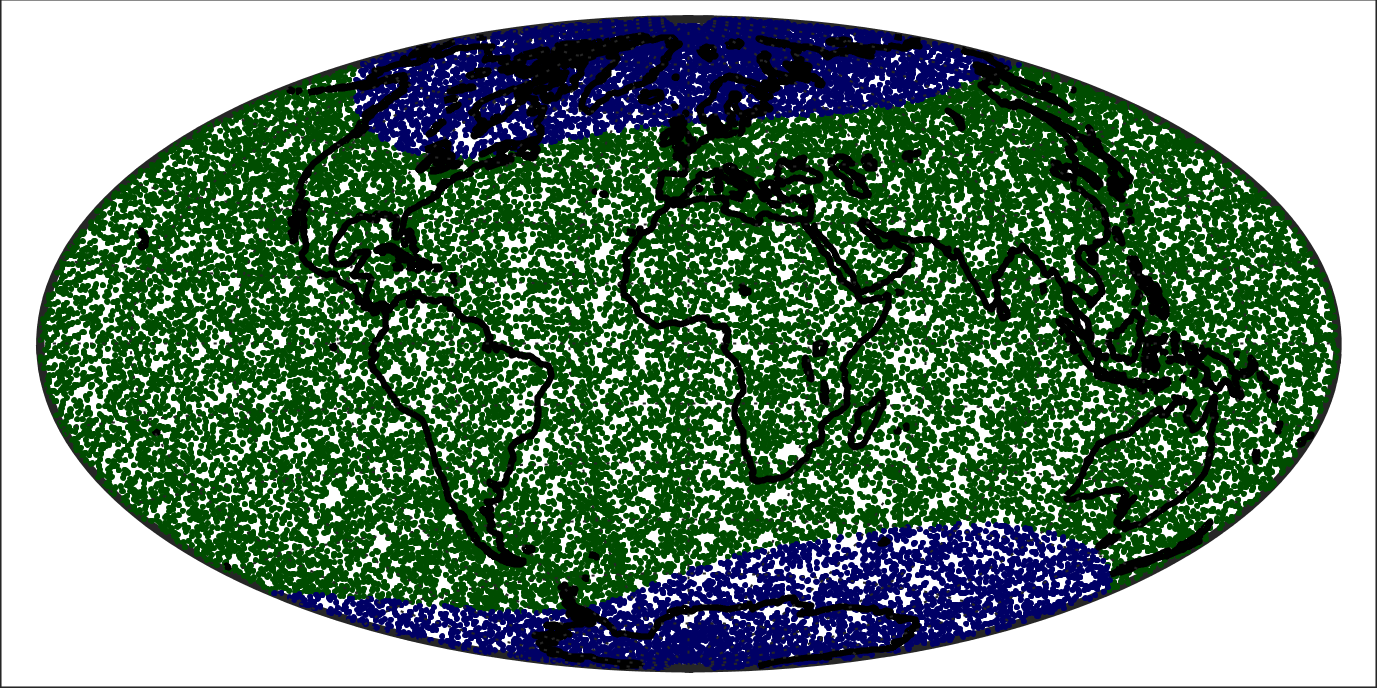}
\caption{Locations of vector and scalar data from the three {\it Swarm} satellites used in deriving CHAOS-7 during the interval September 2018 to September 2019.    Locations of vector data locations are shown as green dots,  loctions of scalar data as blue dots.  
\label{fig:swarm_loc_2019}
}
\end{figure}

Fig.~\ref{fig:IMF} presents histograms showing the distribution of IMF $B_z$, IMF $B_y$ and $E_m$ for the selected satellite data.  The distribution for $B_z$ peaks towards zero while the distributions for $B_y$ in the northern and southern hemispheres peak towards -3 and +3\,nT respectively.   Fig.~\ref{fig:swarm_loc_2019} shows the locations of the selected \textit{Swarm} scalar and vector data between September 2018 and September 2019.  This illustrates the excellent geographical coverage available within a year during the \textit{Swarm} era, despite the rather strict selection criteria applied. 

\begin{figure}[!ht]
\includegraphics[angle=0, width=\textwidth]{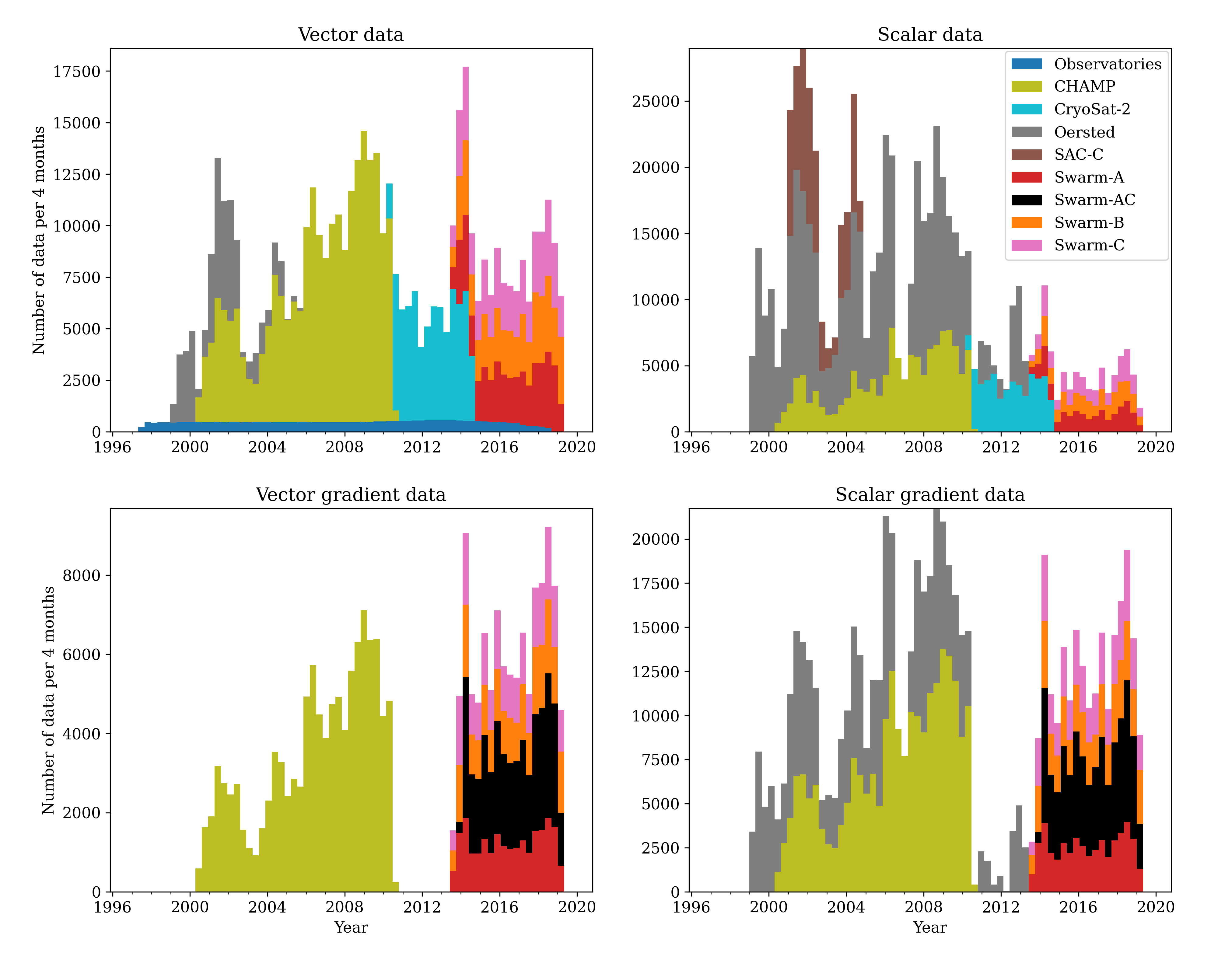}
\caption{Number of data used in CHAOS-7 versus time, collected into 4 monthly bins, colours represent the various contributing data sources. Swarm-AC denotes East-West field differences between  {\it Swarm}  Alpha and Charlie. 
\label{fig:chaos7_temp_dist}
}
\end{figure}

Fig.~\ref{fig:chaos7_temp_dist} shows the number of data contributing to CHAOS-7 as a function of time, separated according to the source.  scalar data is only used at high QD latitudes for \textit{Swarm} (see Fig.~\ref{fig:swarm_loc_2019}), CryoSat-2 and CHAMP.  The variations with time in the number of data reflect the availability of data, especially for the first decade, and variations with solar cycle of the amount of data  satisfying the selection criteria listed above.
\clearpage 

\subsection{{\bf Ground observatory data}}

\begin{figure}[!ht]
\includegraphics[angle=0, width=0.9\textwidth]{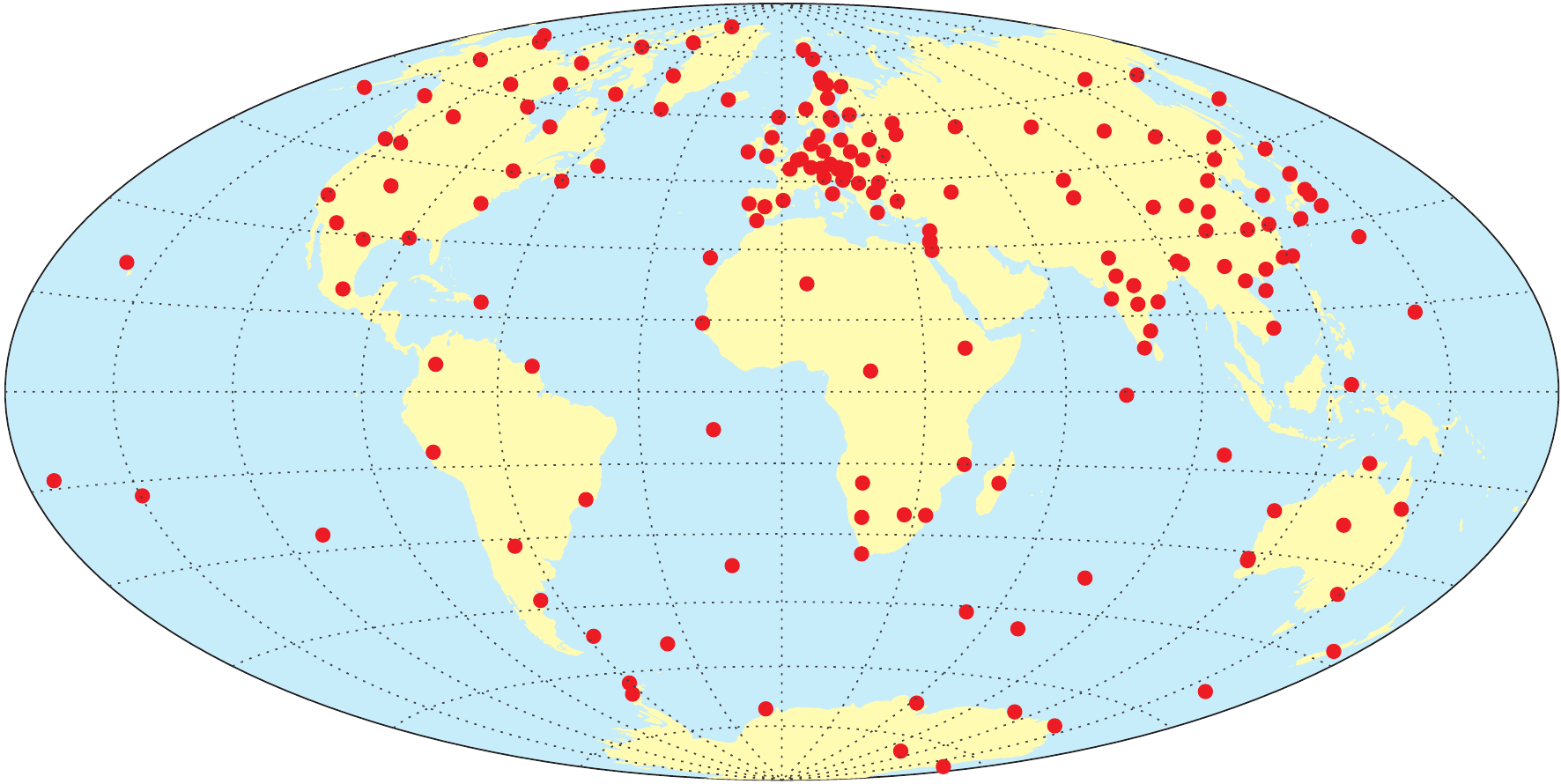}
\caption{Locations of ground magnetic observatories whose data is used in the derivation of CHAOS-7.  IAGA codes for the observatories are AAA, AAE, ABG, ABG, ABK, AIA, ALE, AMS, AMT, API, API, AQU, ARS, ASC, ASP, BDV, BEL, BFE, BFO, BGY, BJN, BLC, BMT, BNG, BOU, BOX, BRW, BSL, CBB, CBI, CDP, CKI, CLF, CMO, CNB, CNH, COI, CSY, CTA, CTS, CYG, CZT, DED, DLR, DLT, DOB, DOU, DRV, EBR, ELT, ESA, ESK, EYR, FCC, FRD, FRN, FUQ, FUR, GAN, GCK, GDH, GLM, GNA, GNG, GUA, GUI, GZH, HAD, HBK, HER, HLP, HON, HRB, HRN, HTY, HUA, HYB, IPM, IQA, IRT, IZN, JAI, JCO, KAK, KDU, KEP, KHB, KIR, KIV, KMH, KNY, KNZ, KOU, KSH, LER, LIV, LMM, LNP, LON, LOV, LRM, LRV, LVV, LYC, LZH, MAB, MAW, MBO, MCQ, MEA, MGD, MIZ, MMB, MNK, MOS, MZL, NAQ, NCK, NEW, NGK, NGP, NMP, NUR, NVS, OTT, PAF, PAG, PBQ, PEG, PET, PHU, PIL, PND, PPT, PST, QGZ, QIX, QSB, QZH, RES, SBA, SBL, SFS, SHL, SHU, SIL, SIT, SJG, SOD, SPT, SSH, STJ, SUA, TAM, TAN, TDC, TEO, TFS, THJ, THL, THY, TIR, TND, TRO, TRW, TSU, TUC, UJJ, UPS, VAL, VIC, VNA, VOS, VSK, VSS, WHN, WIC, WIK, WNG, YAK, YKC.  
\label{fig:obs}
}
\end{figure}

Annual differences of revised monthly means of ground observatory data \citep{Olsen:2014a}  for the time interval January 1997 to July 2019 were utilized as a further source of information on the core field secular variation. Revised monthly means were derived from hourly mean values of 182 observatories, see Fig. \ref{fig:obs},  (including 11 with site changes during the considered time interval) which were checked for trends, spikes and other errors \citep{MacMillanEtAl:2013}.  Monthly means were calculated using a robust method relying on Huber weights \citep{Huber:2004}, from all local times after the removal of hourly estimates of the ionospheric (plus induced) field predicted using the CM4 model \citep{Sabaka:2004} and hourly estimates of the large-scale magnetospheric (plus induced) field, predicted by the CHAOS-6x9 model. \\

\newpage

\section{Field modelling}
\label{sec:Model}
\subsection{{\bf Model parameterization}}
The basic parametrization of the CHAOS-7 model is the same as that of earlier versions in the CHAOS series, with some minor extensions which we describe below.  We assume measurements are collected in a region free from electric currents so, under the quasi-static approximation of electromagnetism, the vector magnetic field ${\bf B}$ can be represented by a scalar potential such that ${\bf B}=-\nabla V$.  The magnetic scalar potential $V=V^\mathrm{int}+V^\mathrm{ext}$ consists of  internal (primarily core and lithospheric) sources, and external (assumed here to be magnetospheric) sources and their internal Earth-induced counterparts.  Both the internal and external parts are expanded in spherical harmonics (SH).  

\subsubsection{Internal potential fields}
\label{sect:Int_sources}
For the internal field, in a geographic Earth-Centered Earth-Fixed (ECEF) coordinate system we adopt a spherical harmonic expansion
\begin{equation}
V^\mathrm{int} =a \sum_{n=1}^{N_\mathrm{int}}\sum_{m=0}^{n}\left( g_{n}^{m}\cos m\phi
+h_{n}^{m}\sin m\phi \right) \left( \frac{a}{r}\right) ^{n+1}P_{n}^{m}\left(
\cos \theta \right)
\label{eq:V-int}
\end{equation}
where $a=6371.2$~km is chosen as the Earth's spherical reference radius, $\left( r,\theta,\phi
\right) $ are spherical polar coordinates, $P_{n}^{m}$ are the 
Schmidt semi-normalized associated Legendre functions \citep{Winch:2005}, $\left\{g_{n}^{m},h_{n}^{m}\right\} $ are the Gauss coefficients
describing internal sources, and $N_\mathrm{int}$ is the maximum
degree and order of the internal expansion.  The internal coefficients $\{g_n^m(t),h_n^m(t)\}$ up to $n=20$ are time-dependent; this dependence is  represented using a basis of order 6 B-splines \citep{deBoor:2001} with a 6-month knot separation and five-fold knots at the endpoints $t=1997.1$ and $t=2020.1$. Internal coefficients for degrees 21 and above are static, a maximum degree of $N_{int}=70$ was used for the parent model estimated here.  For the distributed versions of the CHAOS-7 model, at degree 25 we merge the static field estimated here to the high resolution LCS-1 lithospheric field model \citep{Olsen:2017} which is provided out to degree 185.

\subsubsection{External potential fields}
\label{sect:Ext_sources}

Turning to the external part of the potential, we adopt an expansion in the {\it Solar Magnetic (SM)} coordinate system  (up to $n=2$, with specific treatment of the $n=1$ terms, see below) of the near magnetospheric sources  and in the {\it Geocentric Solar Magnetospheric (GSM)} coordinate system (also up to $n=2$, but restricted to order $m=0$) of remote magnetospheric sources, (e.g., magnetotail and magnetopause currents)
\begin{subequations}
\begin{align}
        V^\mathrm{ext} &= a\sum_{m=0}^{1}\big[q_1^{m,\mathrm{SM}} (t) \cos mT_d + s_1^{m,\mathrm{SM}} (t) \sin mT_d\big]\left(\frac{r}{a}\right) P^m_1(\cos \theta_d\big) \label{eq:Vnm_ext:a}\\
        &+ a\sum_{m=0}^{1}\big[\Delta q_1^{m,\mathrm{SM}}(t)R_{1,c}^{m,\mathrm{SM}}(t, r, \theta, \phi) + \Delta s_1^{m,\mathrm{SM}}(t)R_{1,s}^{m,\mathrm{SM}}(t, r, \theta, \phi)\big] \label{eq:Vnm_ext:b}\\
        &+ a\sum_{m=0}^{2}\big[q_2^{m,\mathrm{SM}}R_{2,c}^{m,\mathrm{SM}}(t, r, \theta, \phi) + s_2^{m,\mathrm{SM}}R_{2,s}^{m,\mathrm{SM}}(t, r, \theta, \phi)\big] \label{eq:Vnm_ext:c}\\
        &+ a\sum_{n=1}^{2}q_n^{0, \mathrm{GSM}}R_n^{0,\mathrm{GSM}}(t, r, \theta, \phi)
        \label{eq:Vnm_ext:d}
\end{align}
\end{subequations}
where $\theta_d$ and $T_d$ are respectively dipole colatitude and dipole local time, where the latter is expressed in units of radians, and $R_{n,c/s}^{m, \mathrm{GSM}}$ and $R_{n,c/s}^{m, \mathrm{SM}}$ are modifications of the solid harmonics (spherical harmonics with the well known radial scalings) in $\mathit{SM}$ and $\mathit{GSM}$ coordinate frames taking account of the induced field based on the diagonal part of the $Q$-matrix \citep{Olsen:1999b} for an assumed 3D Earth conductivity model \--- see below for more details. The degree 1 SM terms have the specific time dependence
\begin{equation}
    \begin{aligned}
        q_1^{0,\mathrm{SM}} (t) &= \hat{q}_1^0 \left[\epsilon(t)+\iota(t)\left(\frac{a}{r}\right)^3\right]\\
        q_1^{1,\mathrm{SM}} (t) &= \hat{q}_1^1 \left[\epsilon(t)+\iota(t)\left(\frac{a}{r}\right)^3\right]\\
        s_1^{1,\mathrm{SM}} (t) &= \hat{s}_1^1 \left[\epsilon(t)+\iota(t)\left(\frac{a}{r}\right)^3\right]
    \end{aligned}
    \label{eq:qnm_scaling}
\end{equation}
where the terms in brackets are designed to describe the magnetic field contribution due to the magnetospheric ring-current and its Earth-induced counterpart.  These are prescribed using the RC index which is derived from ground observatory hourly means \citep{Olsen:2014a}, $RC(t)=\epsilon(t) + \iota(t)$ (see section~\ref{sect:induction} below for further details on the treatment of induced fields).  We estimate the static regression factors $\hat{q}_1^0, \hat{q}_1^1, \hat{s}_1^1$ and the time-varying ``RC baseline corrections'' $\Delta q_1^0, \Delta q_1^1$ and $\Delta s_1^1$ in bins of 30~days. These allow for differences between the ground-based estimate of the degree 1 order 0 external magnetic signal (the RC index) and the degree 1 field seen by low-Earth orbit satellites. 

It should be remembered that the CHAOS-7 magnetospheric field model is designed to represent the field during geomagnetically quiet times that were considered during the model construction.  The magnetospheric field is known to have a more complex structure, in particular related to its local-time dependence, for higher levels of geomagnetic disturbance, as measured for example by the $Kp$ index \cite[see e.g.][]{Luehr:2020}.

\subsubsection{Treatment of induced fields}
\label{sect:induction}

In this section we describe in more detail how magnetic fields induced in the electrically-conducting oceans and interior of the Earth, due to time-changing magnetospheric fields, are represented in CHAOS-7.

In CHAOS-7 induced fields are not parameterized separately in the form of additional internal sources since these are known to be difficult to separate from core field variations.  Instead we consider induced fields calculated based on an assumed Earth-conductivity model, via the $Q$-responses which couple in the time domain internal (induced) and external Gauss coefficients \cite[e.g.][]{Schmucker:1985b, Price:1967}. Generally, for a 3-D Earth conductivity distribution $\sigma(r, \theta, \phi)$, each external coefficient induces infinitely many internal coefficients. In the frequency domain, the relation between external and induced internal coefficients for a given angular frequency $\omega$ reads \citep{Olsen:1999b}
\begin{equation}
\label{eq:qmatrix}
\tilde{\iota}_k^l(\omega; \sigma) = \sum\limits_{n = 1}^{N_{\textnormal{ext}}}\sum_{m = -n}^{n}\tilde{Q}_{kn}^{lm}(\omega; \sigma)\tilde{\varepsilon}_n^m(\omega).
\end{equation}
Here, $\tilde{Q}_{kn}^{lm}$ are transfer functions, that can be arranged into the so-called $Q$-matrix, given by
\begin{equation}
\label{eq:qnmkl}
\tilde{Q}_{kn}^{lm}(\omega; \sigma) = \frac{1}{(k + 1)\| Y_k^l \|^2}\int_S \left( \tilde{B}_{n, r}^m (\vec{r}_a, \omega; \sigma) - B_{n,r}^{m,ext}(\vec{r}_a) \right)Y_k^{l*}(\theta, \phi)\textnormal{d}S,
\end{equation}
where $Y_k^l(\theta, \phi) = P_k^{|l|}(\cos{\theta})\exp{(\textnormal{i}l\phi)}$ is a SH of degree $k$ and order $l$, $^*$ denotes complex conjugation, $\textnormal{d}S$ is an elementary spherical surface area, $\vec{r}_a = (a, \theta, \phi)$ is the position vector at the Earth's surface. $B_{n, r}^m$ is radial magnetic field which is (numerically) computed for a given Earth conductivity model driven by a unit amplitude ($\tilde{\varepsilon}$ = 1) SH source, and 
\begin{equation}
\label{Br_ext}
B_{n,r}^{m,ext}(\vec{r}_a) = -n Y_n^m(\theta, \phi)
\end{equation}
is the corresponding external part of the radial magnetic field. $B_{n, r}^m$ has in this study been numerically computed using a finite element code \citep{dealII92, Grayver:2015, Grayver:2019}.
Formulae for the unit amplitude SH source can be found, for example, in  \cite{Guzavina:2018}. Substituting eq.~(\ref{Br_ext}) into~(\ref{eq:qnmkl}) yields 
\begin{equation}
\label{eq:qnmkl_new}
\tilde{Q}_{kn}^{lm}(\omega; \sigma) = \frac{1}{(k + 1)\| Y_k^l \|^2}\int_S \tilde{B}_{n, r}^m (\vec{r}_a, \omega; \sigma) Y_k^{l*}(\theta, \phi) \, \textnormal{d}S + \frac{n}{n+1}.
\end{equation}

The $Q$-matrix used in CHAOS-7 to couple induced and external fields is derived from an electrical conductivity model consisting of a mantle with 1-D (radial) conductivity distribution overlaid by a surface layer of laterally-variable conductance. The latter approximates heterogeneous oceans and continents. In order to retain the computational advantages of a 1-D approach, we here take only the diagonal part of the $Q$-matrix as an approximation, since the off-diagonal elements of the $Q$-matrix are generally much smaller \citep{Puethe:2014}. 

We use the resulting diagonal $Q$-matrix in both the frequency domain and the time domain to compute the induced counterparts to different parts of our magnetospheric field model, these are discussed separately below. The conversion of eq. (\ref{eq:qmatrix}) into the time domain yields a convolution integral \citep{Maus:2004a, Olsen:2005c, Grayver:2020} which couples induced and external coefficients via impulse responses of the corresponding transfer functions in the frequency domain (given by the $Q$-matrix). For example, for a diagonal $Q$-matrix and degree 1 zonal external field, the external and internal coefficients in time domain are related through a convolution integral as
\begin{equation}
\label{eq:qresponse_td}
\iota_1^0(t; \sigma) = \int_{-\infty}^t Q_{11}^{00}(t - \tau; \sigma) \epsilon_1^0(\tau) \textnormal{d}\tau,
\end{equation}
where $-\infty$ limit is replaced by a finite value of 1 year in practice. 

\begin{figure}[!ht]
\centerline{\includegraphics[angle=0, width=\textwidth]{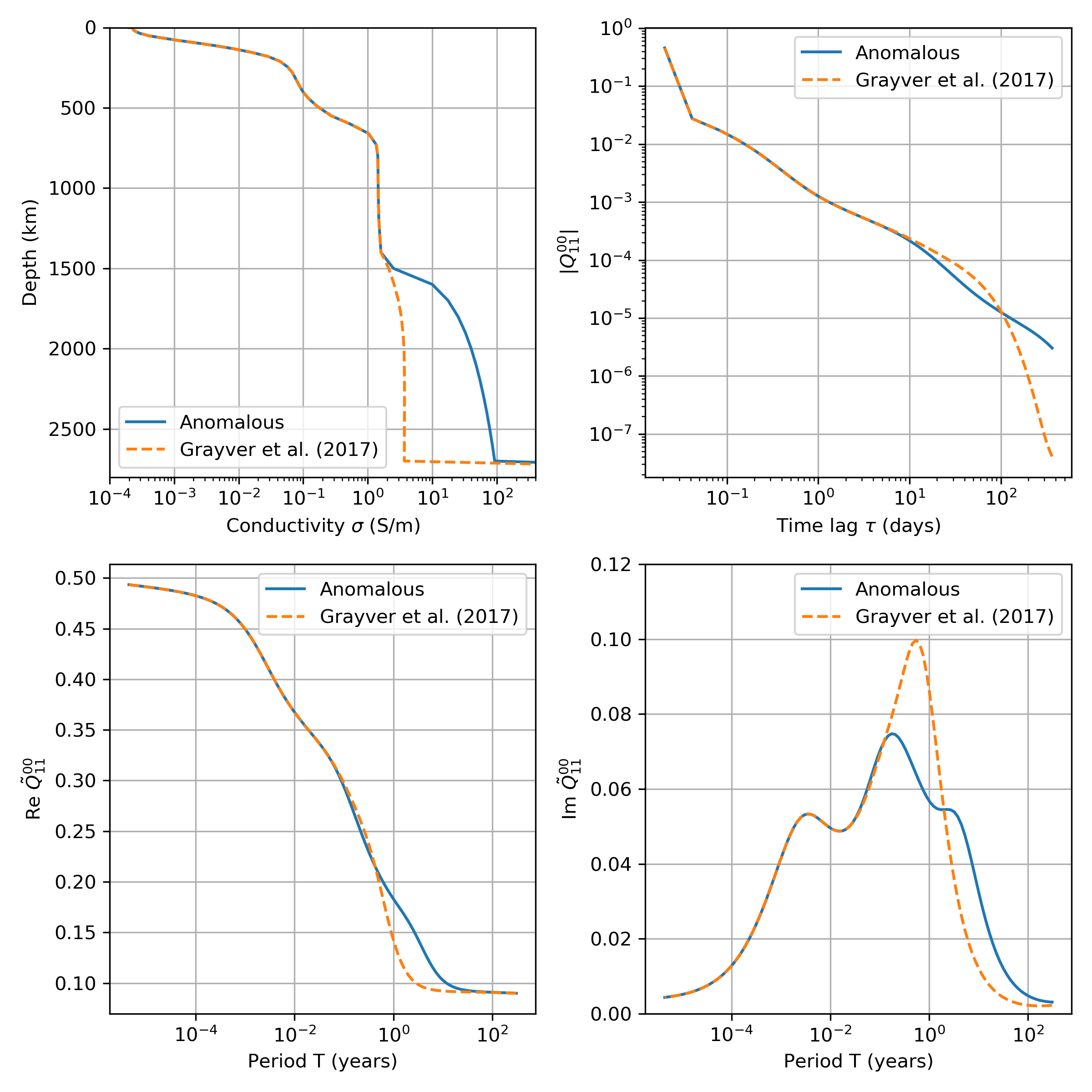} }
\caption{Top row: 1-D best fit conductivity profile (left) from \cite{Grayver:2017} used in this study (red dashes, labelled "Grayver et al. (2017)") and a hypothetical case with increased conductivity in the lower mantle (blue line, labelled "anomalous"). The Earth's core, at depths below 2850 km, is assumed to have a finite conductivity of $10^6$ S/m, which is off the edge of the plot. The magnitude of the corresponding discrete impulse response  (right) for the $\tilde{Q}_{11}^{00}$ transfer function. Bottom row: $\tilde{Q}_{11}^{00}$ transfer function in frequency domain, real part (left) and imaginary part (right).
\label{fig:Induction}}
\end{figure}

The 1-D conductivity model used in CHAOS-7 and the related $Q$-kernels are presented in Fig.~\ref{fig:Induction}, along with a similar profile designed to illustrate a hypothetical global conductivity anomaly in the lowermost mantle. This anomalous case is unlikely for the entire lower mantle \citep{Karato:2013}, but even in this case the real part of the $\tilde{Q}_{11}^{00}$ is still relatively small at the frequencies overlapping with core field secular variation, less than 0.2 at periods of 1 year and longer. 

The induced fields resulting from time variations in the degree 1 external field in SM coordinates, are calculated in the time domain from the RC index \citep{Olsen:2014a} as follows. The input $RC(t)$, which contains both external and induced parts $RC(t)=\epsilon(t) + \iota(t)$, is first detrended, then convolved with a time-domain IIR filter \citep{Maus:2004a, Olsen:2005c} derived from the diagonal elements of the $Q$-matrix, rotated into geomagnetic coordinates as appropriate for working with RC(t), and based on a window of length 1 year.  In performing the convolution we truncate the IIR filter to a length 1 year, so the response to signals older than a year are neglected, although all frequencies are present within the window are considered.  $\epsilon(t)$ is then obtained by subtracting  $\iota(t)$ from the original $RC(t)$.  This is done prior to deriving the CHAOS-7 model in which $\epsilon(t)$ and $\iota(t)$ parameterize the time dependence of the degree 1 SM external and induced counterparts, only the parameters describing static regression  (eq.~\ref{eq:qnm_scaling}) and offset parameters (eqs.~\ref{eq:Vnm_ext:a} --~\ref{eq:Vnm_ext:d})  are solved for during the model estimation.

Induced fields are also accounted for in CHAOS-7 due to the static fields in SM and GSM coordinates which are time-dependent in the Earth-Centered-Earth-Fixed (ECEF) frame, due to the "wobble" between the frames (depending on the solar position, season etc.).  A Fourier decomposition of the time dependence resulting from each SM and GSM coefficient in the ECEF frame is carried out, then the diagonal $Q$-matrix described above, based on the same conductivity model of \cite{Grayver:2017}, is used in the frequency domain to determine the amplitude of the induced internal response to a unit excitation.  Collecting these responses for all frequencies (here we used uniformly sampled frequencies with periods between 1~hour and 4 years, there was little power at periods beyond 4 years) and inverse Fourier transforming provided a corresponding time-dependent induced field that scales linearly with the amplitude of the static coefficients.

To summarize, in CHAOS-7 we have made an effort to account in a consistent fashion for the induced response of all the parameterized magnetospheric sources, with the exception of the time dependence of the offset parameters for SM degree 1 (eq.~\ref{eq:Vnm_ext:a}).  In CHAOS-7 these offsets show variations of up to 2\,nT on timescales close to 1~year and up to 4~nT on a timescales of around  10~years.  Given the real part of the $Q$-response at these periods are likely to be less than 0.2 and 0.1 respectively (see Fig.~\ref{fig:Induction}) the corresponding induced responses are expected to have amplitude less than 0.5\,nT.

Another important driver of induced field variations is ionospheric sources, for example the Sq current system.  In CHAOS-7 we use satellite data only from the night side and do not explicitly model this source or its induced response.  It is therefore possible that we map the night-side induced response into our internal field model \citep{Olsen:2005d}.  This signal is expected to leak predominantly into the zonal terms, particularly the coefficients $g^0_1$ and $g^0_3$.  One way we can assess the magnitude of this effect is through comparisons with other models which do seek to model this process.  Of particular relevance here is the CM6 model \citep{Sabaka:2020}.  Comparisons of $g^0_1(t)$ and $g^0_3(t)$ from CHAOS-7 and CM6 show rms differences between 2000 and 2019 of 2.25\,nT and 1.27\,nT respectively, with largest differences in the years 2002 and 2014.  The SV in the two models generally agrees fairly well for the low degree secular variation with rms differences in $dg^0_1/dt$ and $dg^0_3/dt$ of 0.71\,nT/yr and 0.20\,nT/yr respectively. Differences in the high degree SV between CHAOS-7 and CM6 are primarily due to differences in their core field temporal regularization schemes. Other models, such that recently described by \cite{Ropp:2020}, which seek to directly co-estimate the induced field arrive at rather different results regarding for example $dg^0_1/dt$ (see Sec.\ref{Sect:SV_comp}).  Further work is needed on this topic.  

\subsubsection{Magnetometer alignment and in-flight calibration parameters}
\label{sect:Euler_cal}
In addition to the above spherical harmonic representation of internal and external potential fields, we co-estimate Euler angles describing the rotation between the vector magnetometer frame and the star tracker frame for {\O}rsted, CHAMP, CryoSat-2 and the three \textit{Swarm} satellites. For {\O}rsted, for historical reasons, we employed two sets of constant Euler angles, implementing a break point at 00.00 on 25th January 2000.  This takes account of an update of the onboard software of the star tracker that took place during 25th January 2000 at 04.05:26  \citep[see also][but note the date was incorrectly given there as 22nd January 2000]{Olsen:2002b}.  For CHAMP, CryoSat-2 and each {\it Swarm} satellite we solve for Euler angles in bins of 10 days. 

In order to use the uncalibrated CryoSat-2 data we co-estimate 9 standard calibration parameters: 3 scale factors, 3 non-orthogonality angles and 3 offsets, \cite[see, for example,][]{Olsen:2003a},  in a series of bins of length 21~days.  In each bin, these parameters relate the measured vector field in engineering units $\mathbf{E}$ to the calibrated magnetic field $ \mathbf{B}$ in \mbox{units of nT} as follows 
\begin{equation}
        \mathbf{B} = \mathbf{\underline{\underline{P}}}^{-1}\mathbf{\underline{\underline{S}}}^{-1}(\mathbf{E}-\mathbf{b}),
\end{equation}
where the matrix describing the non-orthogonalities is 
        \begin{equation}
        \mathbf{\underline{\underline{P}}}=
        \begin{pmatrix}
        1 & 0 & 0 \\ 
        -\sin u_1 & \cos u_1 & 0 \\ 
        \sin u_2 & \sin u_3 & \sqrt{1-\sin^2 u_2-\sin^2 u_3}
        \end{pmatrix},
        \end{equation}
 that describing the scale factors for the magnetometer sensors in the three directions  is
        \begin{equation}
        \mathbf{\underline{\underline{S}}}=
        \begin{pmatrix}
        S_1 & 0 & 0 \\ 
        0 & S_2 & 0 \\ 
        0 & 0 & S_3
        \end{pmatrix},
        \end{equation}
while the vector containing the offsets is
        \begin{equation}
        \mathbf{b}=
        \begin{pmatrix}
        b_1 \\ 
        b_2 \\ 
        b_3
        \end{pmatrix}.
        \end{equation}
Further details can be found in \cite{Olsen:2020a} and \cite{Kloss:2020}.
        
\subsubsection{Summary of parameters defining the model setup}
\label{sect:params}

Details of the chosen SH truncation levels and of the temporal parameterization of the various parts of the model are summarized in table~\ref{table:model_params}.   In all the model consists of 31757 parameters that are simultaneously estimated from 4007404 magnetic field observations (counting each and gradient vector component separately).

\begin{table*}[!ht]
	\caption{Summary of parameters defining the model setup in CHAOS-7.
		\label{table:model_params}}
	\vspace{2truemm}
	{\scriptsize
\begin{tabular}{c|lr|c}
	                                 Setup  Parameter &   Description &   &   \\ 
	                                          \hline
$N_\text{tdep}$           &  Maximum SH degree of time-dependent internal field & & 20   \\ 
$J$       &  Order of B-Splines  & &6   \\
$\Delta t_k$ & B-spline knot spacing & & 0.5 yr \\ 
$t_\text{start}$ & Start time of spline basis & & 1997.1 \\ 
$t_\text{end}$ & End time of spline basis & & 2020.1 \\ 
$N_\text{int}$       &  Maximum SH degree of static internal field  & &70   \\ 
$N_\text{SM}$       &  Maximum SH degree of SM external field  & &2   \\ 
$\Delta T_\text{SM1}$ & Bin size for degree 1 SM offsets   &&   30 days \\
$N_\text{GSM}$       &  Maximum SH degree of GSM external field  & &2 (only $m=0$ terms)   \\ 
$\Delta T_\text{Euler}$ & Bin size for Euler angle determination   &&   10 days \\
$\Delta T_\text{CAL}$ & Bin size for calibration parameters  &&   21 days \\
\hline
\end{tabular} 
}
\end{table*}

\subsection{{\bf Model estimation}}
\label{sec:estim}

The model parameters $\mathbf{m} = [\mathbf{p}^\mathrm{T}, \mathbf{q}^\mathrm{T}, \mathbf{e}^\mathrm{T}]^\mathrm{T}$, where $\mathbf{p}$ represents the spherical harmonic coefficients comprising the field model, $\mathbf{q}$ are the Euler angles and $\mathbf{e}$ is a vector of the calibration parameters, are determined by iteratively minimizing the following cost function using a Newton-type algorithm
\begin{equation}
        \Phi(\mathbf{m}) = \big[\mathbf{g}(\mathbf{p})-\mathbf{d}(\mathbf{q}, \mathbf{e})\big]^\mathrm{T}\underline{\underline{\mathbf{C}}}_d^{-1}\big[\mathbf{g}(\mathbf{p})-\mathbf{d}(\mathbf{q}, \mathbf{e})\big] 
        + \mathbf{m}^\mathrm{T} \underline{\underline{\Lambda}} \mathbf{m}
\end{equation}
where $\mathbf{g}(\mathbf{p})$ are model predictions based on field model coefficients, $\mathbf{d}(\mathbf{q}, \mathbf{e})$ are the data, rotated to the geocentric frame using the model Euler angles  $\mathbf{q}$ and calibrated (relevant only for CryoSat-2) using the model calibration parameters $\mathbf{e}$ \citep{Kloss:2020}.  $\underline{\underline{\mathbf{C}}}_d$ is the data error covariance matrix constructed as in previous versions of the CHAOS model series based on a-priori data error estimates for each satellite, with the vector error estimates specified  in the frame of the star tracker which allows the allocation of anisotropic pointing errors. Details of the regularisation matrix $\underline{\underline{\Lambda}}$ are given below. Additional data weights proportional to $\sin \theta$ were implemented for the satellite data in order to approximate an equal area data distribution. Huber data weights \citep{Huber:2004,Constable:1988a} were calculated after each iteration and used to re-weight the data; this enables robust estimation in the presence of long-tailed error distributions.  Data error estimates for the ground observatory SV data were derived from the residuals to a previous model CHAOS-6x9, after detrending and taking account of data error covariances between the components of the vector triples.

 In order to calibrate the CryoSat-2 magnetometer data we include data from both sunlit and dark conditions, but only the dark data contribute to the determination of the spherical harmonic coefficients of the field model.  A vector calibration is carried out at mid and low latitudes and a scalar calibration at polar latitudes \citep{Olsen:2003a,Olsen:2020a}.  More details about our approach to co-estimate calibration parameters will appear in a forthcoming study \citet{Kloss:2020}.  
  
 Since scalar data are included, and because Euler angles and calibration parameters are co-estimated, the relation between the model parameters and the data is nonlinear.  The cost function above is therefore iteratively minimized using a Newton-type descent method, with the Huber data weights updated at each step.  The starting model was chosen to be a static internal field from a previous field model, CHAOS-6x8, evaluated in May 2015.   The external field was initialized to zero.  The Euler angles were initialized to the values determined in pre-flight tests, implemented via a pre-rotation step. The calibration offset and non-orthogonality parameters for CryoSat-2 were initialized to zero, while the scale factors were initialized to values of 1.   Nine iterations from this starting model were carried out by which stage we judged that the model had converged to a satisfactory level; the maximum percentage change in a model parameter during the final iteration was 0.1425~\%.  There was no noticeable change in predictions of the internal field at Earth's surface (i.e.\ the IGRF relevant part of the model) during the final three iterations.

$ \underline{\underline{\Lambda}}$ is a block diagonal temporal regularization matrix which was derived by adding contributing sub-matrices, each of which implements a quadratic measure of the temporal complexity of a certain aspect of the model.  These are  $\underline{\underline{\Lambda}}_{i3}$, which implements a quadratic measure of the 3rd time derivative of the internal radial field integrated over the core surface and throughout the model timespan, $\underline{\underline{\Lambda}}_{i2e}$ which implements a quadratic measure of the second time derivative of the internal radial field integrated over the core surface but only at the model endpoints $t_{start}=1997.1$ and $t_{end}=2020.1$,  $\underline{\underline{\Lambda}}_{sm}$ which implements a quadratic measure of the time derivative (approximated by bin-to-bin differences) of the offset terms in the SM expansion of the magnetospheric field at Earth's surface integrated throughout the timespan, and  $\underline{\underline{\Lambda}}_{cs}$, $\underline{\underline{\Lambda}}_{cu}$, $\underline{\underline{\Lambda}}_{cb}$ implementing quadratic measures of the time derivative (again using bin-to-bin differences) of the CryoSat-2 calibration scale factors, non-orthogonality angles and offsets, respectively.  Each of these temporal regularization sub-matrices are scaled by regularization parameters, denoted by $\lambda_{i3}$, $\lambda_{i2e}$, $\lambda_{sm}$, $\lambda_{cs}$, $\lambda_{cu}$, $\lambda_{cb}$.

There is a special treatment for $\lambda_{i3}$, which we allow to vary with the spherical harmonic degree and order $(n,m)$.  As was already the case in CHAOS-5 and 6, the zonal ($m=0$) terms are regularized more strongly than the non-zonal terms; in CHAOS-7  $\lambda_{i3}(n,0) = 10 \lambda_{i3}(n, m>0)$.  A summary of the regularization parameters used in CHAOS-7 is given in Table \ref{tab:reg_params} Preliminary test models showed that the large value of $ \lambda_{i3}$ required to ensure stability at low degree (where there is leakage from external fields) results in strongly suppressed time dependence of the higher degree coefficients. The inability to retrieve information concerning the time-dependence of small length scales is particularly disappointing during the past six years when high quality \textit{Swarm} data is available.  In order to relax the temporal regularization at higher degrees a degree dependence of $\lambda_{i3}$ was implemented. It takes its largest value $\lambda_{i3}(n_{low}, m)$ at low degree, $n_{low} < n_\text{tpmin}$,  then gradually reduces, eventually by a factor $5\cdot10^{-3}$ by degree $n_\text{tpmax}$. We set $n_{tpmin}=3$ and $n_{tmax}=11$ and implemented the reduction with degree using a Tukey cosine taper
\begin{equation}
    \lambda_{i3}(n,m) = \lambda_{i3}(n_{low},m)
    \left\{
    \begin{aligned}
        & 1, \quad n < n_{tpmin}\\
        & \tau(n) , \quad n_{tpmin} \leq n \leq n_{tpmax} \\
        & 0.005, \quad n > n_{tpmax}
    \end{aligned}
    \right.
    \label{eqn:taper}
\end{equation}
with
\begin{equation}
    \tau(n) = \frac{0.995}{2} \left[ 1 + \cos\pi \left( \frac{n - n_{tpmin}}{n_{tpmax} - n_{tpmin}} \right) \right] + 0.005.
\end{equation}

\begin{table*}[!ht]
	\caption{Choice of regularization parameters in CHAOS-7.
		\label{tab:reg_params}}
	\vspace{2truemm}
	{\scriptsize
\begin{tabular}{c|ccc}
	                                 Regularization  Parameter &   Value      \\ 
	                                          \hline
$\lambda_{i3} (n_{low},m > 0) $           &   1  (nT\, days$^{-3})^{-2}$  \\ 
$\lambda_{i3} (n_{low},m=0) $           &   10 (nT\, days$^{-3})^{-2}$  \\ 
$\lambda_{i3}(n_{high}, m > 0) $           &   0.005   (nT\, days$^{-3})^{-2}$ \\ 
$\lambda_{i3}(n_{high}, m = 0) $           &   0.05   (nT\, days$^{-3})^{-2}$ \\ 
$\lambda_{i2e}$       &  100  (nT\, days$^{-2})^{-2}$ \\ 
$\lambda_{sm}$      &   900 000  (nT\, days$^{-1})^{-2}$ \\ 
$\lambda_{cs}$      &   441  ((eu/nT)\,days$^{-1})^{-2}$ \\ 
$\lambda_{cu}$      &  4410  (arcsec\,days$^{-1})^{-2}$ \\ 
$\lambda_{cb}$      &  4.41   (eu\, days$^{-1})^{-2}$\\ 

\hline 
\end{tabular} 
}
\end{table*}


\section{Results}

\label{sec:Results}
\subsection{{\bf Fit to satellite data}}
\label{sec:sat_fit}
We begin reporting results by presenting the fit of CHAOS-7 to its dominant contributing data source, the satellite data.  Tables~\ref{tab:statistics_oer} to~\ref{tab:statistics_swarm} collect the Huber weighted means and rms residuals for different categories of data for each satellite. Overall, CHAOS-7 provides a satisfactory fit to the contributing satellite data, with no evidence for large biases and with residual histograms compatible with the assumed long-tailed error distributions.\\

\begin{table*}[!ht]
	\caption{Model statistics of misfit between CHAOS-7 and {\O}rsted data.  Mean and rms refer to Huber weighted mean and rms values in units of~nT.  $\delta F_\mathrm{AT}$ denotes along-track field differences calculated at 15~sec spacing.
		\label{tab:statistics_oer}}
	\vspace{2truemm}
{\scriptsize
\begin{tabular}{r|rrr}
	                                          & \multicolumn{3}{c}{{\O}rsted}  \\
	                                          &    $N$ &  mean &      rms  \\ 
	                                          \hline
$F_\mathrm{polar}$ [nT]           &   134139  &   0.92 &     3.02  \\ 
$F_\mathrm{non-polar}$ [nT]       &   261614  &   0.53 &     1.93 \\ 
$B_r$ [nT]                        &   47841  &   0.01 &     4.04\\ 
$B_\theta$ [nT]                   &   47841  &    -0.07 &     4.73 \\ 
$B_\phi$ [nT]                     &   47841  &    0.05 &     4.80 \\ 
$\delta F_\mathrm{AT, polar}$ [nT]     &    68097  &   -0.00 &     0.35\\ 
$\delta F_\mathrm{AT, non-polar}$ [nT] &   142801  &    0.00 &     0.19\\ 
\hline
\end{tabular} 
}
\end{table*}

\begin{table*}[!ht]
	\caption{Model statistics of misfit between CHAOS-7 and SAC-C data.  Mean and rms refer to Huber weighted mean and rms values in units of~nT.
		\label{tab:statistics_sacc}}
	\vspace{2truemm}
	{\scriptsize
\begin{tabular}{r|rrr}
	                                          & \multicolumn{3}{c}{SAC-C}  \\
	                                          &    $N$ &  mean &      rms  \\ 
	                                          \hline
$F_\mathrm{polar}$ [nT]           &   26711  &   0.10 &     3.49  \\ 
$F_\mathrm{non-polar}$ [nT]       &   48804  &   0.18 &     2.43 \\ 
\hline
\end{tabular} 
}
\end{table*}

\begin{table*}[!ht]
	\caption{Model statistics of misfit between CHAOS-7 and CHAMP data.  Mean and rms refer to Huber weighted mean and rms values in units of~nT. $\delta F_\mathrm{AT}$  and $\delta B_\mathrm{AT}$ denote along-track field differences calculated at 15~sec spacing.
		\label{tab:statistics_CHAMP}}
	\vspace{2truemm}	
{\scriptsize
\begin{tabular}{r|rrr}
	                                          & \multicolumn{3}{c}{CHAMP}  \\
	                                          &    $N$ &  mean &      rms  \\ 
	                                          \hline
$F_\mathrm{polar}$ [nT]           &   127529 &   -0.86 &     4.26  \\ 
$F_\mathrm{non-polar}$ [nT]       &   223744  &   -0.53 &     1.85 \\ 
$B_r$ [nT]                        &   223744  &   0.05 &     1.80\\ 
$B_\theta$ [nT]                   &   223744  &    0.29 &     2.48 \\ 
$B_\phi$ [nT]                     &   223744  &    0.04 &     2.08 \\ 
$\delta F_\mathrm{AT, polar}$ [nT]    &   77693  &   0.00 &     0.75 \\ 
$\delta F_\mathrm{AT, non-polar}$ [nT] &   154347  &    0.00 &     0.26\\ 
$\delta B_{r,\mathrm{AT}}$ [nT] &   111546  &   -0.00 &     0.38\\ 
$\delta B_{\theta,\mathrm{AT}}$ [nT] &  111546  &   -0.01 &     0.38 \\ 
$\delta B_{\phi,\mathrm{AT}}$ [nT] &  111546  &   -0.00 &     0.40 \\ 
\hline
\end{tabular} 
}
\end{table*}

\begin{table*}[!ht]
	\caption{Model statistics of misfit between CHAOS-7 and CryoSat-2 data.  Mean and rms refer to Huber weighted mean and rms values in units of~nT.  Only the misfit to data from dark regions used to determine the field model coefficients are reported here.
		\label{tab:statistics_cryosat}}
	\vspace{2truemm}
{\scriptsize
\begin{tabular}{r|rrr}
	                                          & \multicolumn{3}{c}{CryoSat-2}  \\
	                                          &    $N$ &  mean &      rms  \\ 
	                                          \hline
$F_\mathrm{polar}$ [nT]           &   16761  &   0.17 &     5.98  \\ 
$F_\mathrm{non-polar}$ [nT]        &    31322  &   0.02 &     4.21 \\ 
$B_r$ [nT]                        &   31322  &   0.08 &     4.08\\ 
$B_\theta$ [nT]                   &   31322  &    -0.07 &     5.23 \\ 
$B_\phi$ [nT]                      &   31322  &    -0.22 &     4.08\\ 
\hline
\end{tabular} 
}
\end{table*}

\begin{table*}[!ht]
	\caption{Model statistics of misfit between CHAOS-7 and {\it Swarm} data.  Mean and rms refer to Huber weighted mean and rms values in units of~nT.   $\delta F_\mathrm{AT}$  and $\delta B_\mathrm{AT}$ denote along-track field differences calculated at 15~sec spacing.   $\delta F_\mathrm{EW}$  and $\delta B_\mathrm{EW}$ denote EW field differences between  {\it Swarm}  Alpha and Charlie.
		\label{tab:statistics_swarm}}
	\vspace{2truemm}
{\scriptsize
\begin{tabular}{r|rrr|rrr|rrr|rrr}
	                                          & \multicolumn{3}{c|}{SW-A} & \multicolumn{3}{c|}{SW-B} & \multicolumn{3}{c|}{SW-C} & \multicolumn{3}{c}{\ SW-A -- SW-C} \\
	                                          &    $N$ &  mean &      rms &    $N$ &  mean &      rms &    $N$ &  mean &      rms &    $N$ &  mean &               rms \\ \hline
$F_\mathrm{polar}$ [nT]           &   23636  &   -0.06 &     3.61&    23128  &   0.05 &     3.39&   23863  &   0.07 &     3.59 & & & \\ 
$F_\mathrm{non-polar}$ [nT]       &   44992  &   -0.07 &     1.81&   46652  &   -0.07 &     1.84&   45531  &   0.00 &     1.80 & & & \\ 
$B_r$ [nT]                         &   44992  &   -0.04 &     1.52&   46652  &   -0.07 &     1.49&   45531  &   -0.04 &     1.54 & & & \\ 
$B_\theta$ [nT]                    &   44992  &    0.09 &     2.38&   46652  &    0.07 &     2.44&   45531  &    0.01 &     2.37 & & & \\ 
$B_\phi$ [nT]                     &   44992  &     0.00 &     1.91&   46652  &    -0.02 &     1.98&   45531  &    -0.02 &     1.94 & & & \\ 
$\delta F_\mathrm{AT, polar}$ [nT]    &   15600  &   0.01 &     0.57&   15456  &   0.00 &     0.51&   15735  &   0.00 &     0.58& & & \\ 
$\delta F_\mathrm{AT, non-polar}$ [nT] &   30570  &    -0.00 &     0.14&   31658  &    -0.00 &     0.12&   30599  &    -0.00 &     0.14& & & \\ 
$\delta B_{r,\mathrm{AT}}$ [nT] &   22469  &   -0.00 &     0.23&   23178   &   0.00 &     0.22&    22958 &   -0.00 &     0.24& & & \\ 
$\delta B_{\theta,\mathrm{AT}}$ [nT] &   22469  &   -0.00 &     0.24&   23178   &   -0.00 &     0.23&    22958  &   0.00 &     0.25& & & \\ 
$\delta B_{\phi,\mathrm{AT}}$ [nT] &   22469  &   0.00 &     0.31&   23178   &   0.00 &     0.30&    22958  &   -0.00 &     0.32& & & \\ 
$\delta F_\mathrm{EW, polar}$ [nT]     & & & & & & & & & &   28738  &   -0.14 &     0.57\\ 
$\delta F_\mathrm{EW, non-polar}$ [nT]    & & & & & & & & & &   55954  &    -0.08 &     0.35\\ 
$\delta B_{r, \mathrm{EW}}$ [nT]   & & & & & & & & & &   40617  &    -0.00 &     0.40\\ 
$\delta B_{\theta,\mathrm{EW}}$ [nT]     & & & & & & & & & &   40617 &   -0.00 &     0.46\\ 
$\delta B_{\phi,\mathrm{EW}}$[nT]     & & & & & & & & & &   40617 &   0.00 &     0.53\\ 
\hline
\end{tabular} 
}
\end{table*}

Compared to CHAOS-6, the Huber weighted residuals between CHAOS-7 and its contributing data are slightly lower, by  between 0.2 and 0.69\,nT considering the non-polar scalar data and the vector components, with the largest improvement in East-West (EW) components of the vector data, for example the weighted rms residual for {\it Swarm A} is 1.91\,nT in CHAOS-7 compared to 2.49\,nT in CHAOS-6.  There was also a small decrease in the residuals to the along-track (AT) and EW differences, on the order of 0.05\,nT.  We attribute the slightly lower rms residuals in CHAOS-7 compared to CHAOS-6 to our stricter data selection, and to the relaxation of the temporal regularization in CHAOS-7.

Of particular interest is the fit to the new dataset provided by CryoSat-2.  CHAOS-7 is able to fit this data, along with simultaneous ground observatory SV data, to a Huber weighted rms level of 4.21\,nT for non-polar scalar data, with vector components fit to between 4.0 and 5.25\,~nT.  Mean residuals are less than 0.25\,nT indicating little evidence for remaining biases. 

\begin{figure}[!ht]
\centerline{\includegraphics[angle=0, width=0.75\textwidth]{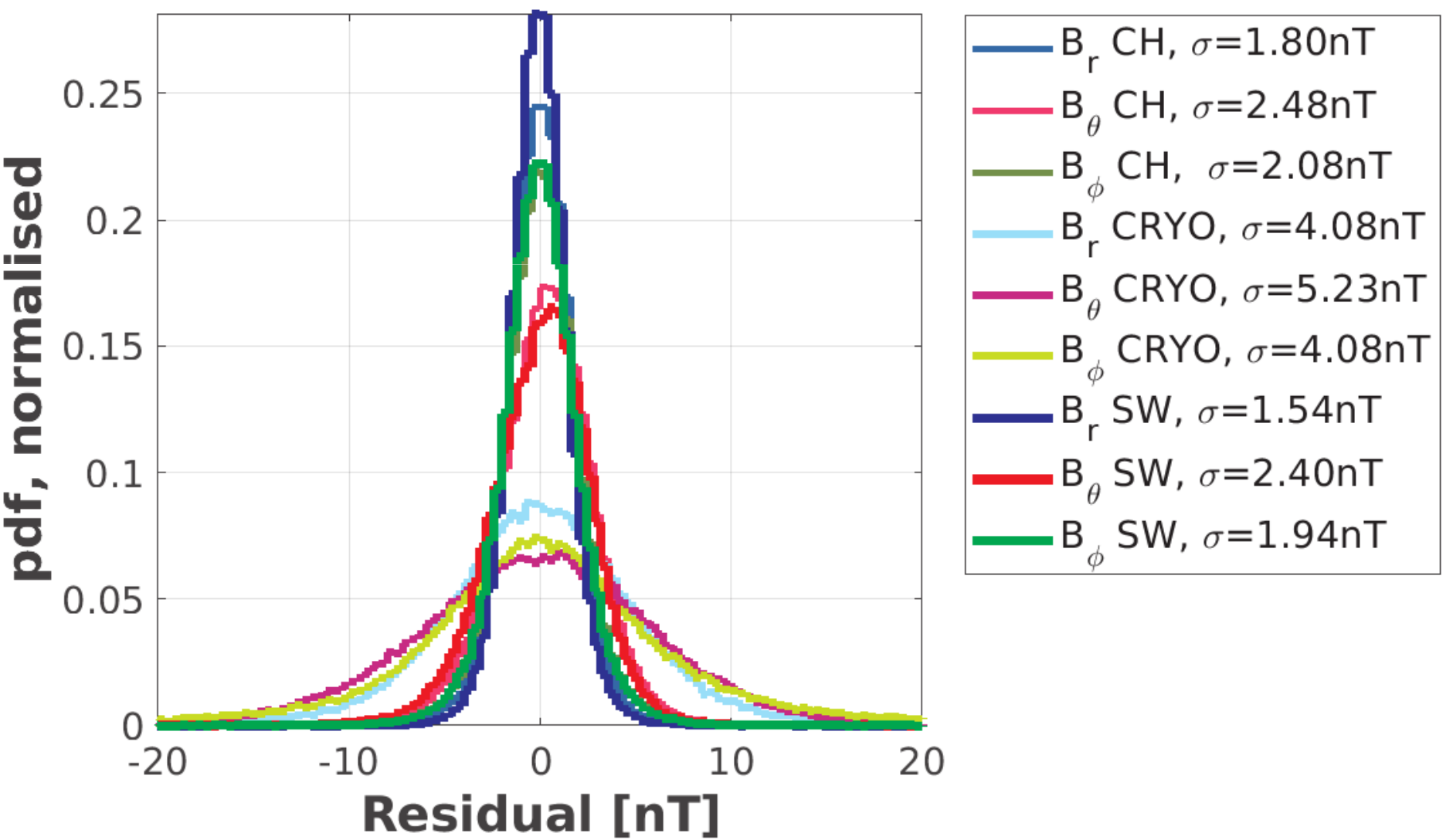}}
\vspace{0.2cm}
\centerline{\includegraphics[angle=0, width=0.75\textwidth]{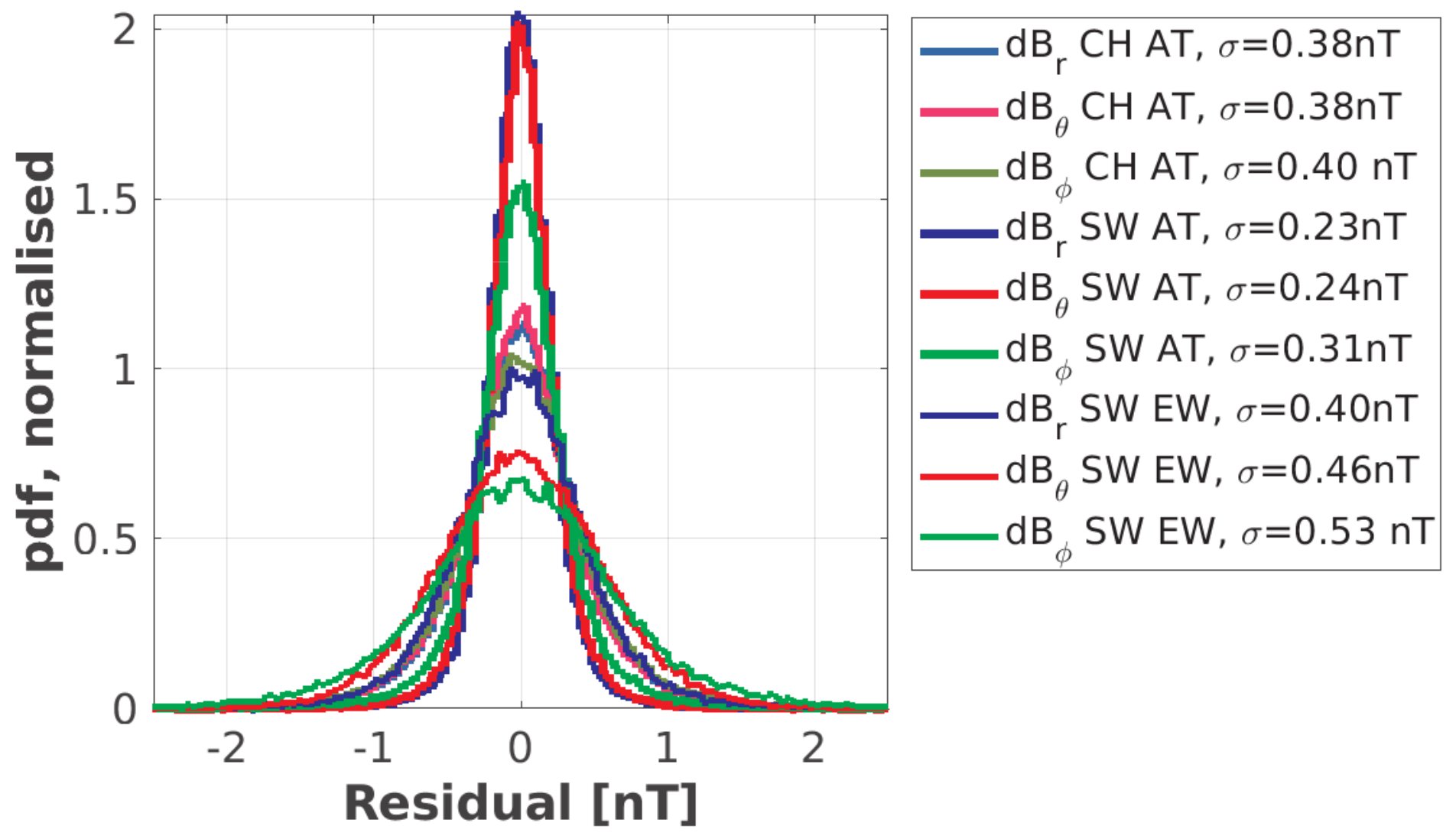} }
\caption{Histograms of residuals between CHAOS-7 and selected contributing datasets.  Top: vector field components from CHAMP, CryoSat-2 and {\it Swarm}, using a histogram bin width of 0.4~nT.  Bottom: vector field differences, CHAMP along-track (AT), {\it Swarm} along-track (AT), and {\it Swarm} East-West (EW), using a histogram bin width of 0.04~nT.  Huber weighted rms residuals are noted in the legend.
\label{fig:Hist_resids}}
\end{figure}

Fig.~\ref{fig:Hist_resids} presents histograms of residuals between CHAOS-7 predictions and vector field data from CHAMP, CryoSat-2 and \textit{Swarm} (top panel) and AT and EW differences of vector field (bottom panel) from CHAMP and {\it Swarm}. Considering the vector data, histograms for the \textit{Swarm} data are most peaked, followed by CHAMP and then CryoSat-2. In each case the radial components of the data are best fit. The southward field components have noticeable biases, at around 0.1-0.3\,~nT these biases are much smaller than the rms levels, and are likely due to imperfectly modelled ring current variations.  Considering AT difference residuals, the histograms for the \textit{Swarm} data are again are again most peaked, with the rms residuals values of only 0.23 to 0.31\,nT.  The histograms for \textit{Swarm} EW difference residuals show a slightly larger dispersion, as expected when taking differences between two different instruments. 

\subsection{{\bf Fit to ground observatory secular variation}}
\label{sec:fit_mm}

Table~\ref{tab:statistics_obs} presents the Huber weighted mean and rms residuals to annual differences of the revised monthly mean ground observatory data between 1997 and 2020, that were used in constructing CHAOS-7, considering all latitudes.  Also shown for reference are similar statistics for the CM6 model \citep{Sabaka:2020}, model A of \citet{Alken:2020} and model MCO\_SHA\_2Y, version 0101, an early version of the core field part of the MCM model described by \citet{Ropp:2020}.  These three  models cover similar time spans to CHAOS-7 but adopt different data selection and modelling strategies. Note that only CHAOS-7 was directly constrained by this dataset, the other models used fits to hourly mean or daily mean observatory data rather than to annual differences of revised monthly means.   The misfits levels for the four models agree to within 0.15\,nT/yr for the radial and southward components, with CHAOS-7 fitting the radial component slightly better and the MCO\_SHA\_2Y model fitting the southward component slightly better. For the eastward component the misfits agree very closely, to within 0.01 nT/yr.

\begin{figure}[!ht]

\includegraphics[angle=0, width=0.3\textwidth]{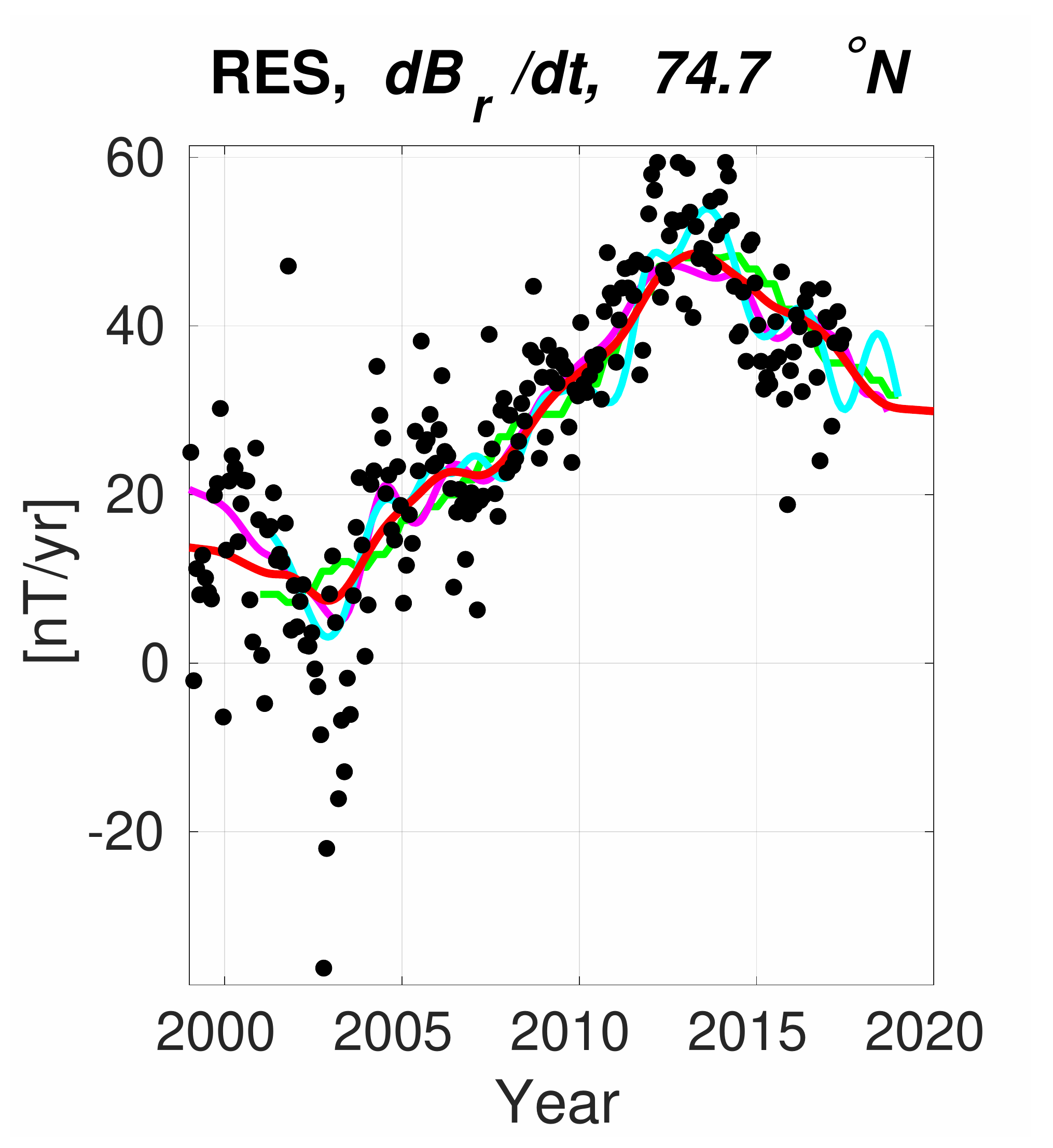}
\includegraphics[angle=0, width=0.3\textwidth]{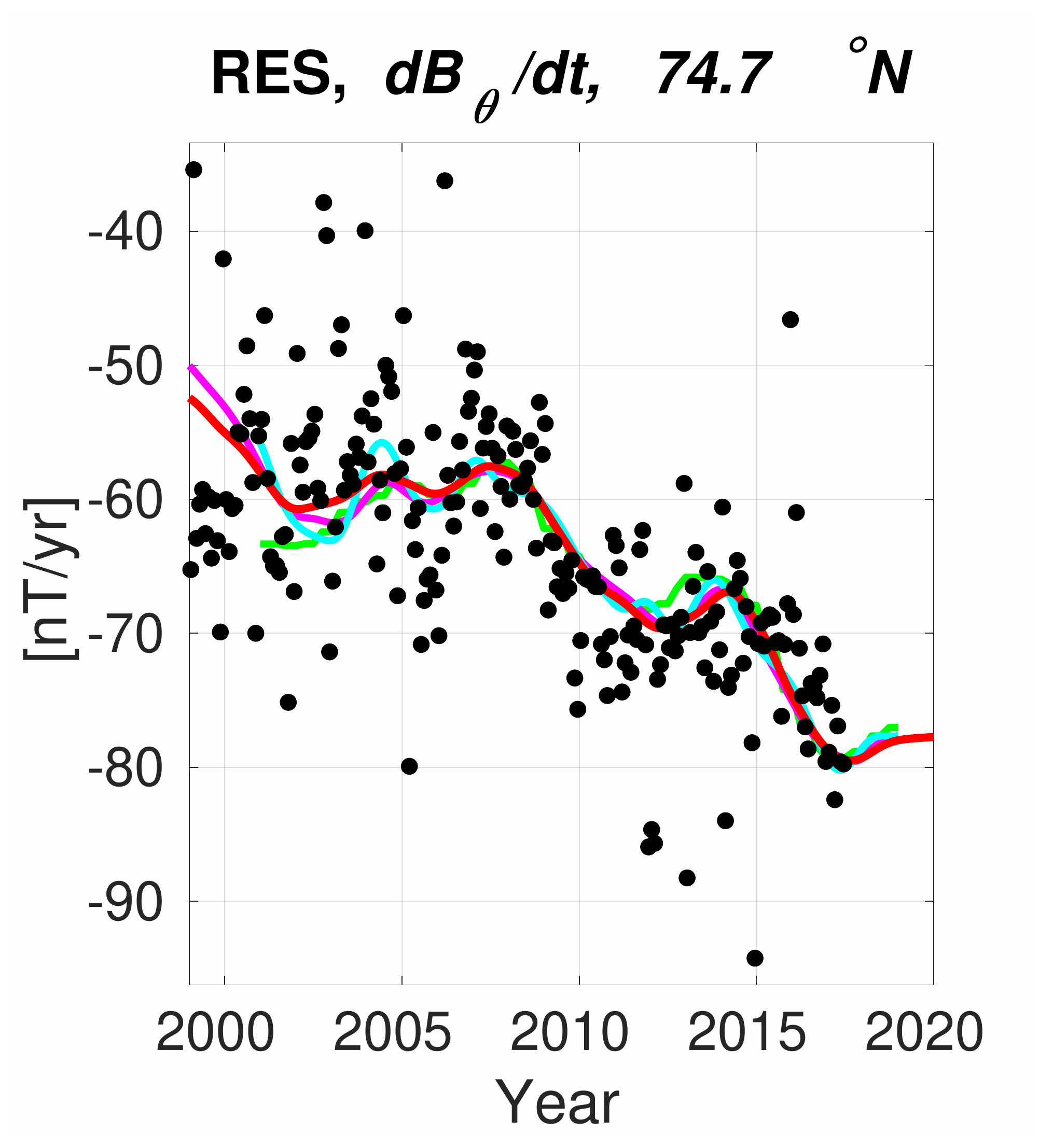}
\includegraphics[angle=0, width=0.3\textwidth]{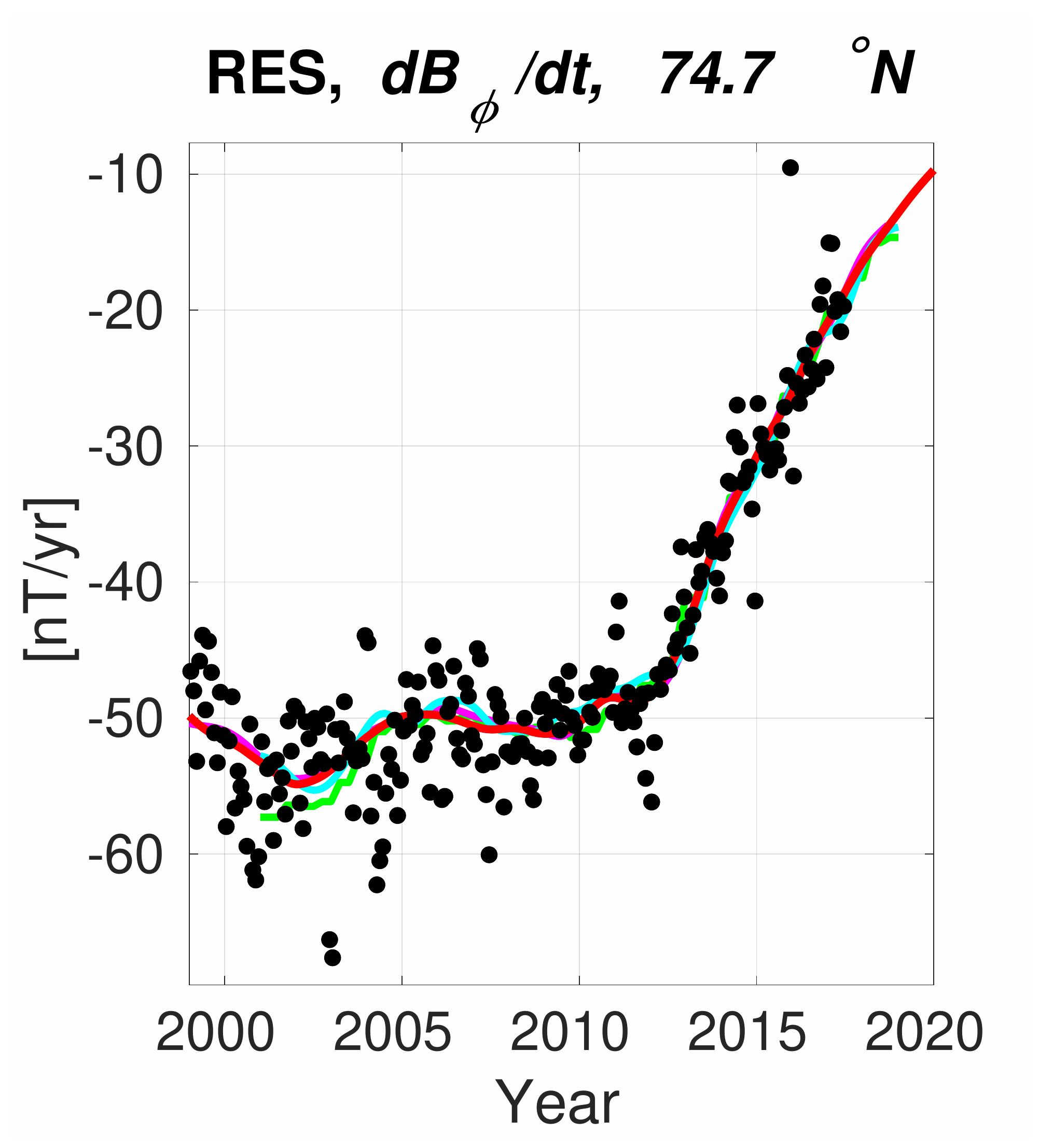}\\

\includegraphics[angle=0, width=0.3\textwidth]{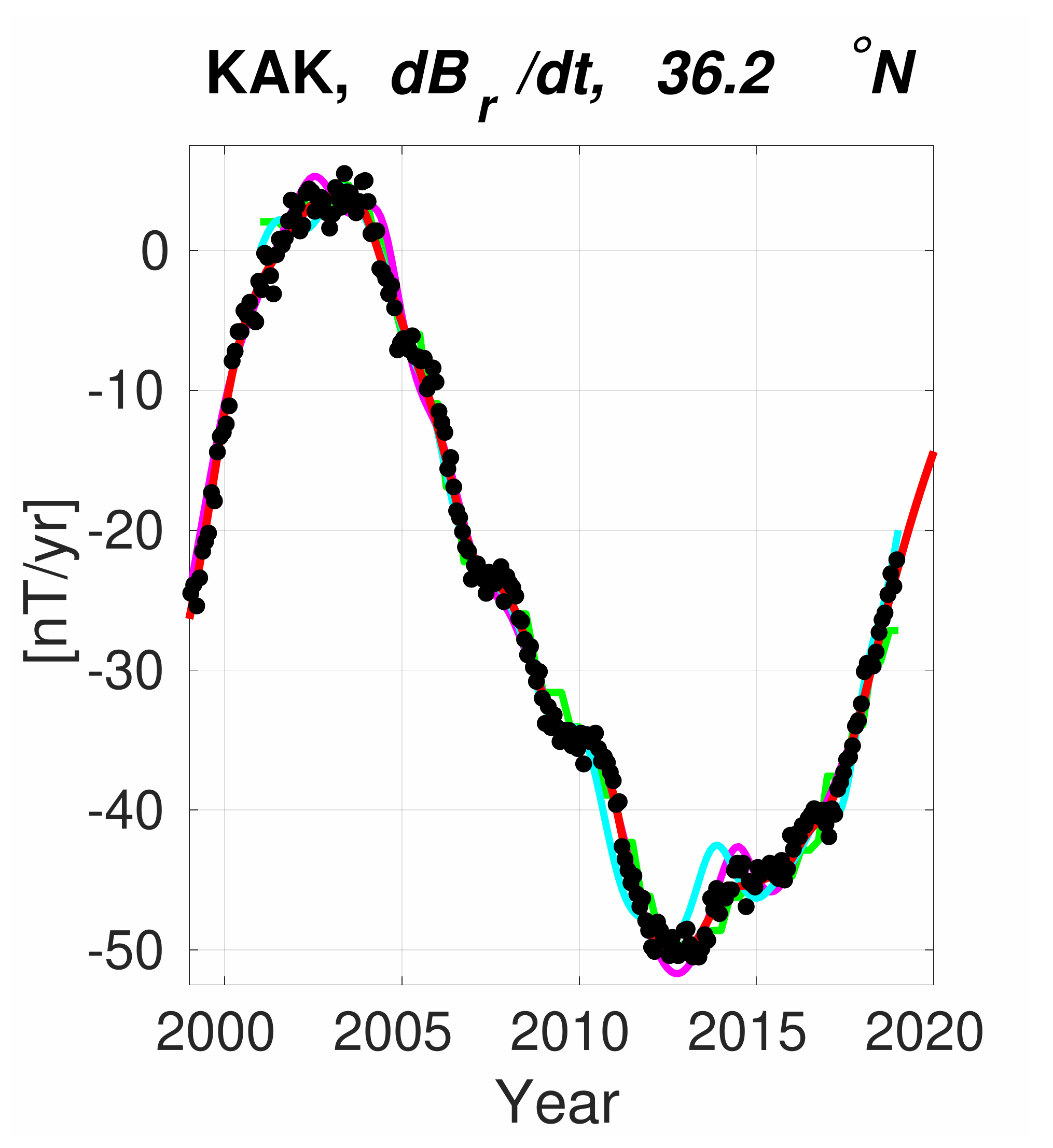}
\includegraphics[angle=0, width=0.3\textwidth]{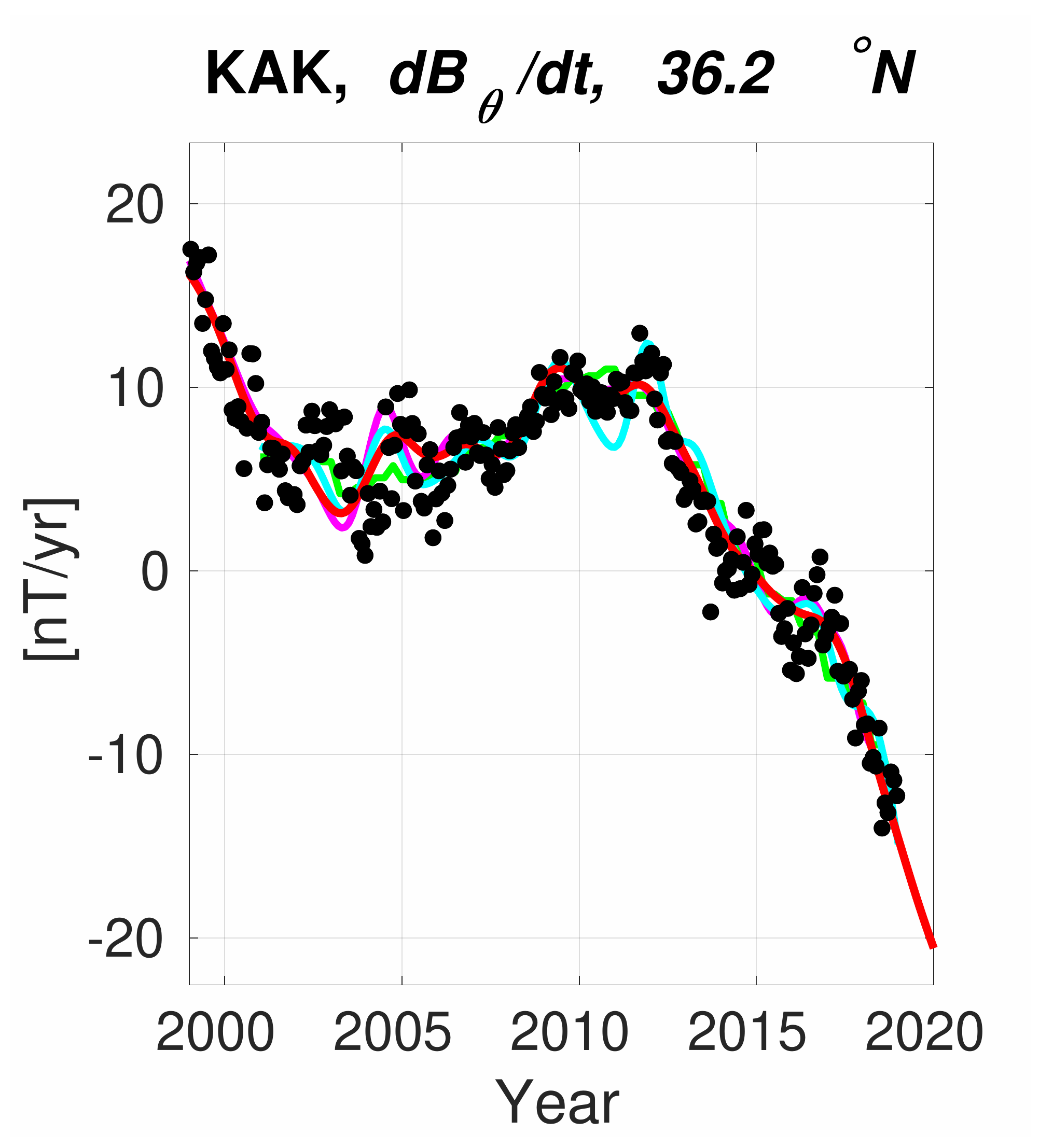}
\includegraphics[angle=0, width=0.3\textwidth]{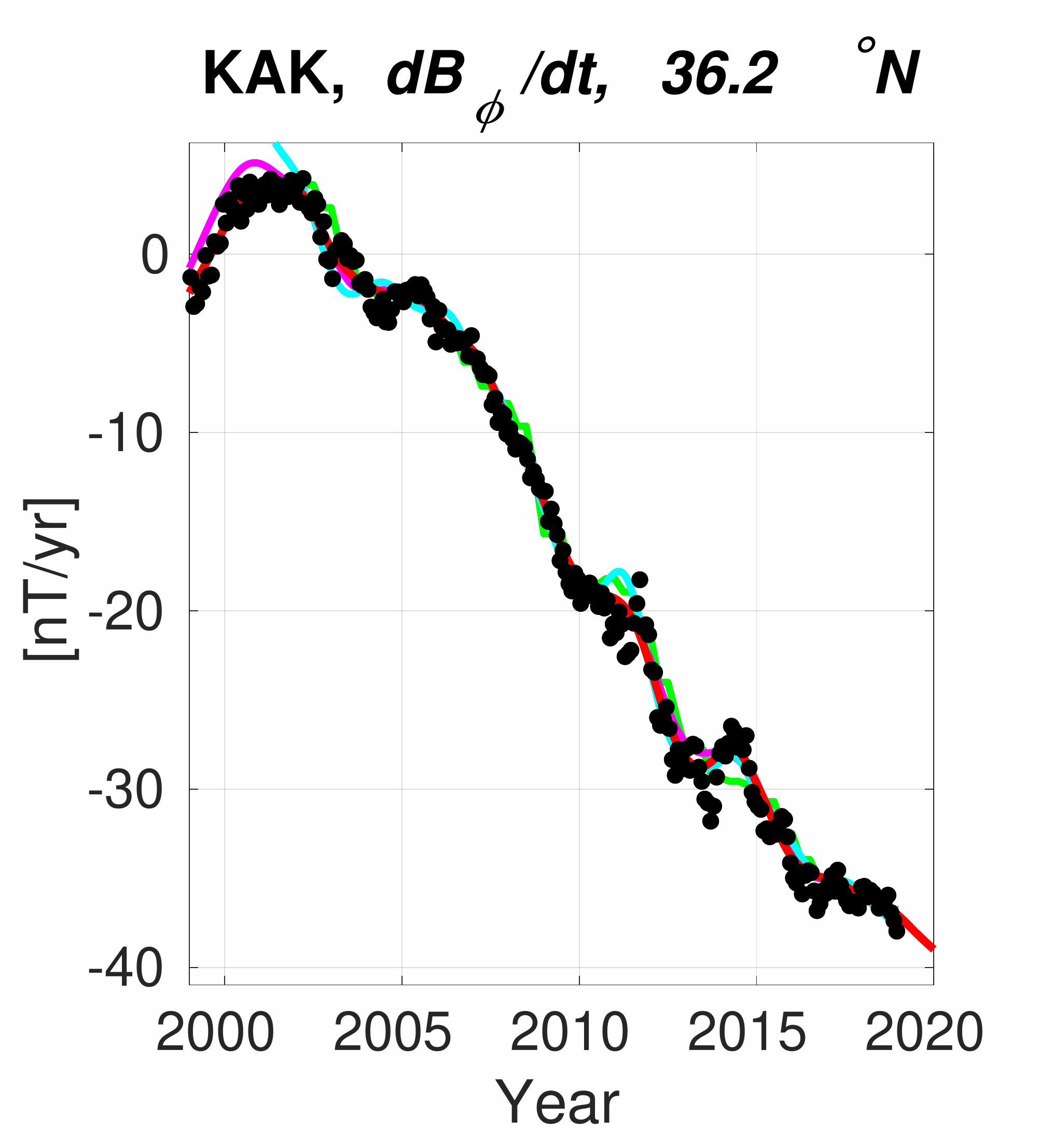}\\

\includegraphics[angle=0, width=0.3\textwidth]{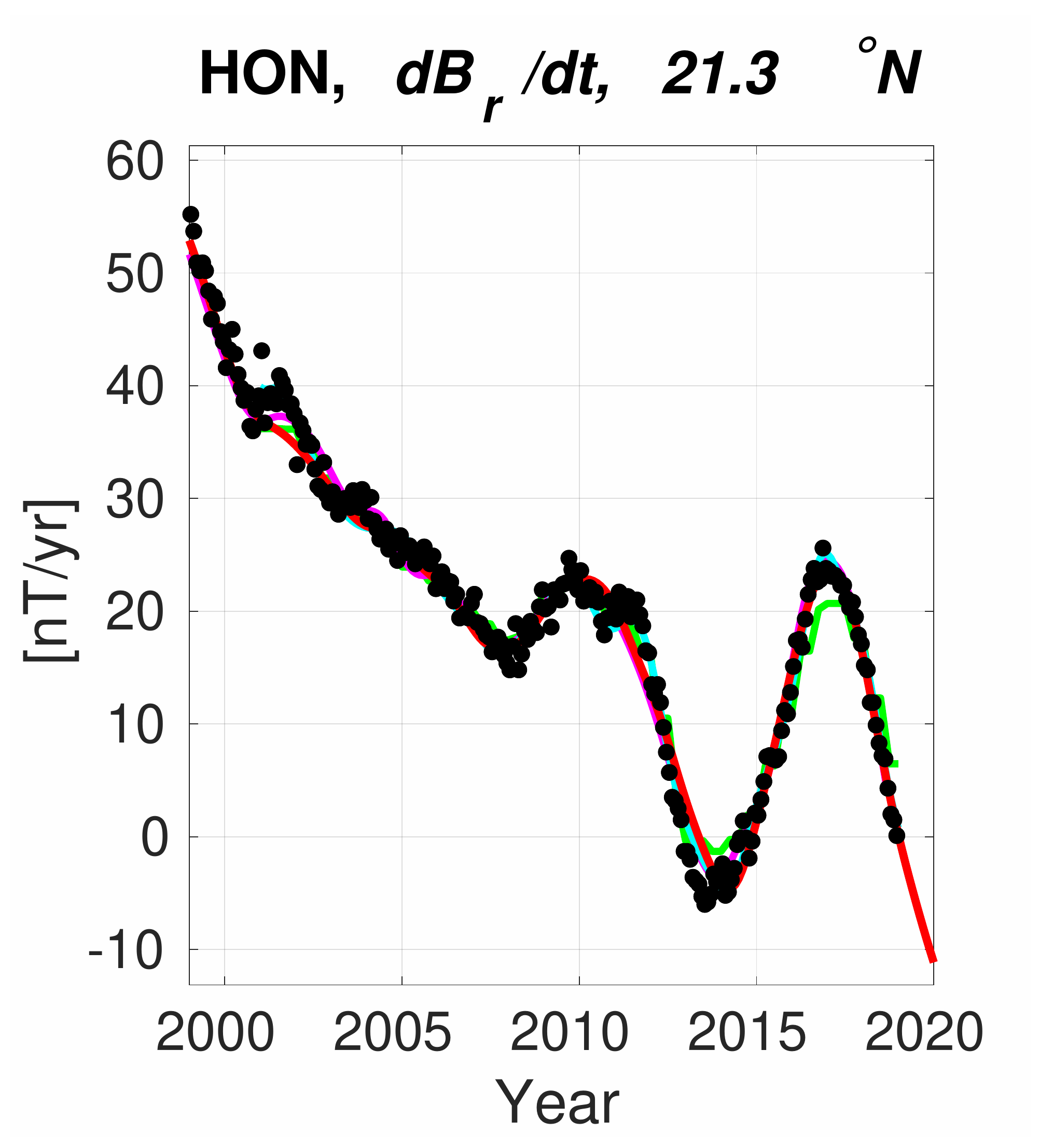}
\includegraphics[angle=0, width=0.3\textwidth]{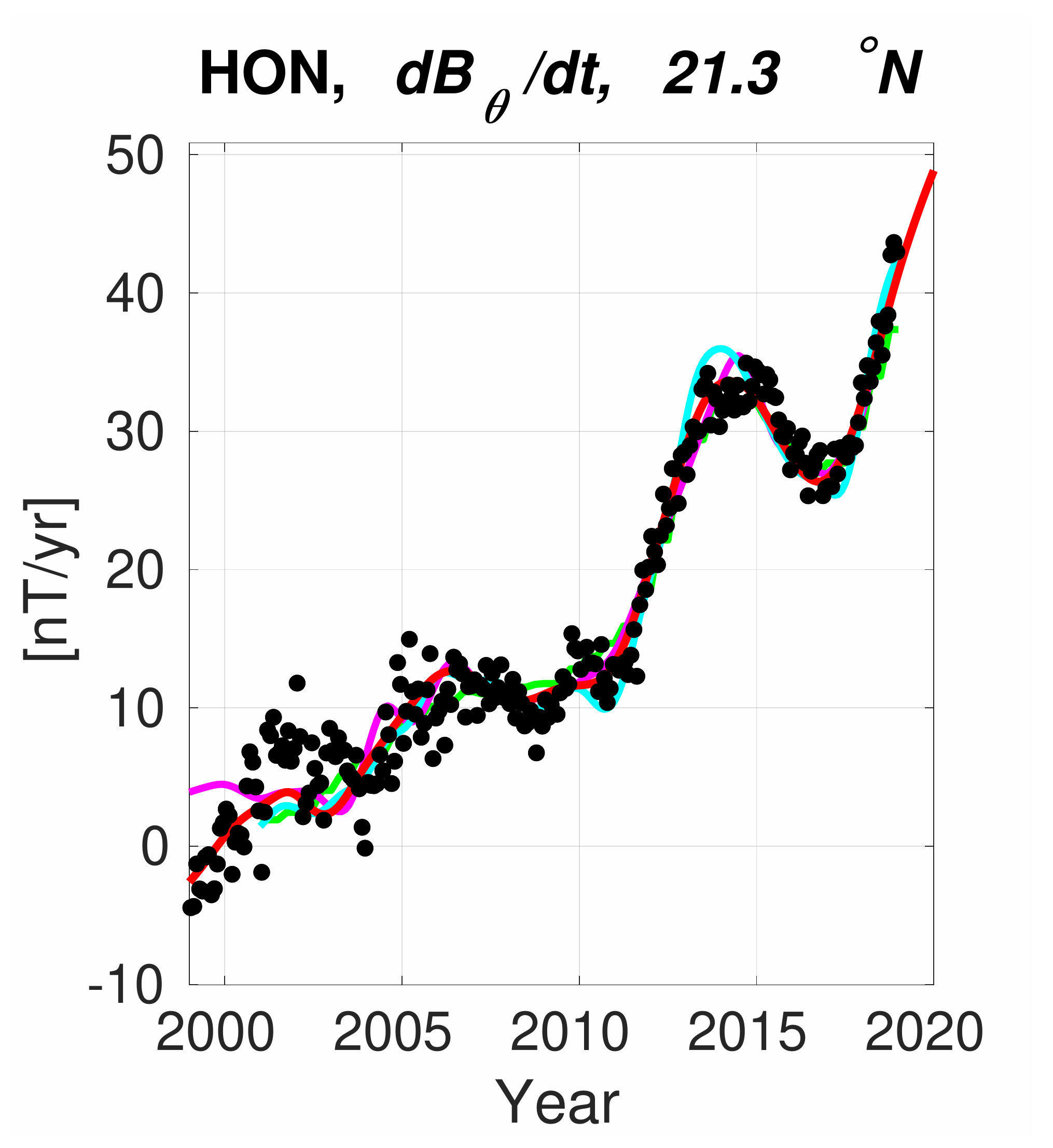}
\includegraphics[angle=0, width=0.3\textwidth]{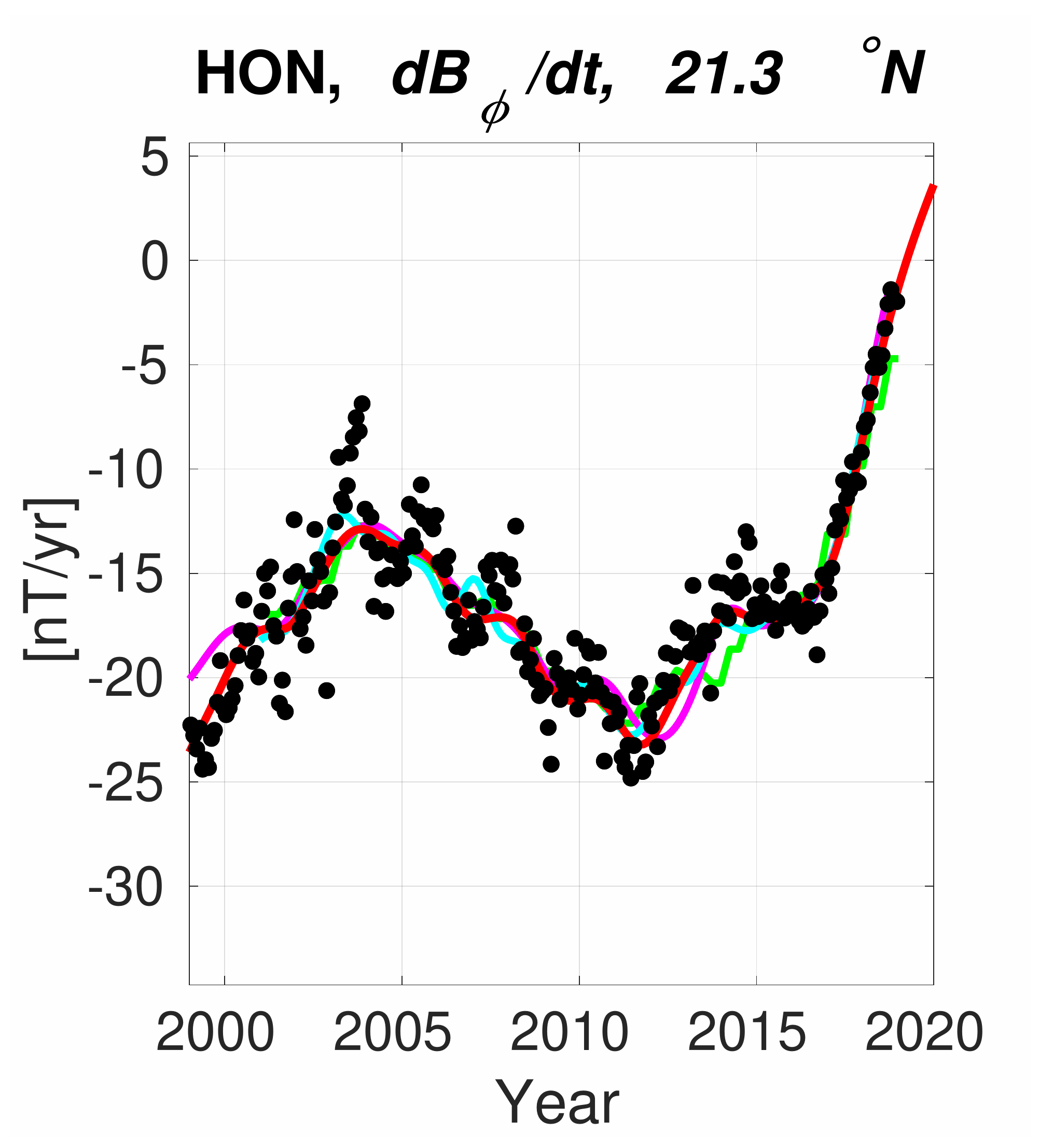}\\

\includegraphics[angle=0, width=0.3\textwidth]{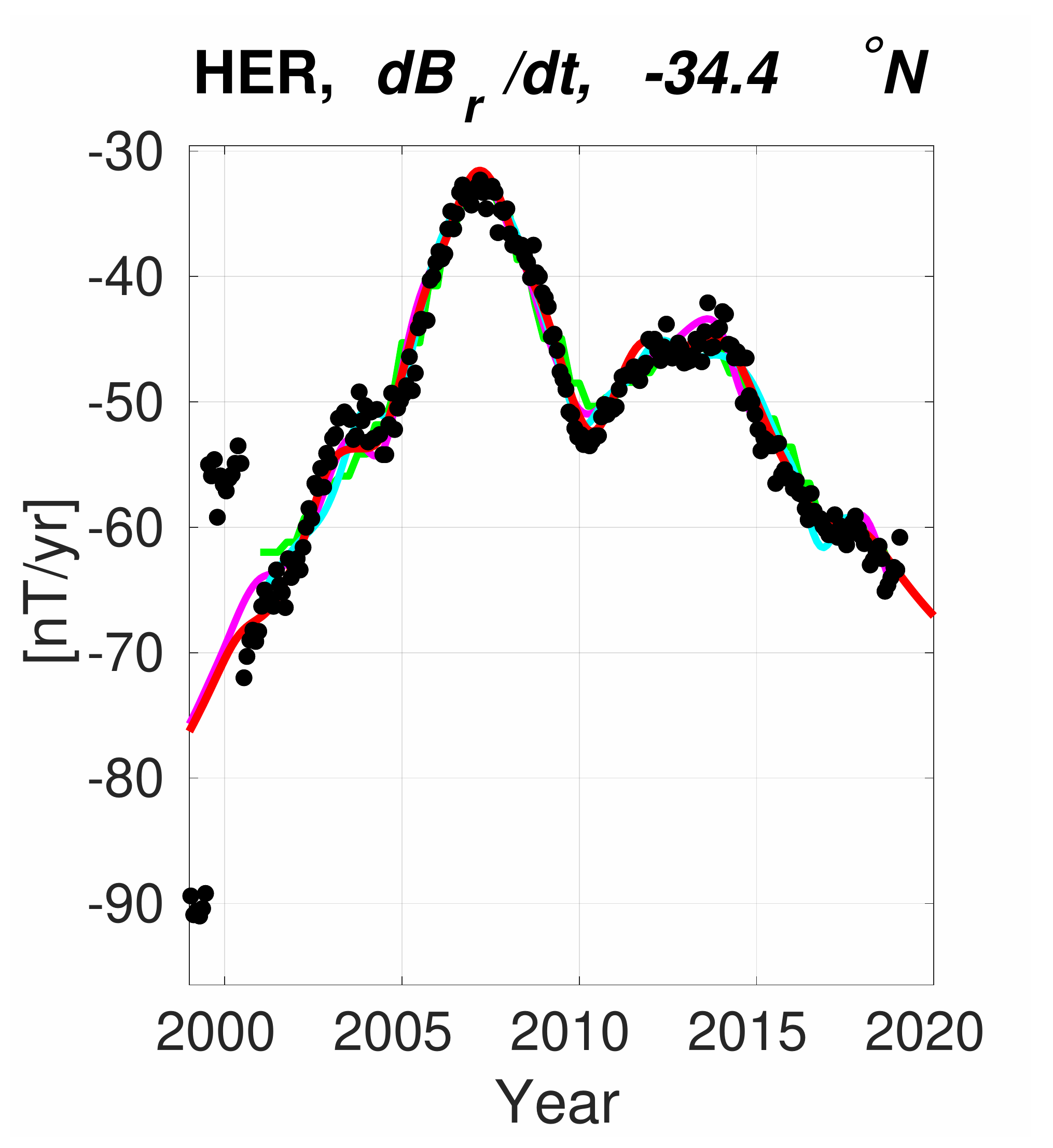}
\includegraphics[angle=0, width=0.3\textwidth]{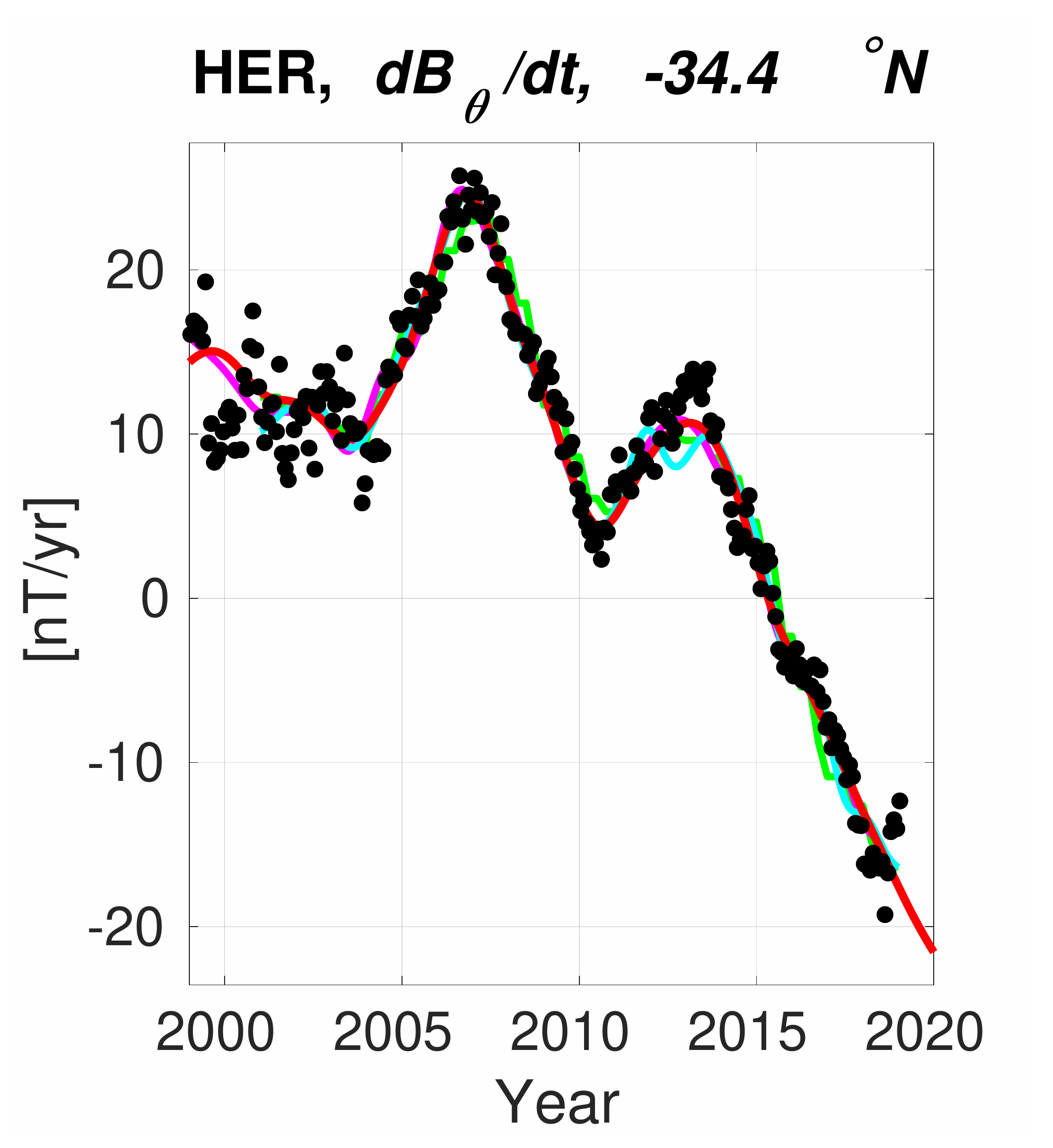}
\includegraphics[angle=0, width=0.3\textwidth]{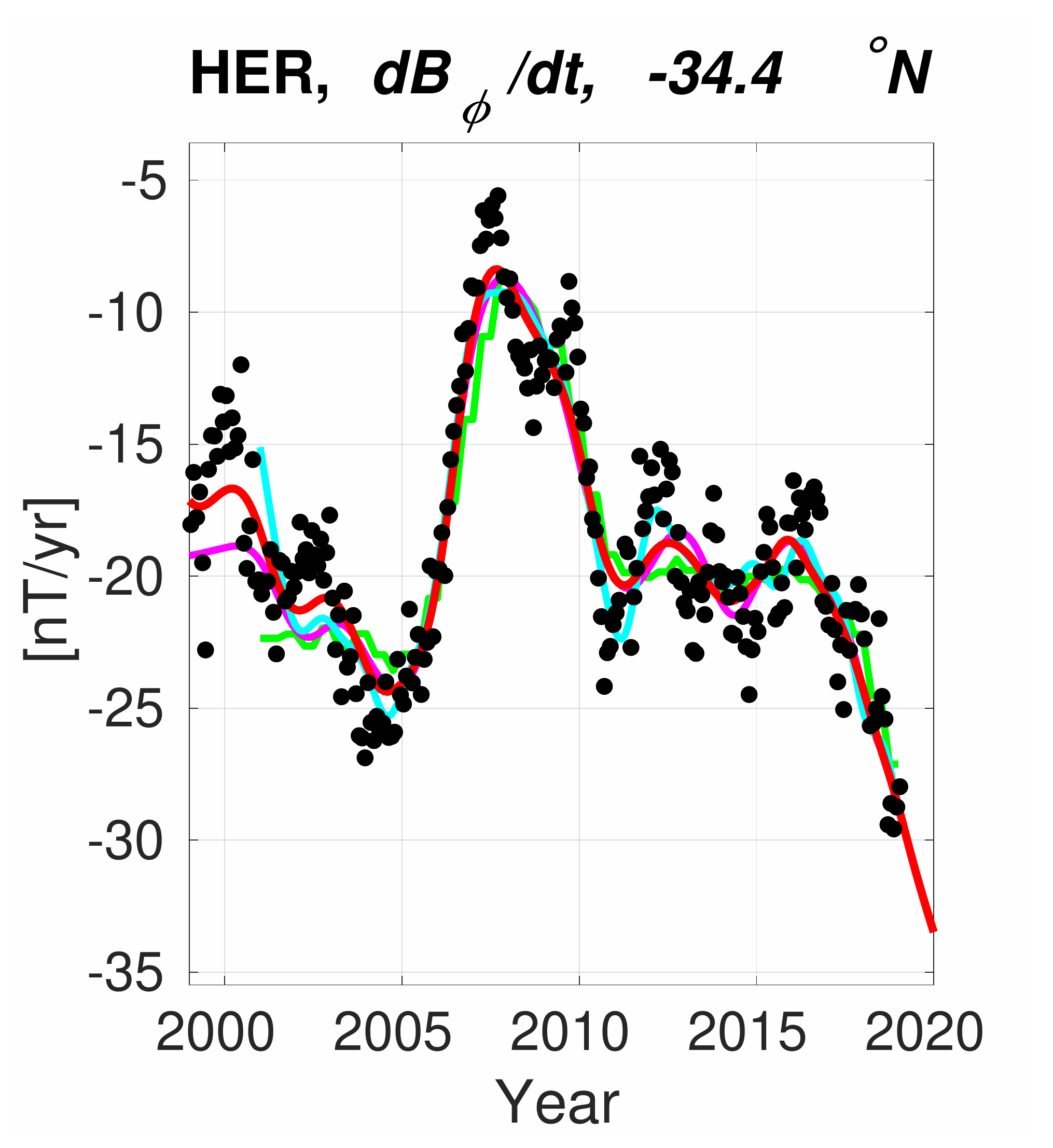}\\

\includegraphics[angle=0, width=0.3\textwidth]{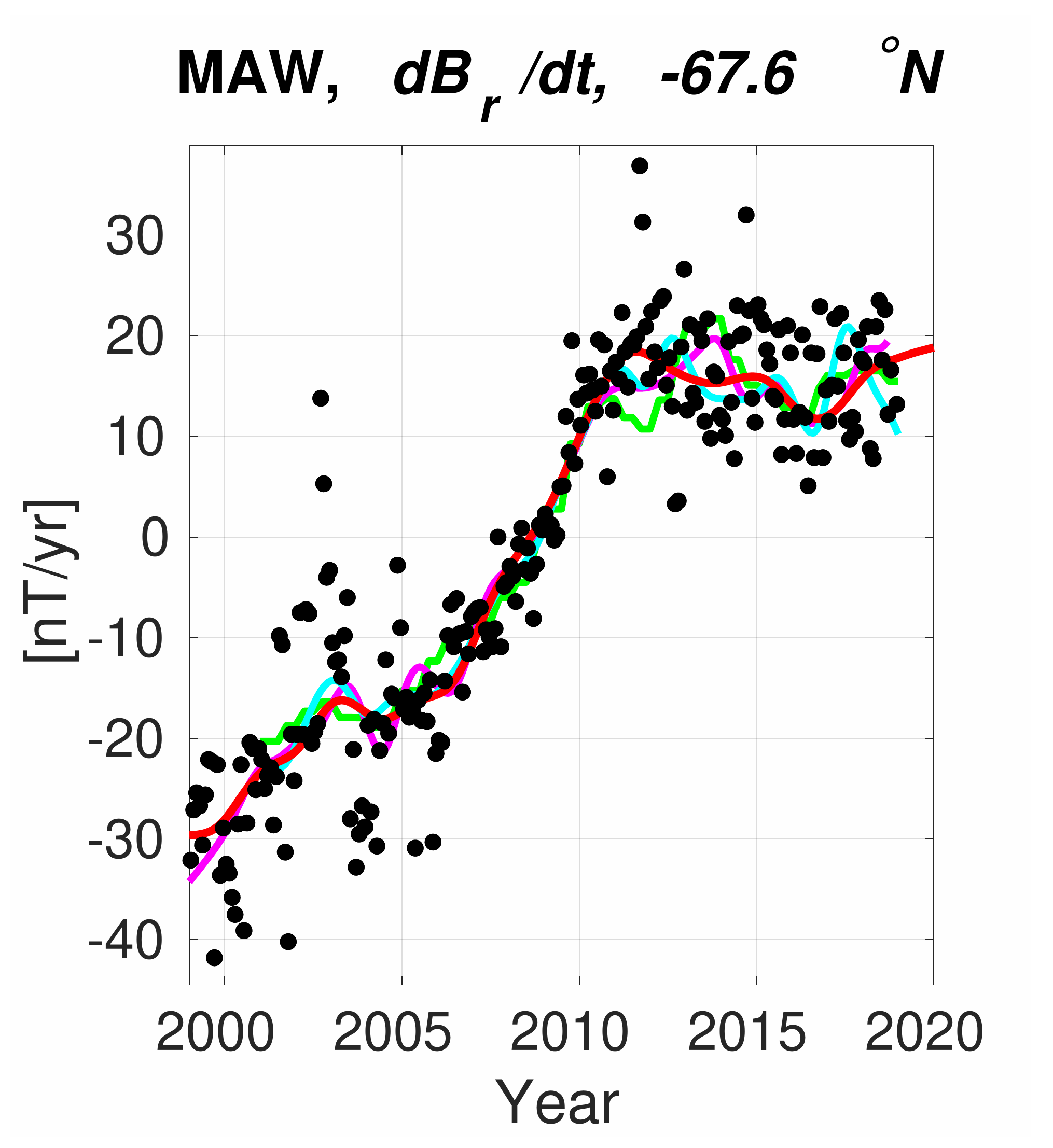}
\includegraphics[angle=0, width=0.3\textwidth]{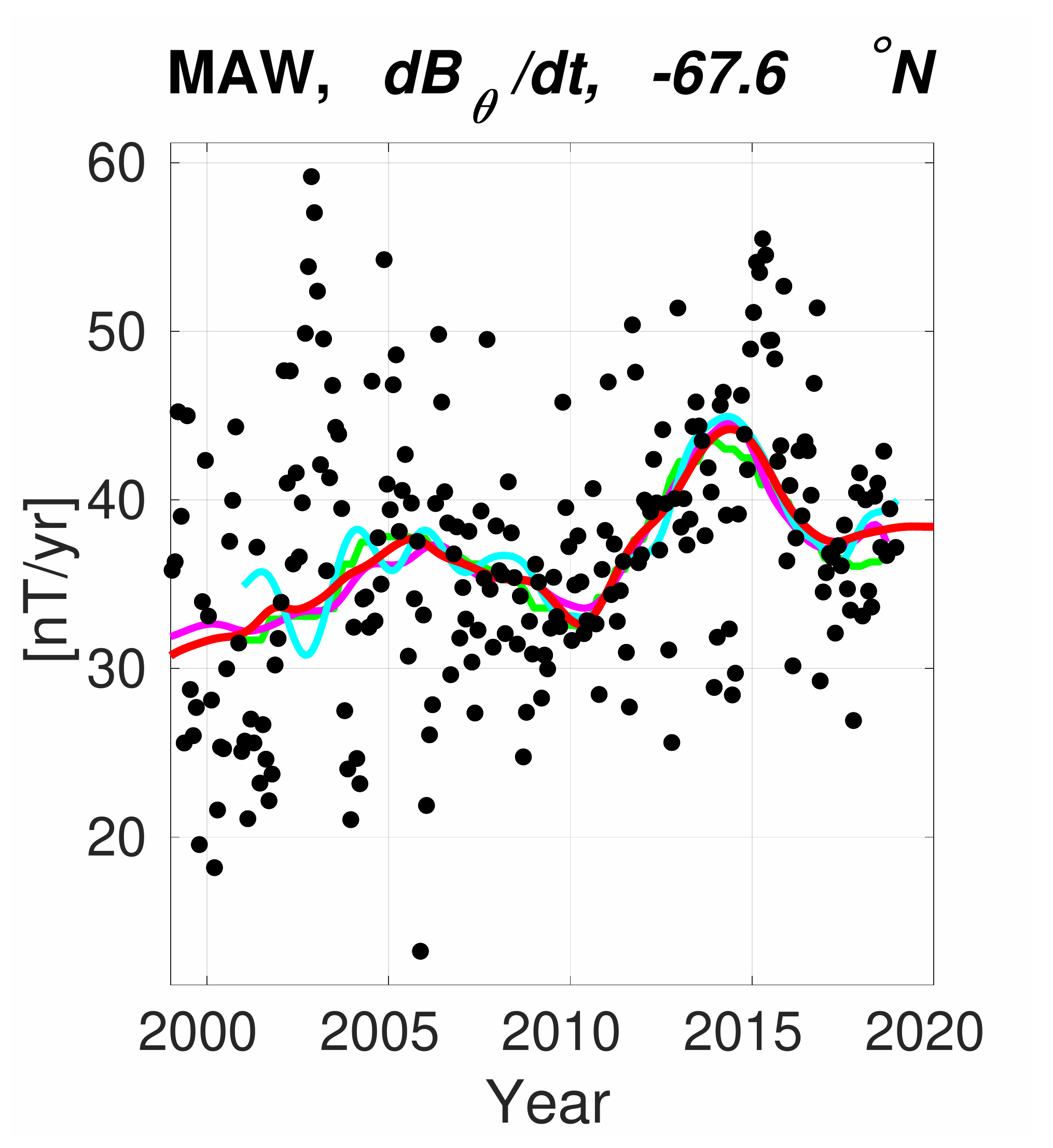}
\includegraphics[angle=0, width=0.3\textwidth]{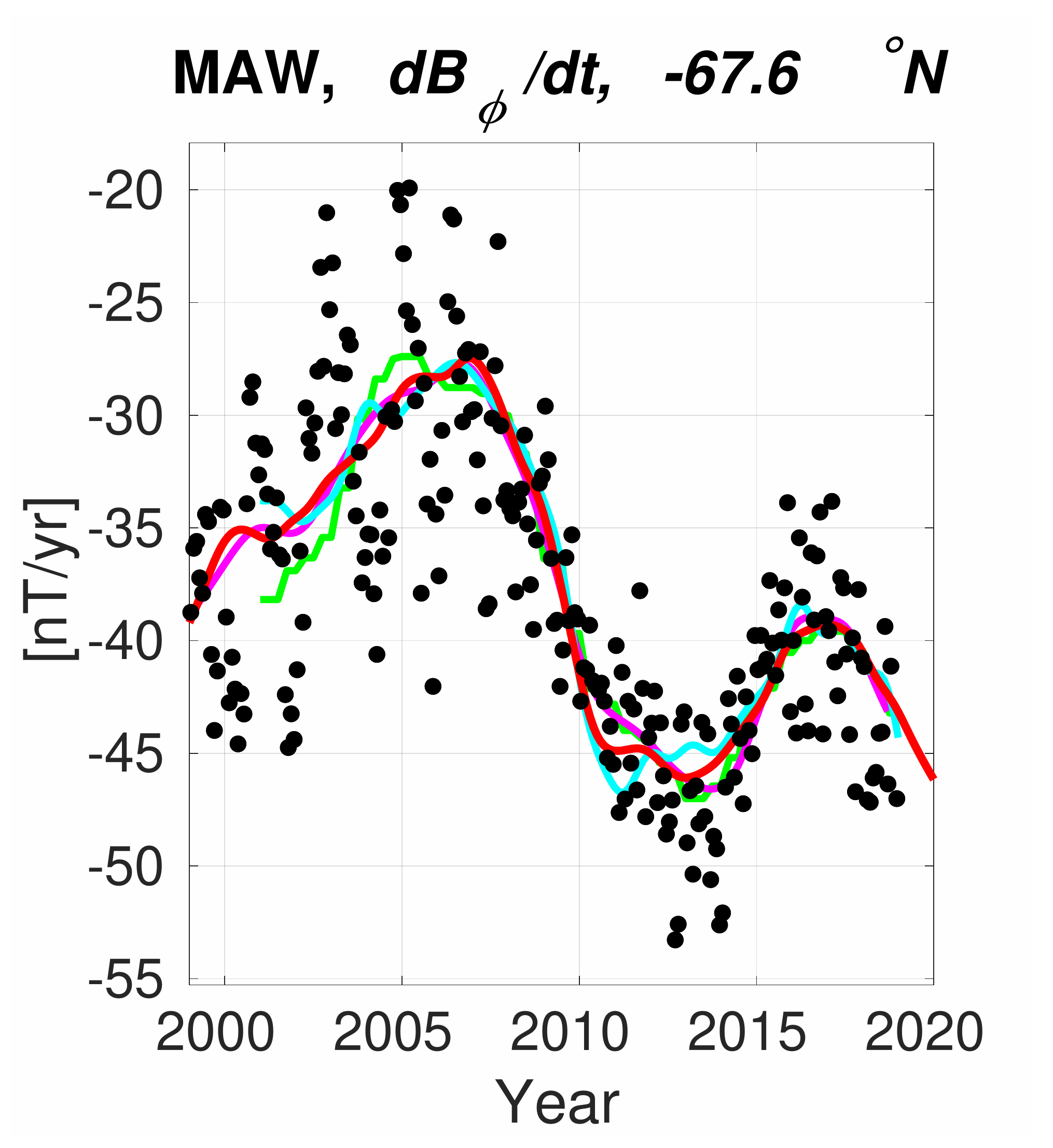}

\caption{Fit of the CHAOS-7 model (red line) to secular variation data, annual differences of revised monthly means (black dots), at example ground observatories (three letter IAGA codes given). Also shown are the MCO\_SHA\_2Y model, version 0101, \citep{Ropp:2020} (green line), the CM6 model of \citet{Sabaka:2020} (magenta line), and Model A of \cite{Alken:2020} (cyan line).
\label{fig:SV_obs}
}
\end{figure}

\begin{table*}[!ht]
	\caption{Model statistics of the misfit  between CHAOS-7 and other field models and 30448 vector triples of ground observatory SV (annual differences of revised-monthly mean) data, between 1997.5 and 2019.5. Mean and rms refer to Huber weighted values in units nT/yr. Similar misfits to other field models, the CM6 model \citep{Sabaka:2020}, model A20 \citep{Alken:2020} and the MCO\_SHA\_2Y model \citep{Ropp:2020} are also reported for reference; note these did not directly use revised monthly means in their derivation. For CM6 28,986 SV vector triples between 1999.0 and 2019.0 were considered for the comparison, while for A20 and MCO\_SHA\_2Y 26,159 vector triples between 2001.0 and 2019.0 were considered.
			\label{tab:statistics_obs}}
	\vspace{2truemm}
{\scriptsize
\begin{tabular}{r|rr||rr|rr|rr}
	                                          & \multicolumn{8}{c}{Misfit to SV at Ground observatories}  \\ \\
	                                          & \multicolumn{2}{c}{{\bf CHAOS-7}} & \multicolumn{2}{c}{CM6} & \multicolumn{2}{c}{A20} & \multicolumn{2}{c}{MCO\_SHA\_2Y}  \\
	                                          &  mean &      rms  &  mean &      rms &  mean &      rms &  mean &      rms \\ 
	                                          \hline
$\mathrm{d}B_r/\mathrm{d}t$ [nT/yr]                       &  {\bf 0.11} &      {\bf 3.73} & -0.05 & 3.84  & 0.14 & 3.81 & 0.11 & 3.88 \\ 
$\mathrm{d}B_\theta/\mathrm{d}t$ [nT/yr]                 &    {\bf -0.21 }&      {\bf 3.59} & -0.27 & 3.67  & -0.12 & 3.66 & -0.12 & 3.51 \\ 
$\mathrm{d}B_\phi/\mathrm{d}t$ [nT/yr]                &    {\bf 0.01} &     {\bf 3.31} & -0.03 & 3.32  & 0.00 & 3.32 & 0.03 & 3.31 \\ 
\hline
\end{tabular} 
}
\end{table*}


It is also of interest to revisit the CHAOS-6 model in this context, to ascertain the extent to which the changes in the satellite data selection criteria and temporal regularization in CHAOS-7 have affected the fit to the ground observatory SV. The Huber weighted rms misfits of CHAOS-6x9 to the ground observatory SV data were 3.78\,nT/yr,  3.62\,nT/yr and 3.33\,nT/yr  for the radial, southward and eastward field components respectively. The close agreement with the CHAOS-7 misfit levels, to within 0.05\,nT/yr, indicates the good agreement between the time-dependent large-scale internal field in CHAOS-6 and CHAOS-7, despite the differences in their construction.

Fig.~\ref{fig:SV_obs} presents the fit of CHAOS-7 and the three other field models introduced above to time series of annual differences of revised monthly means at some example observatories.  Fits to the three geocentric vector components are shown in the three columns; the latitude of the selected observatories moves from high northern latitudes down through the equator to high southern latitudes going down the rows. The model predictions are generally in agreement regarding the long-term trends but there are differences on timescales of one to two years.  The model of \cite{Alken:2020} generally shows more high frequency fluctuations. The CM6 model \citep{Sabaka:2020} and CHAOS-7 are generally in good agreement, despite the different treatment of induced fields on the nightside in CM6.  Differences between the models are most prominent at high latitudes where there is also much larger scatter in the observatory data.  Overall there is encouraging agreement between the models and it is difficult to prefer one model over another based on these comparisons. 
\clearpage

\subsection{Co-estimated magnetometer calibration parameters for CryoSat-2}
\begin{figure}[!ht]
\centerline{\includegraphics[angle=0, width=\textwidth]{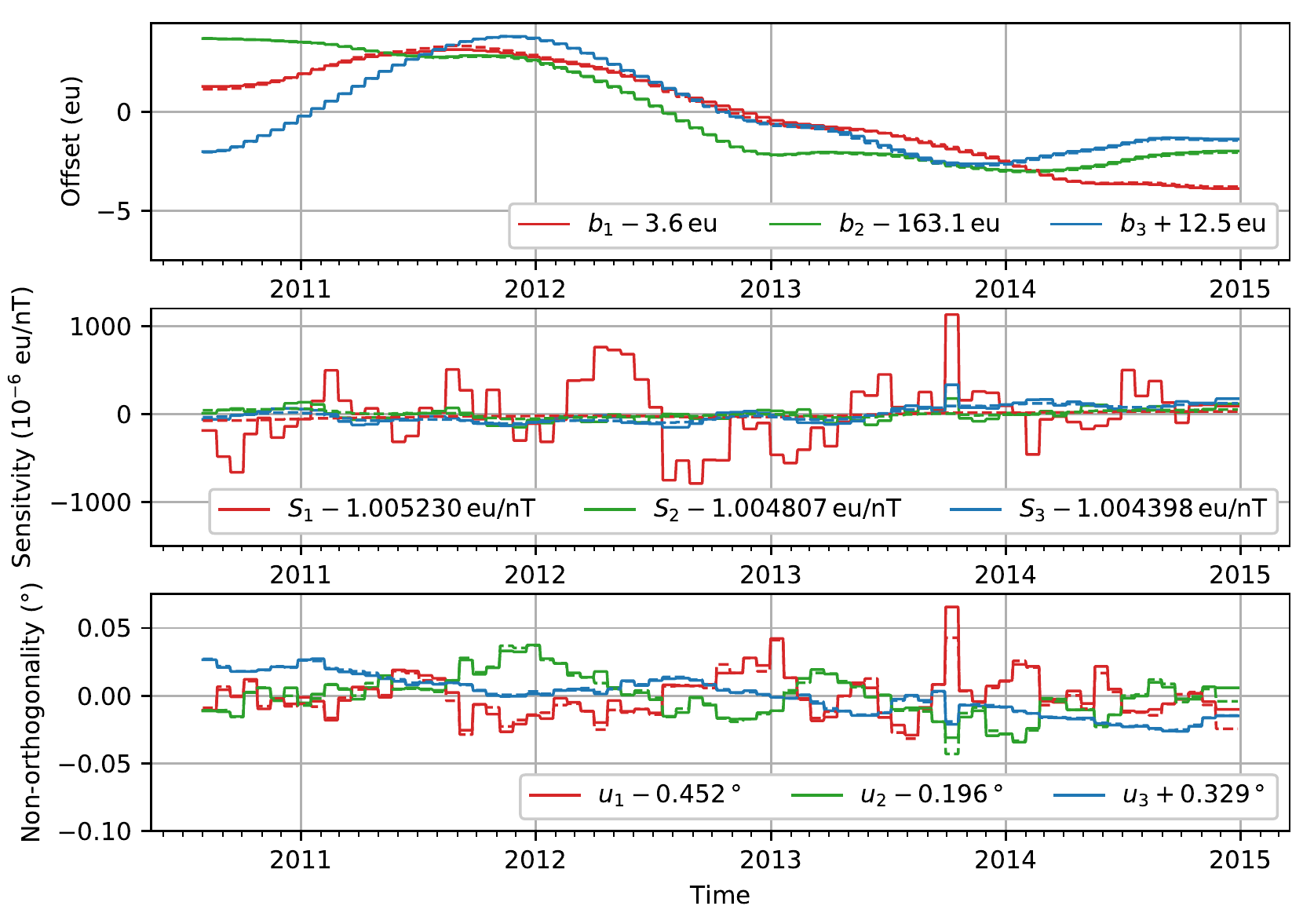}}
\caption{CryoSat-2 co-estimated magnetometer calibration parameters estimated as part of CHAOS-7 (solid lines) and CHAOS-7.2 (dashed lines) as a function of time. The panels show the offsets (top), sensitivities (middle) and non-orthogonality parameters (bottom).  Heavier regularization of the sensitivities was used in CHAOS-7.2.}
\label{fig:cal}
\end{figure}

Fig.~\ref{fig:cal} documents the co-estimated CryoSat-2 calibration parameters as a function of time.  The non-orthogonalities are rather stable throughout the four years used.  The offsets slowly vary by up to 4\,~nT.  Largest variations are seen in the scale value of sensor one, $S_1$, where especially in earlier years there are variations over about nine months.  Such time variations are not found when this instrument is calibrated against a fixed field model, CHAOS-6-x9 \citep{Olsen:2020a}, suggesting that an increase in the regularization of the sensitivities may be required.  This has been implemented in the most recent version of the CHAOS model, CHAOS-7.2 which was released in April 2020; the calibration parameters for CHAOS-7.2 are shown as dashed lines in Fig.~\ref{fig:cal}.

\subsection{Validation tests against independent ground and satellite data}
\begin{figure}[!ht]
\centerline{\includegraphics[angle=0, width=0.49\textwidth]{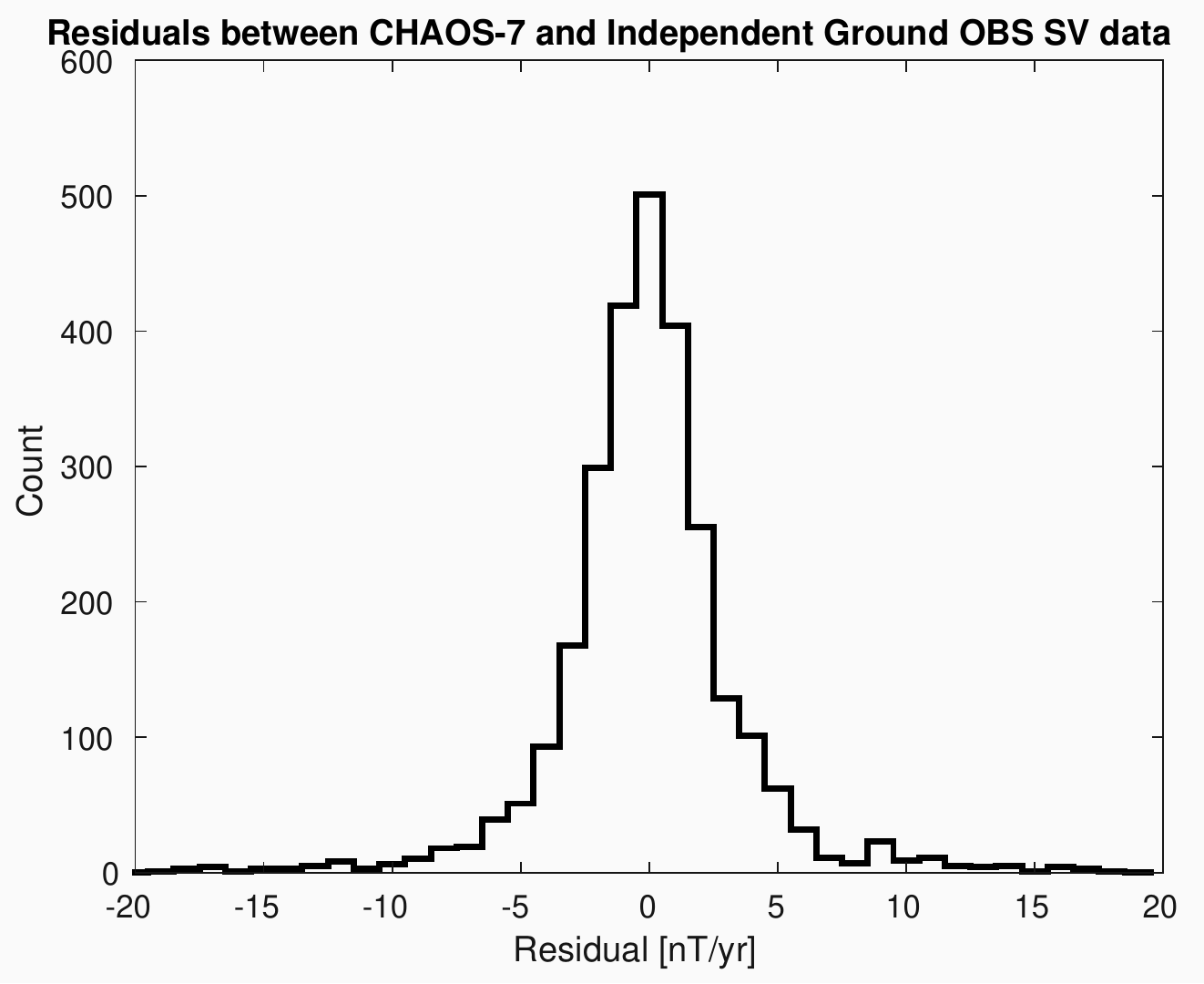}\includegraphics[angle=0, width=0.49\textwidth]{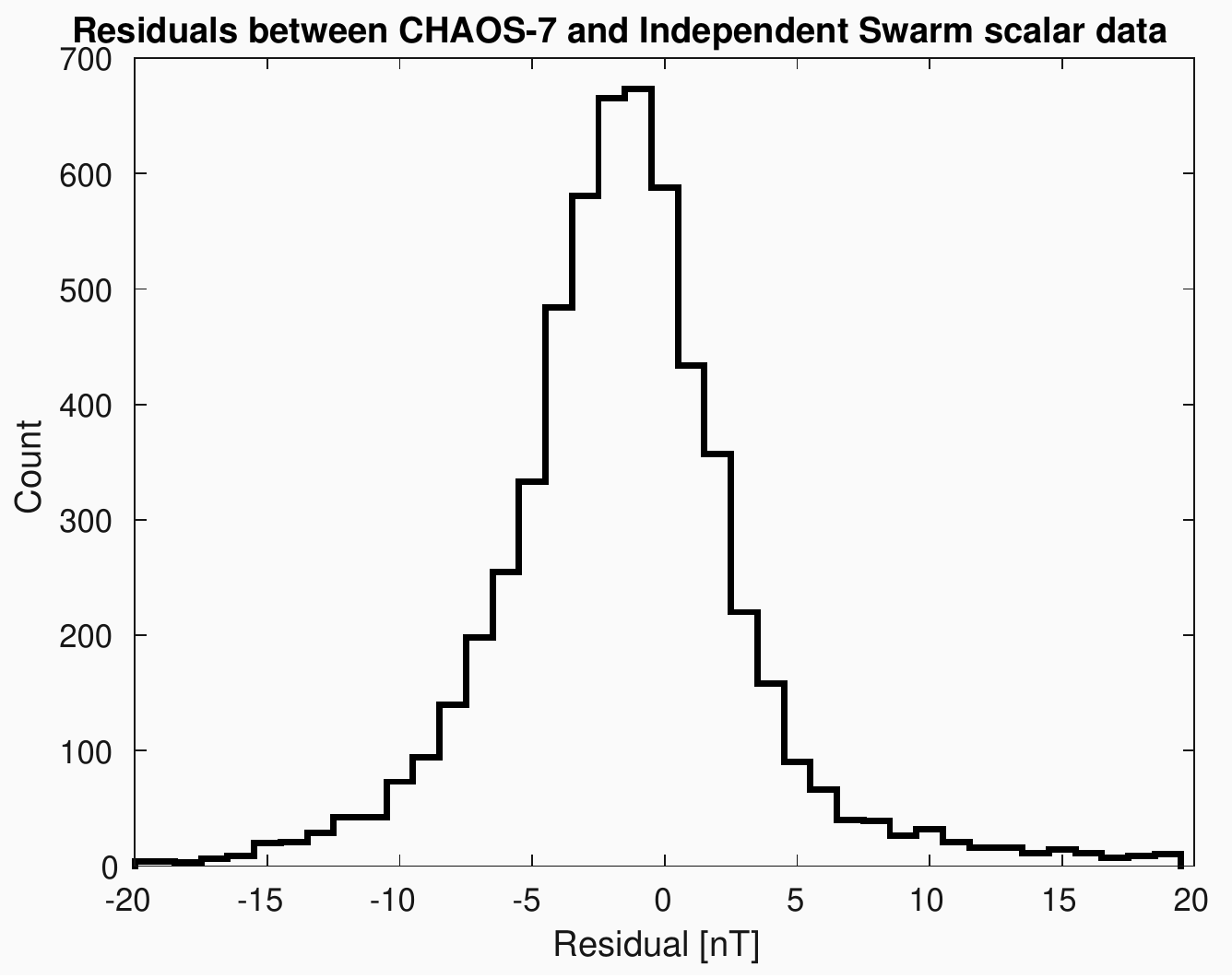}}
\caption{Histograms showing distribution of residuals between CHAOS-7 predictions and independent validation data up to February 2020, not included in the construction of CHAOS-7.}
\label{fig:valid}
\end{figure}

In an attempt to test the performance of CHAOS-7 we have performed comparisons with independent data collected up to February 2020, more than 5~months after the construction of the model.  These consist of (i) new secular variation data from ground observatories: annual differences of revised monthly means, derived from newly reported hourly mean values, using version 0122 of the ground observatory hourly mean database AUX\_OBS prepared by BGS \citep{MacMillanEtAl:2013} \--  an earlier version 0121 was used in the construction of CHAOS-7, and (ii) new data collected from the \textit{Swarm} satellites between September 2019 and February 2020.  We consider only satellite data that fulfills the same selection criteria used in the construction of CHAOS-7 and for simplicity we focus here on scalar data. In making predictions for the new satellite data we use both the internal part of CHAOS-7, with a linear extrapolation after the formal end of the model in 2020, and the external part of CHAOS-7 with input from the an updated version of the RC index (see \url{http://www.spacecenter.dk/files/magnetic-models/RC/}, and note that RC tapers to Dst for real-time values so is less accurate in this test for times after September 2019).  Histograms of the residuals for the newly reported ground observatory secular variation data and the newly collected \textit{Swarm} scalar data are presented in Figure~\ref{fig:valid} and misfit statistics (unweighted mean and rms residuals) are given in Table~\ref{tab:valid}, similar unweighted statistics for data actually used in the construction of CHAOS-7 are presented for reference.

\begin{table*}[!ht]
	\caption{Statistics of the misfit between CHAOS-7 predictions and independent validation data, that was not used in the CHAOS-7 model construction. Mean and rms residuals are calculated without Huber-weighting, in contrast to earlier tables. Similar statistics for the data used in the construction of the CHAOS-7 model are also shown for reference.
			\label{tab:valid}}
	\vspace{2truemm}
{\scriptsize
\begin{tabular}{r|r|rrr|rr}
& & \multicolumn{5}{c}{ $\quad$Validation dataset  $\quad$   CHAOS-7 dataset }  \\
	                                          &  &  $N$ &  mean &      rms  & mean & rms \\ 
	                                          \hline
Ground Obs RMM SV & $\mathrm{d}B_r/\mathrm{d}t$ [nT/yr]                       &   909  &   0.01 &     3.95 & 0.13 & 4.30 \\ 
    & $\mathrm{d}B_\theta/\mathrm{d}t$  [nT/yr]                  &   909  &  -0.35 &     3.79 & -0.24 & 4.17 \\ 
           & $\mathrm{d}B_\phi/\mathrm{d}t$  [nT/yr]                    &   909 &    0.09 &     3.31 & 0.04 & 3.93 \\ 
\hline
Swarm Scalar & $F$        [nT]                 &  5989  &   -1.37 &     5.76 & 0.19 & 4.18 \\ 
 & $F_{polar}$        [nT]                 &   3771  &   -1.67 &     6.49 & 0.61 & 6.62 \\ 
  & $F_{non-polar}$        [nT]                 &   2127  &   -0.84 &     4.16 & -0.04 & 1.96 \\ 
\hline
\end{tabular} 
}
\end{table*}

We find that the independent ground secular variation data are fit to within 4~nT/yr,  as good as the fit to the data actually used in the model construction.  However it should be acknowledged that the selection of new data is not random, only 313 of the 909 newly reported vector triples of SV data were recorded after the date of the last ground SV data used in CHAOS-7 (the remainder fill gaps in data coverage at earlier times), and there is certainly a bias towards observatories that regularly report new data.  In addition there is a different distribution of station latitudes in the validation dataset compared with the original dataset.  Nevertheless, the histograms of residuals for the SV validation data in Fig.\ref{fig:valid} shows no evidence for systematic bias and there are few residuals greater than 10~nT/yr. There is however a trend towards slightly larger mean residuals for the new SV data collected after 2019.0 (mean residuals post-2019.0 are 0.74, -1.18, -0.70 nT/yr for the radial, southward and eastward components respectively); this is not a consequence of the impact of spurious offsets at a small number of observatories.

The rms misfit to independent \textit{Swarm} scalar data, considering all latitudes is 5.76\,~nT, which is 1.58\,~nT higher than for the data used in the model construction. There is also a noticeable negative skew to the residual distribution.  This is to be expected, given the limitations of using the CHAOS-7 external field model in predictive mode, due to RC merging with D$_{st}$ at the end of the RC series, and because time-dependent SM offset parameters (the "RC baseline correction") are not available.  

Despite their limitations, we are encouraged by these tests that CHAOS-7 does a satisfactory job in predicting the observed magnetic field values on ground and at satellite altitude up to 5~months after the model construction, a length of time comparable to that between consecutive CHAOS model updates.

\subsection{Extraction of IGRF candidate models}
DTU's candidate models for IGRF-13 were obtained from the parent model CHAOS-7 parent as follows:
\begin{itemize}
\item {\bf DGRF, epoch 2015.0}\\
The parent model CHAOS-7 was evaluated at epoch 2015.0 and the spherical harmonic coefficients for the internal field up to degree and order 13 were output to 0.01~nT.
\item {\bf IGRF, epoch 2020.0}\\
CHAOS-7 was evaluated at epoch 2019.75 the last date for which satellite data were available for constructing the model.  These coefficients were then propagated forward to epoch 2020.0, using the linear SV from  CHAOS-7 in epoch 2019.0 (this epoch was chosen to avoid spline-model end effects and because this was the time of the final annual differences of monthly means of ground observatory data that directly constrain the SV), for all spherical harmonics up to degree 13, as follows:
\begin{equation}
g^m_n(t_{2020.0})=g^m_n(t_{2019.75}) + 0.25\text{ yr} \cdot \dot{g}^m_n(t_{2019.0})
\label{eqn:IGRF}
\end{equation}
Here $g^m_n$ represents each of the Gauss coefficients $\{g_n^m,h_n^m\}$, $\dot{g}^m_n$ represents the SV coefficients $\{\dot{g}_n^m,\dot{h}_n^m\}$ in~nT/yr.
The resulting spherical harmonic coefficients for the internal field in epoch 2020.0 up to degree and order 13 were output to 0.01\,nT.\\
\item {\bf Prediction for the average SV, 2020.0 to 2025.0}\\
Due to possible spline-model end effects in the secular acceleration, we chose to evaluate the SV from CHAOS-7 at epoch 2019.0, rather than in 2020.0.  No extrapolation was attempted.  
 The spherical harmonic coefficients for the SV (the first derivative in time of the spline-based CHAOS-7 model) for the internal field in epoch 2019.0, up to degree and order 8 were output to 0.01\,nT/yr.  
\end{itemize}

We did not provide uncertainty estimates along with our candidate models since we are not able to calculate such estimates in a rigorous fashion; the formal error will be unrealistically small as data error correlations are ignored and the model is incomplete.  The largest errors will be due to biases caused by sources not modelled (e.g.\ due to the polar electrojet).\\

\subsection{Time-dependence of SV coefficients}
\label{Sect:SV_comp}
\begin{figure}[!ht]
\centerline{\includegraphics[angle=0, width=0.33\textwidth]{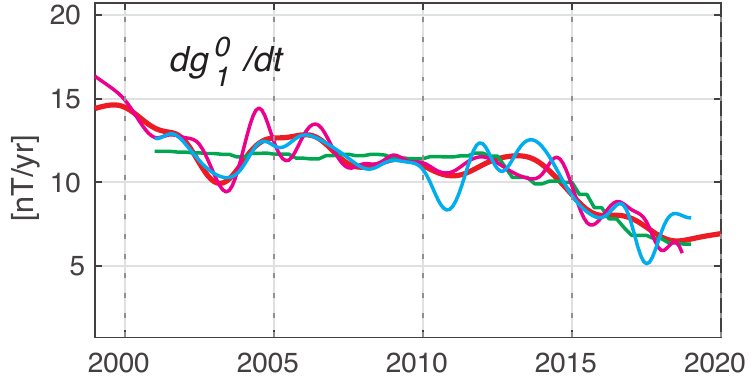}\includegraphics[angle=0, width=0.33\textwidth]{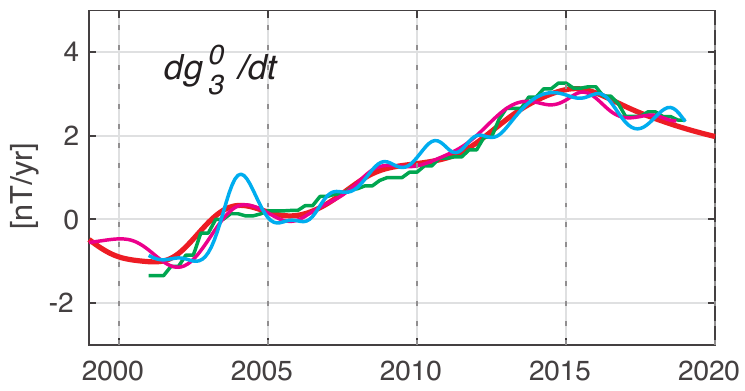}\includegraphics[angle=0, width=0.33\textwidth]{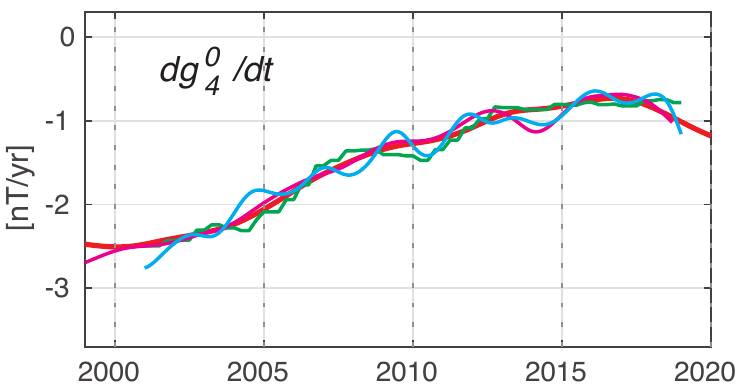}}
\centerline{\includegraphics[angle=0, width=0.33\textwidth]{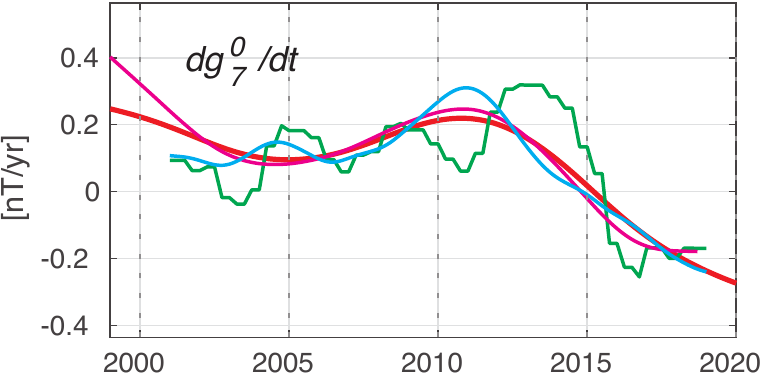}\includegraphics[angle=0, width=0.33\textwidth]{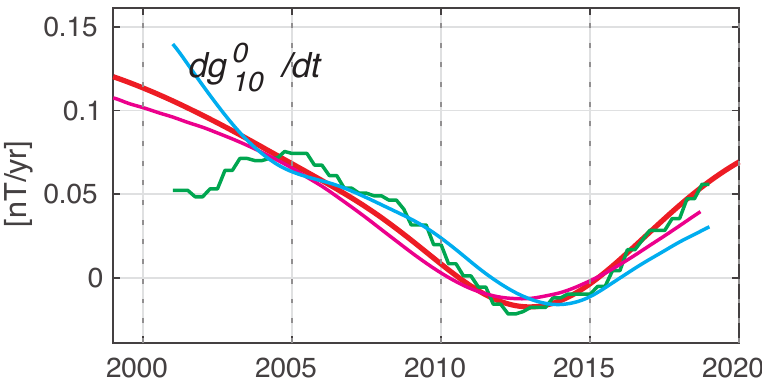}\includegraphics[angle=0, width=0.33\textwidth]{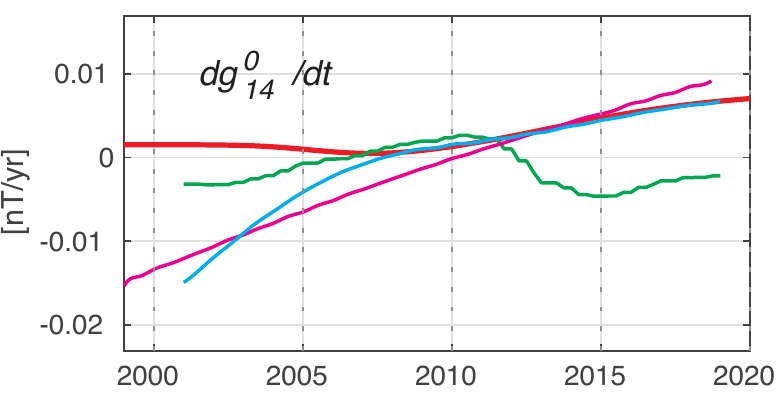}}
\centerline{\includegraphics[angle=0, width=0.33\textwidth]{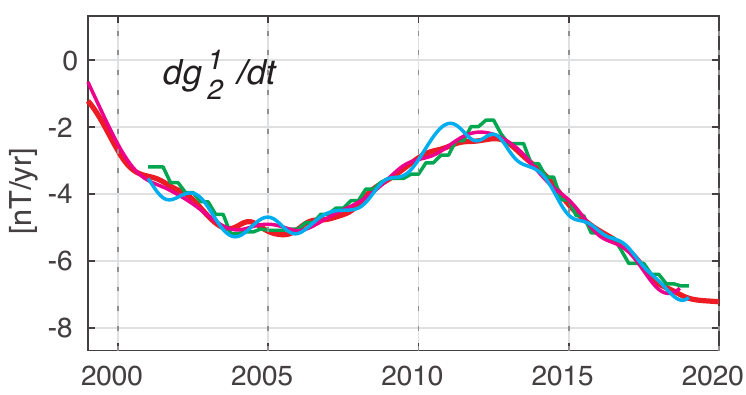}\includegraphics[angle=0, width=0.33\textwidth]{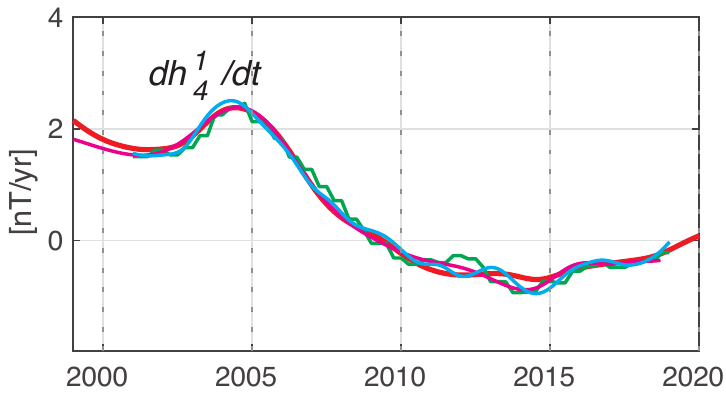}\includegraphics[angle=0, width=0.33\textwidth]{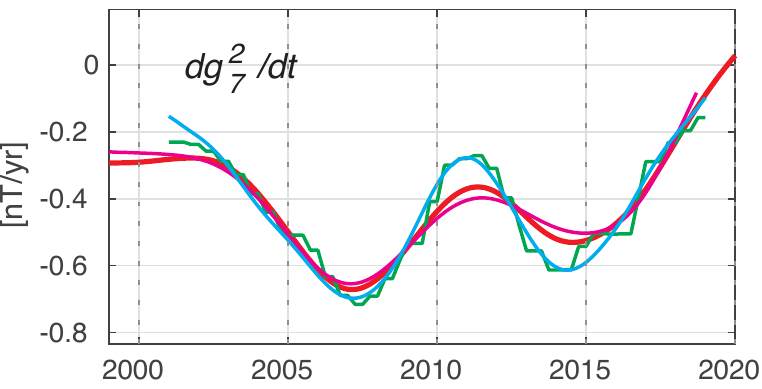}}
\centerline{\includegraphics[angle=0, width=0.33\textwidth]{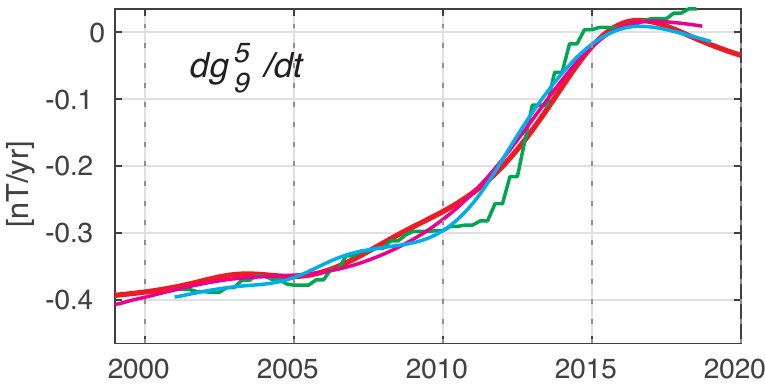}\includegraphics[angle=0, width=0.33\textwidth]{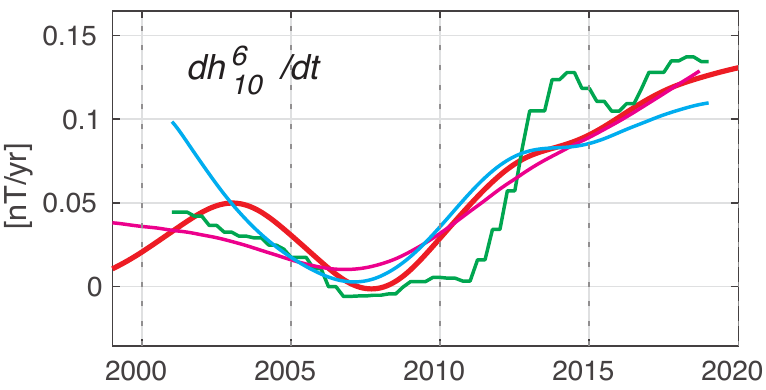}\includegraphics[angle=0, width=0.33\textwidth]{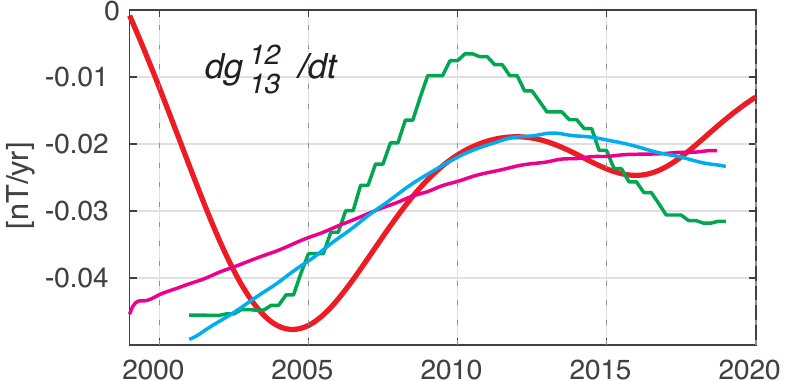}}
\centerline{\includegraphics[angle=0, width=0.33\textwidth]{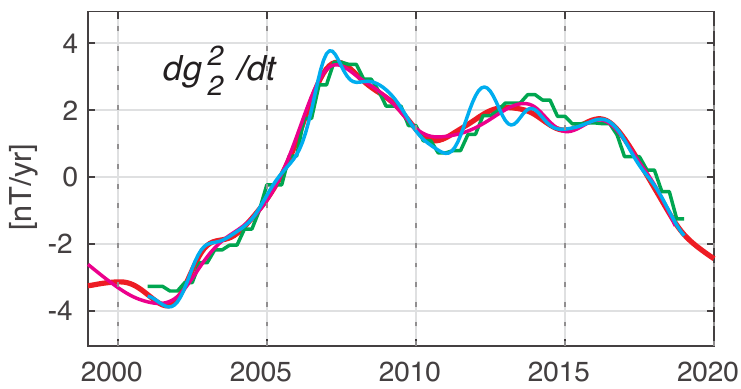}\includegraphics[angle=0, width=0.33\textwidth]{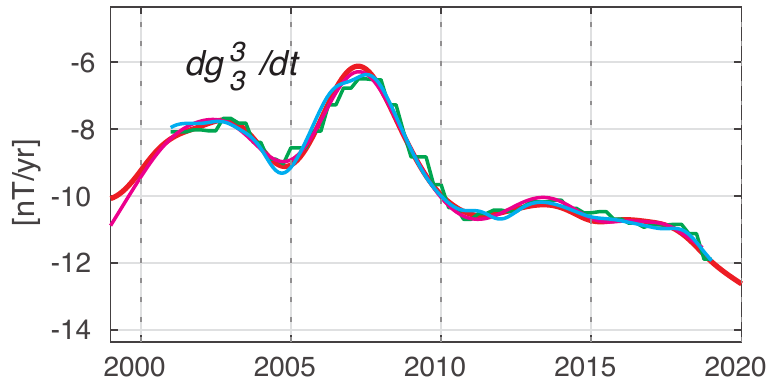}\includegraphics[angle=0, width=0.33\textwidth]{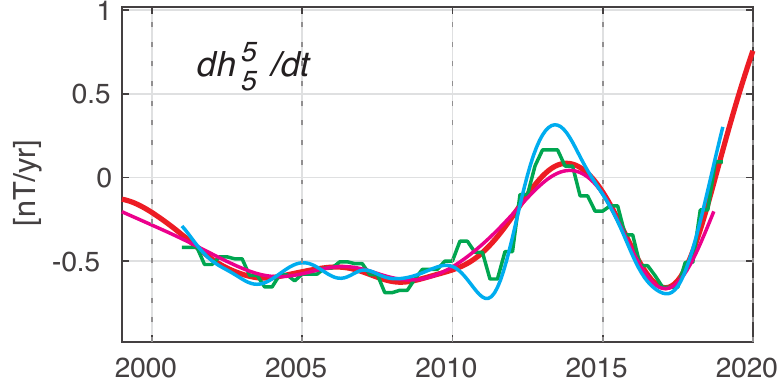}}
\centerline{\includegraphics[angle=0, width=0.33\textwidth]{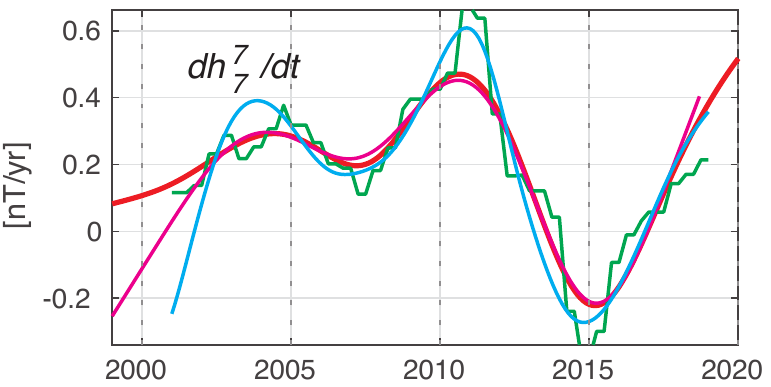}\includegraphics[angle=0, width=0.33\textwidth]{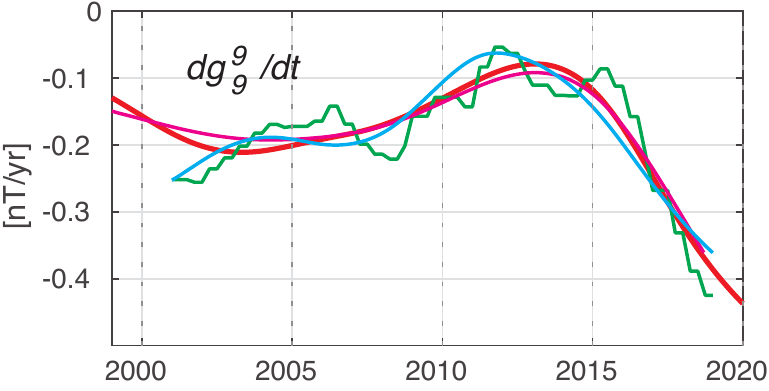}\includegraphics[angle=0, width=0.33\textwidth]{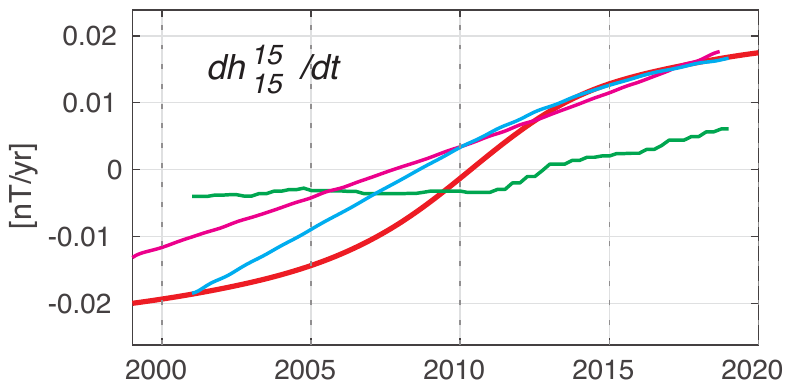}}
\caption{Time-dependence of example spherical harmonic coefficients of the internal field SV from CHAOS-7 (solid red line). Also shown are the CM6 model of \cite{Sabaka:2020}(magenta line), MCO\_SHA\_2Y a preliminary model following the approach of \citep{Ropp:2020} (green line) and model A of \cite{Alken:2020} (cyan line).  Top two rows are zonal coefficients, bottom two rows are sectoral coefficients, middle two rows are tesseral coefficients.
\label{fig:SV_COEF}}
\end{figure}

Having established the ability of CHAOS-7 to represent adequately magnetic measurements made onboard satellites and at ground observatories over the past 21 years, we now proceed to present it's predictions concerning the structure of the global geomagnetic field and its evolution during this period.  We begin by presenting in Fig.~\ref{fig:SV_COEF} time series of the first time derivatives of the Gauss coefficients $dg^m_n/dt$ and $dh^m_n/dt$ between 1999 and 2020.  For reference purposes we also present similar series for model A of \citet{Alken:2020}, model MCO\_SHA\_2Y version 0101 \-- an early version of the model described by \citet{Ropp:2020}, and the CM6 model of \citet{Sabaka:2020}.  The top two rows show a selection of zonal harmonics, the bottom two rows examples of sectoral harmonics and the middle two rows selected tesseral harmonics, in each case increasing in degree within the two rows.

Due to the difficulties in accurately modelling the rapidly-varying magnetospheric field, and in separating externally-driven signals induced in the electrically-conducting Earth from core field variation on timescales of months to years, the estimated SV of the zonal terms show interesting differences between the four models.  Considering the axial dipole, they show differences of 1-2~nT/yr.  Compared to CHAOS-7, the MCO\_SHA\_2Y model \citep{Ropp:2020}, which seeks to co-estimate induced signals, shows less time variation in its presented core field part between 2000 and 2012; during this time its axial dipole SV is almost constant. On the other hand CM6 \citep{Sabaka:2020}, which seeks to estimate induction using an a-priori conductivity model, shows prominent oscillations on periods around 1 year between 2003 and 2008.  Model A of \cite{Alken:2020} shows larger oscillations than CHAOS-7, with periods of 1 \--- 2 years between 2010 and 2019. Here it should be remembered that CHAOS-7 was constructed using enhanced temporal regularization of the zonal terms; this is not the case for the other models. The MCO\_SHA\_2Y model displays an oscillation in  $dg_7^0/dt$ with a period 3 to 5 years which is not present in the other models.  After 2005 there is good agreement for $dg^{0}_{10}/dt$ between the four models.

For the tesseral components there is typically rather good agreement across the four models at least up to degree 10, with the MCO\_SHA\_2Y model  and model A of \cite{Alken:2020} showing relatively more variability and CM6 being relatively smoother as the spherical harmonic degree increases.  Turning to the sectoral harmonics, there is impressive agreement between the models for large scale harmonics such as $dg^3_3/dt$. The  MCO\_SHA\_2Y model and model A of \cite{Alken:2020} show sharper variations in $dh^7_7/dt$, all models show similar trends in  $dh^{15}_{15}/dt$ over the past twenty years.  There are larger differences in the time-dependence of the SV across the models on going to high degree; the time changes seen in CHAOS-7 are generally slightly stronger than those seen in CM6 and weaker than those seen in the  MCO\_SHA\_2Y model and in Model A of \citet{Alken:2020}.  These differences reflect differences in the selected data, the level to which the modellers seek to fit the data and the prior assumptions or regularization applied in order to control the field's time-dependence.  Having a variety of different modelling approaches to compare thanks to the impetus of IGRF-13 is very informative, providing insight into which aspects are robust across the various approaches and which aspects are more challenging.


\subsection{Spherical harmonic spectra of the field and its time derivatives}
\label{sec:surf_spect}
\begin{figure}[!ht]
\centerline{CHAOS-7, Earth's surface}
\includegraphics[angle=0, width=0.65\textwidth]{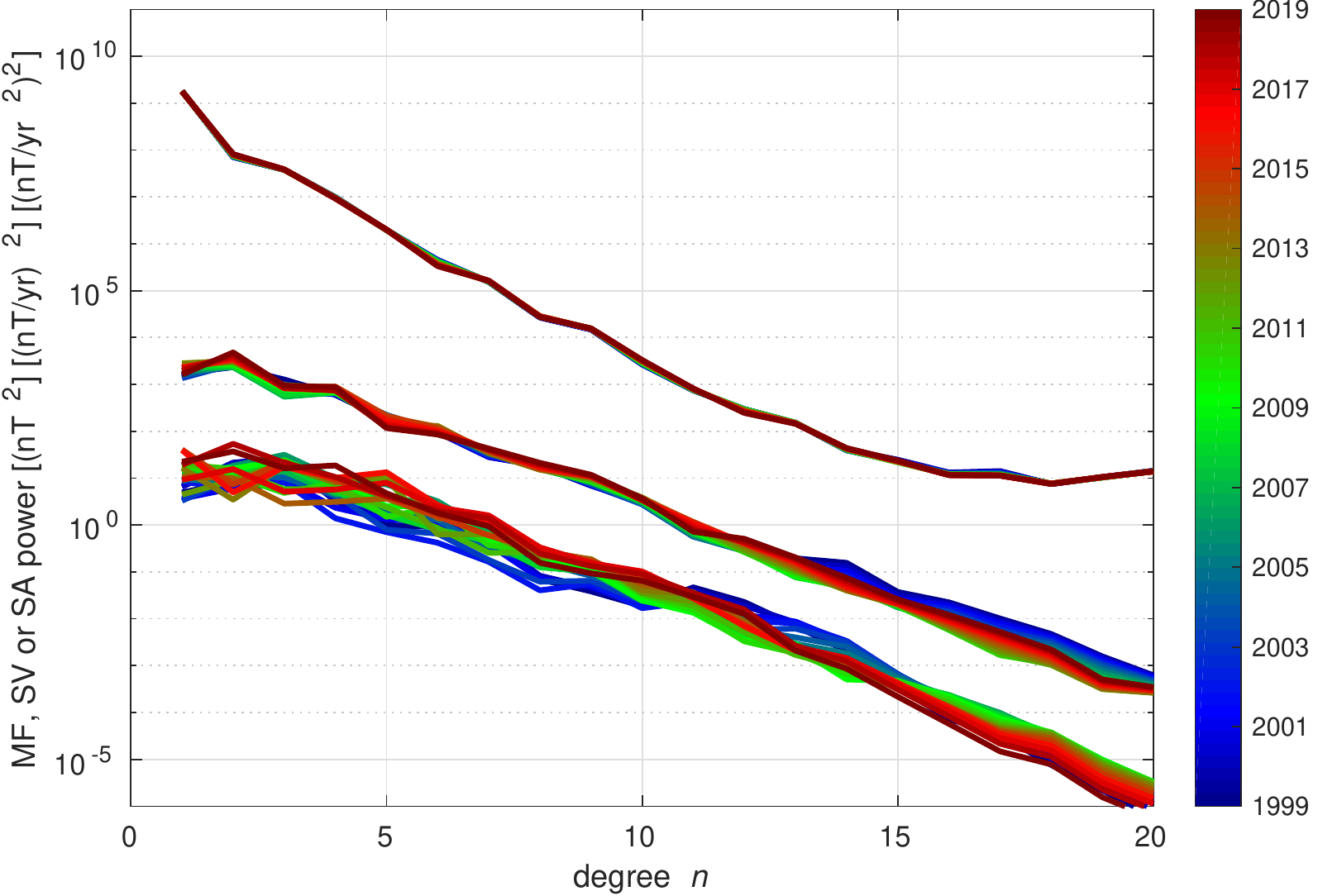} \\
\centerline{CHAOS-6-x9, Earth's surface}

\includegraphics[angle=0, width=0.65\textwidth]{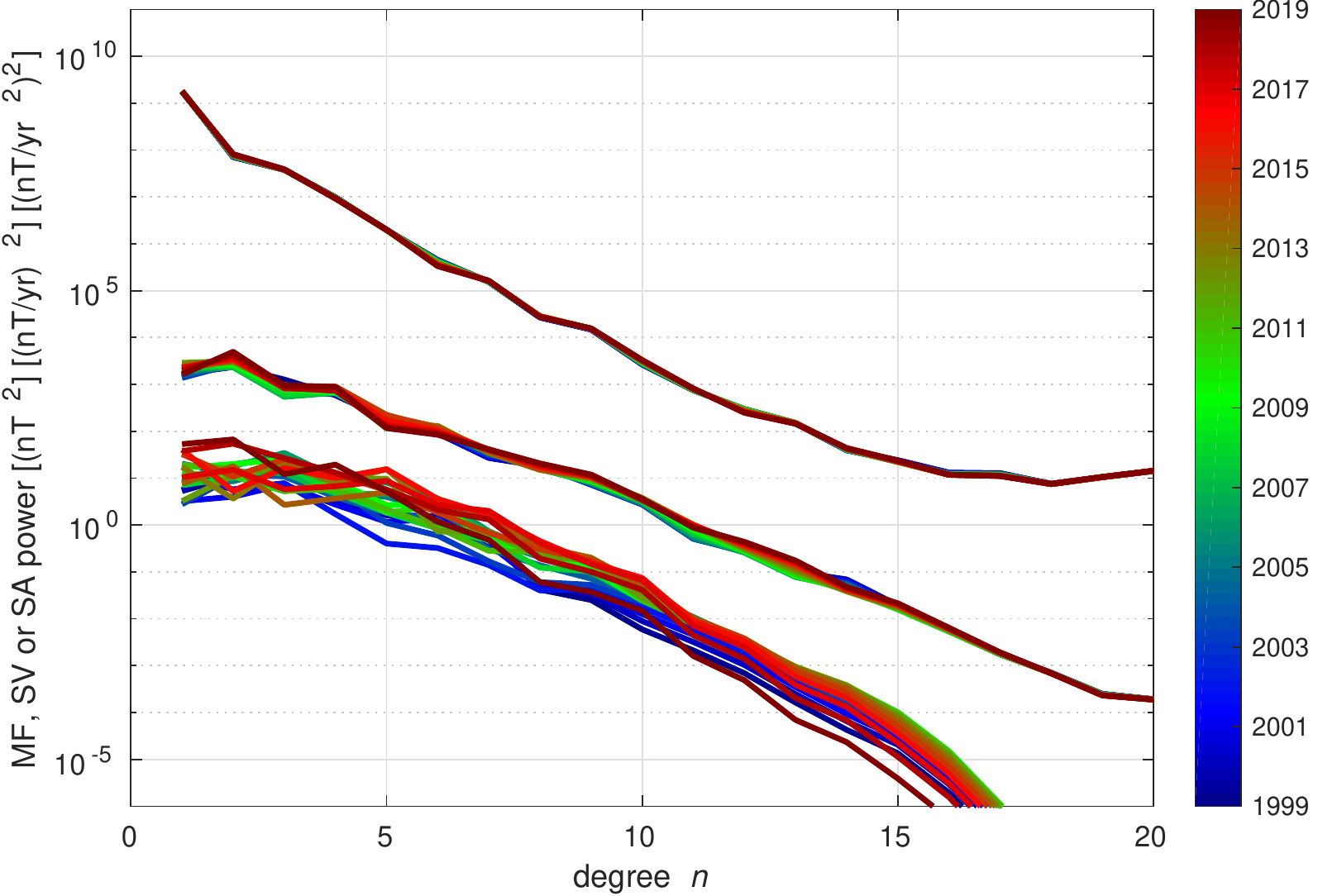}

\caption{Spherical harmonic spectra of the time-dependent mean square internal vector magnetic field, its Secular Variation, and Secular Acceleration at the Earth's surface, up to spherical harmonic degree 20.  Colours indicate the epoch, blue colours for 1999 to 2004, green for 2005 to 2013, red for 2014 to 2020.  Top is for CHAOS-7, bottom is for reference CHAOS-6-x9.
\label{fig:Spectra}}
\end{figure}

\begin{figure}[!ht]
\centerline{CHAOS-7, Core-mantle boundary}

\includegraphics[angle=0, width=0.65\textwidth]{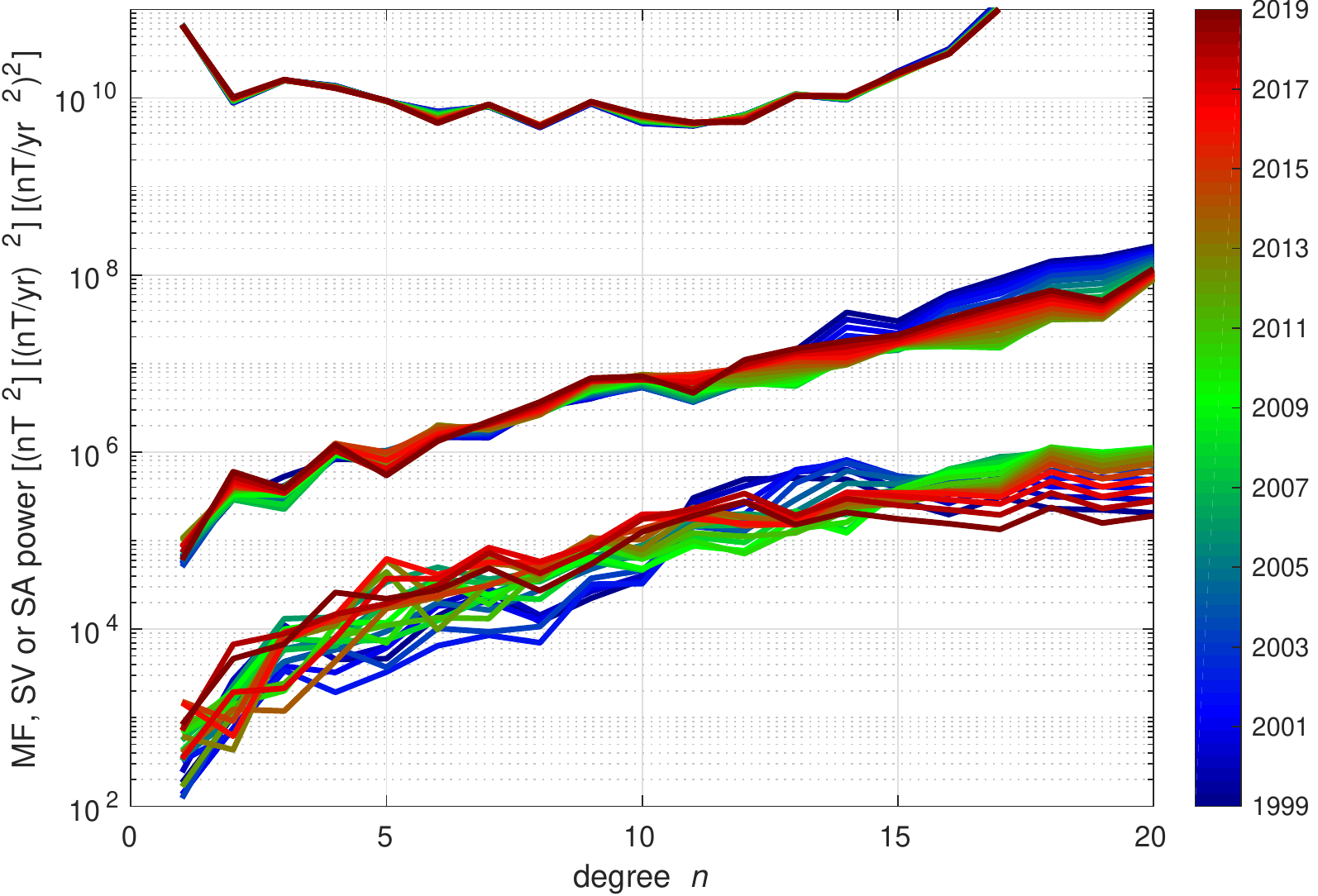} 

\centerline{CHAOS-6-x9, Core-mantle boundary}
\includegraphics[angle=0, width=0.65\textwidth]{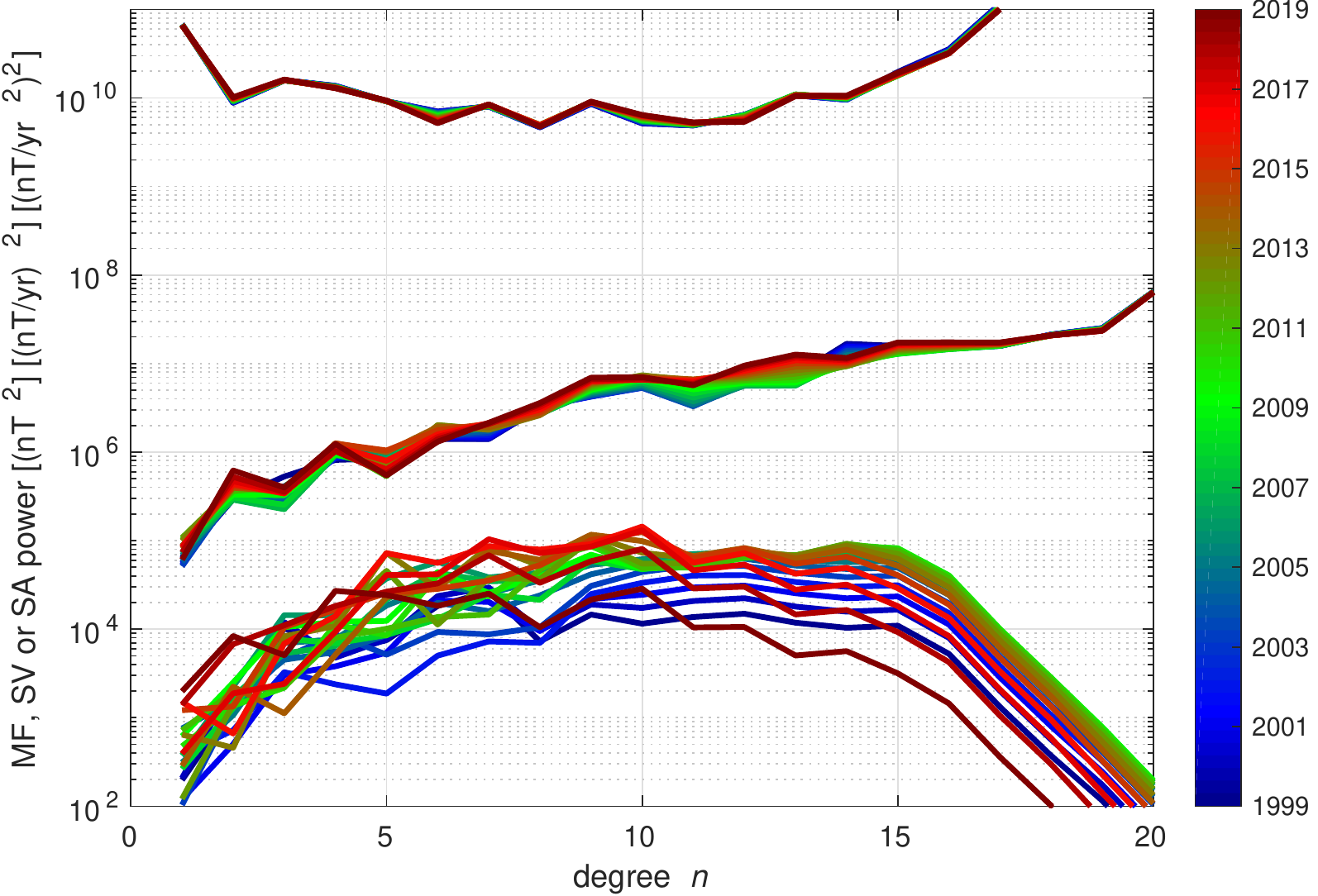}
\caption{Spherical harmonic spectra of the time-dependent mean square internal vector magnetic field, its Secular Variation and Secular Acceleration at the core-mantle boundary, up to spherical harmonic degree 20.  Colours indicates the epoch, blue colours for 1999 to 2004, green for 2005 to 2013, red for 2014 to 2020.  Top is for CHAOS-7, bottom is for reference CHAOS-6-x9.
\label{fig:Spectra_CMB}}
\end{figure}

Spherical harmonic spectra for the internal field, its first time derivative (the SV) and its second time derivative (the SA) are displayed, colour coded by epoch, at the Earth's surface and at the core-mantle boundary in Fig.~\ref{fig:Spectra} and Fig.~\ref{fig:Spectra_CMB} respectively. CHAOS-7 and the final CHAOS-6 model, CHAOS-6-x9, constructed used data up to May 2019, are shown for comparison.  
The spectra for the main field are very similar, but CHAOS-7 shows more time-dependence in its high degree SV, particularly at degree 14 and above.  The SA spectra decrease more gradually at the Earth's surface in CHAOS-7, and at the core-mantle boundary they continue to trend upwards in contrast to the sharp decrease in the SA spectrum above degree 14 seen in CHAOS-6-x9.

The reason for the different behaviour of the high degree SA in CHAOS-7 is the decrease in the strength of its high degree temporal regularization compared to CHAOS-6x9.  The relaxation of the temporal regularization at high degree was made in order to enable the study of high degree SA during the {\it Swarm} era; there was a concern the strong regularization imposed at high degree in CHAOS-6 was preventing such changes from being recovered.  This change had the desired result: the high degree SA since 2014, when {\it Swarm} data begin to provide constraints, and which was the focus for CHAOS-7 in order to provide candidate models for IGRF-13, is stable and reasonable.  However, prior to 2005 the SA at degrees above 10 shows evidence for instability.  In a more recent updates of the CHAOS-7 model, attempts have been made to remedy this by increasing the upper limit to the tapering of the temporal regularization from degree 11 up to 15.

\clearpage

\subsection{{\bf Maps of the field, secular variation and secular acceleration}}
\label{sec:core_maps}
\begin{figure}[ht]
\centerline{\includegraphics[angle=0, width=0.75\textwidth]{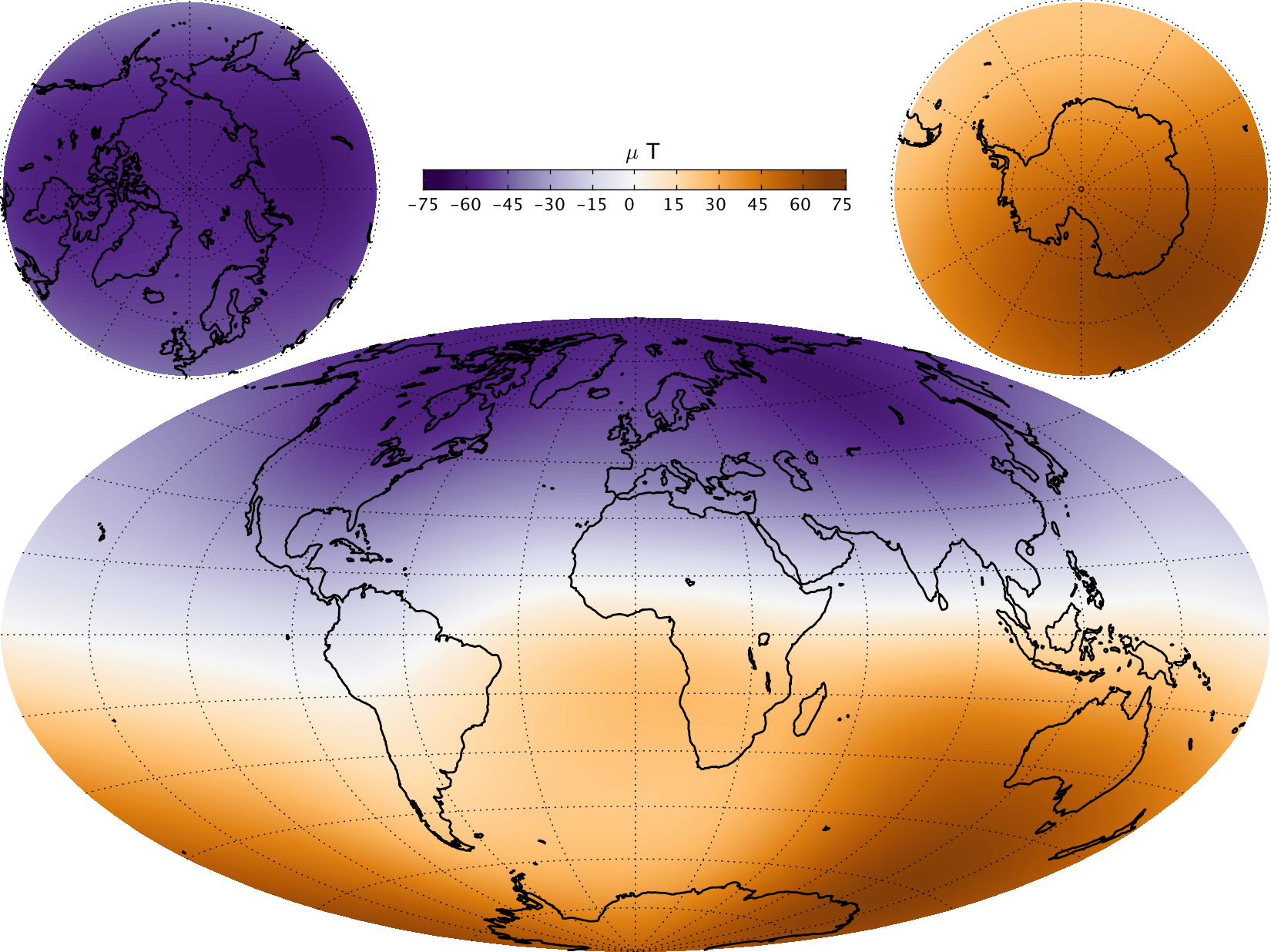} }
\centerline{\includegraphics[angle=0, width=0.75\textwidth]{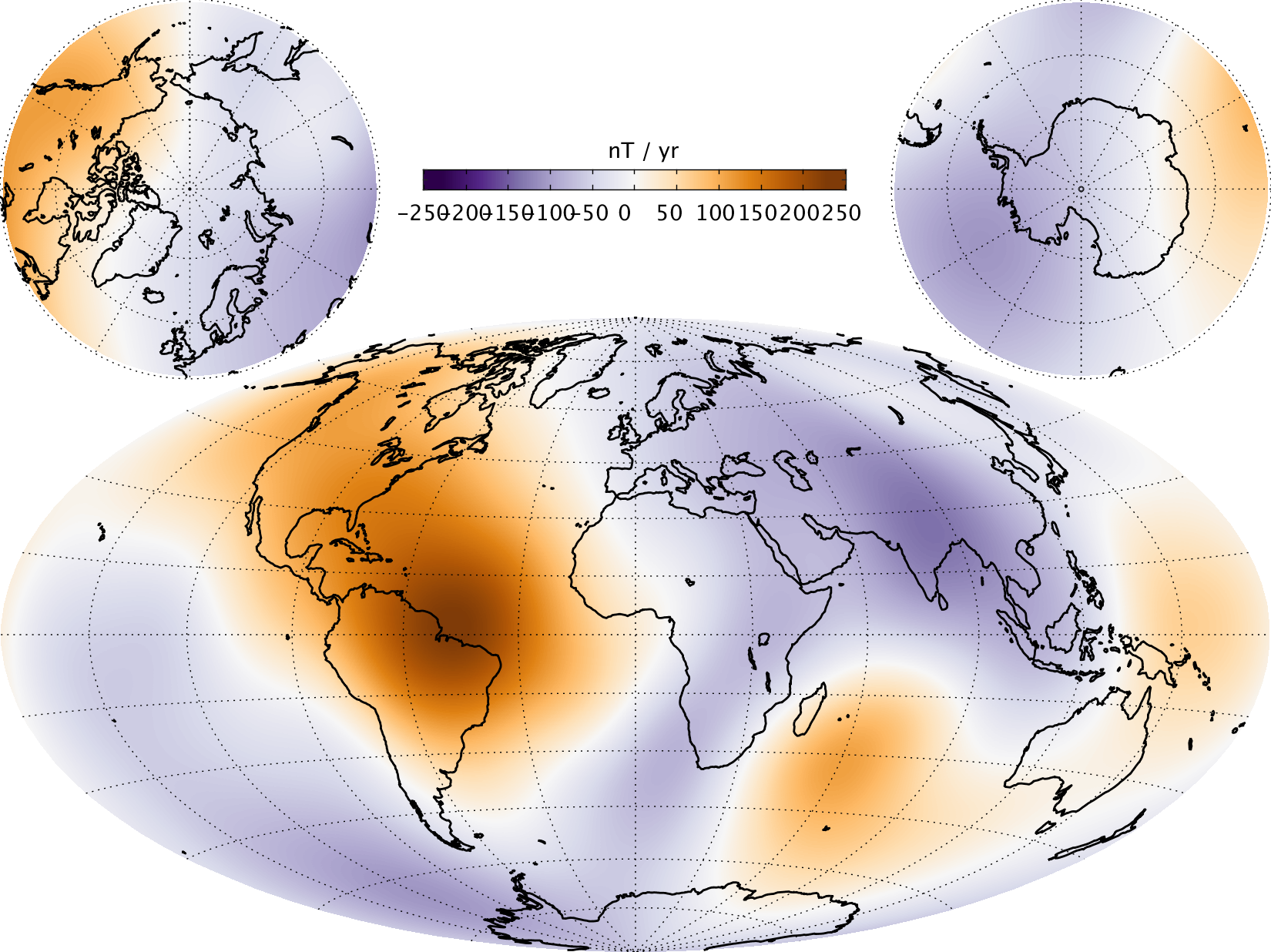} }
\centerline{\includegraphics[angle=0, width=0.75\textwidth]{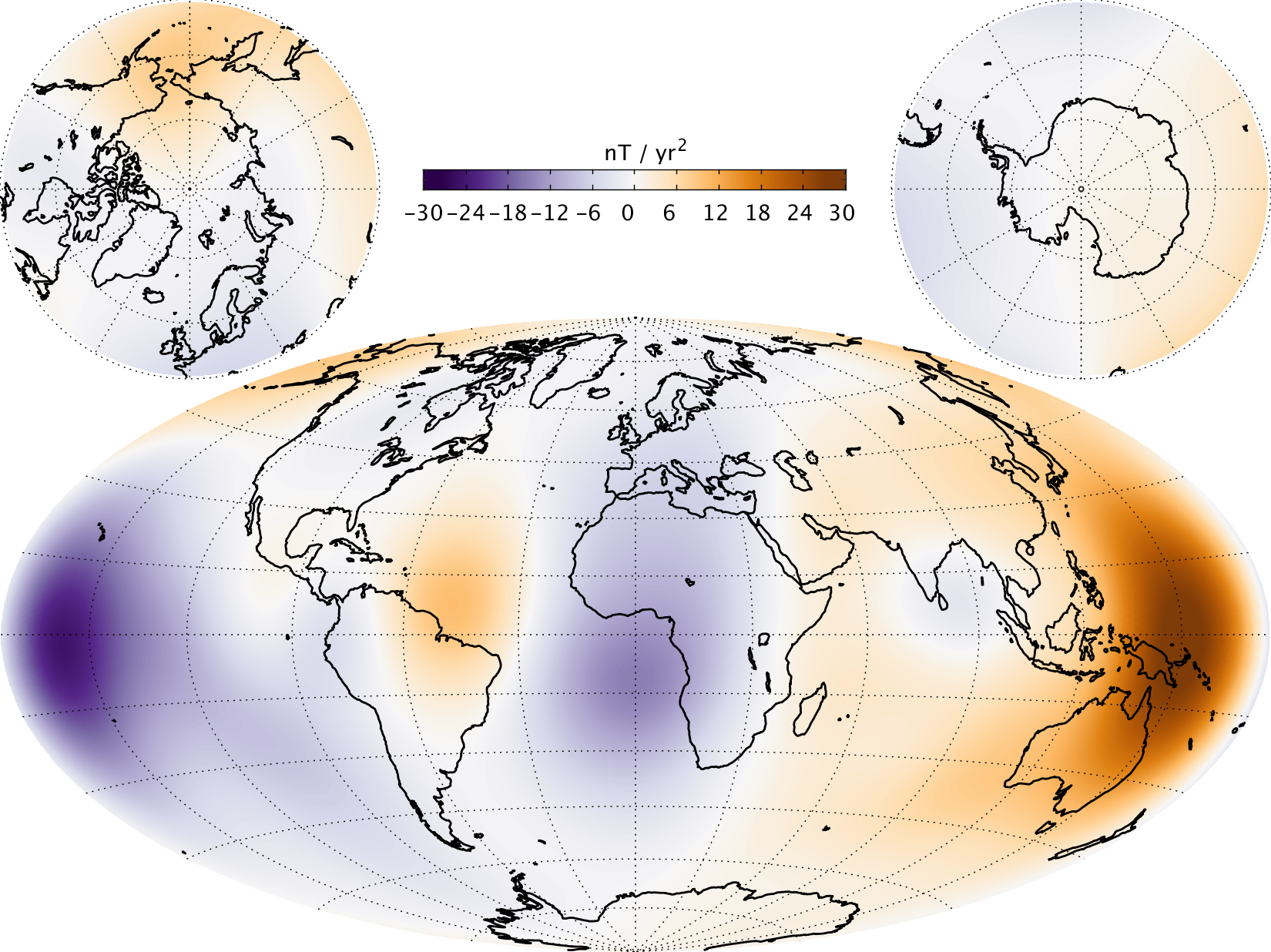} }
\caption{Maps of the radial magnetic field (MF, top row), its first time derivative (SV, middle row) and second time derivative (SA, bottom row) at the Earth's surface in 2019.0, up to SH degree 20. 
\label{fig:Maps}}
\end{figure}

\begin{figure}[ht]
\centerline{\includegraphics[angle=0, width=0.75\textwidth]{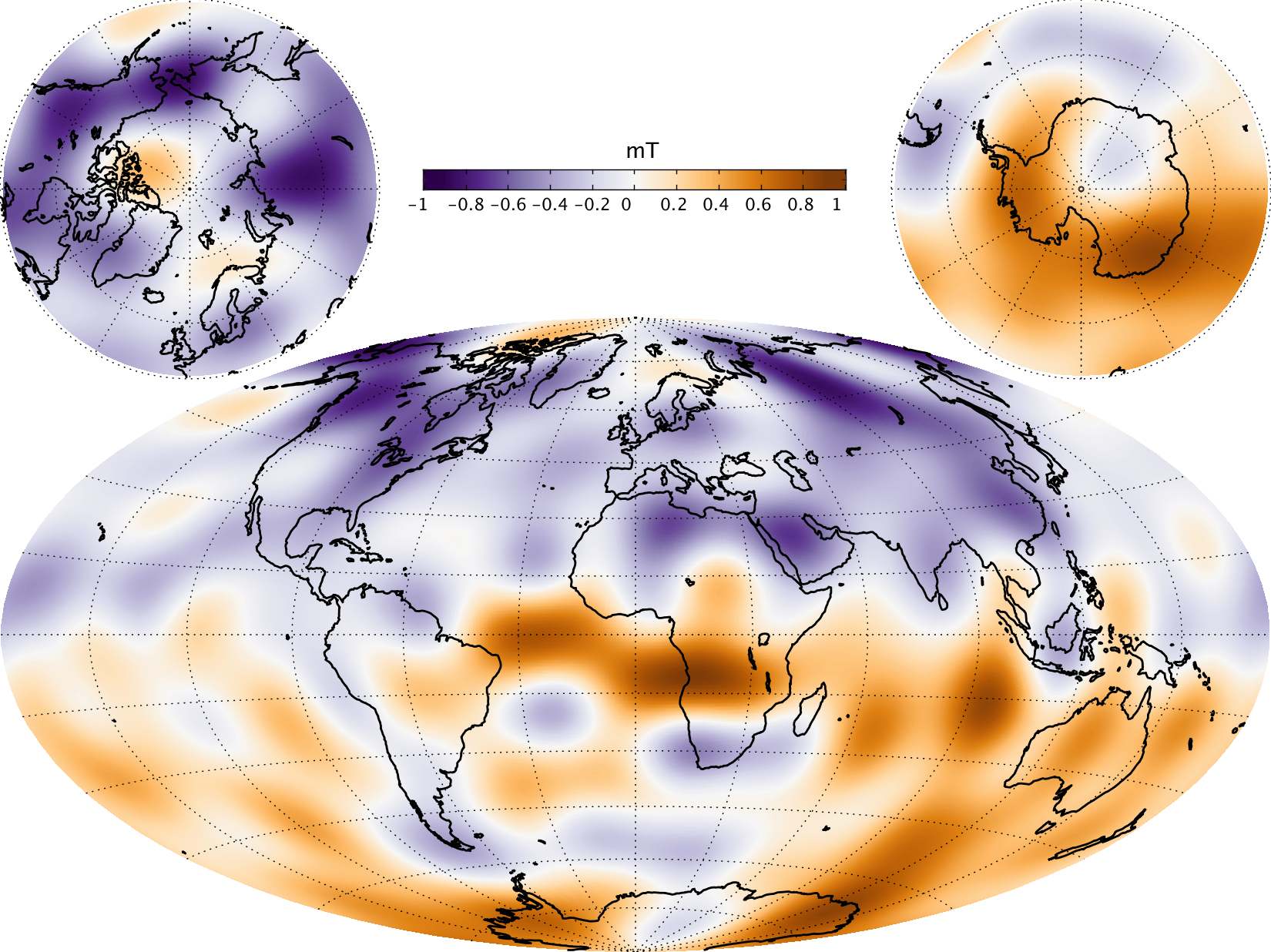}}
\centerline{\includegraphics[angle=0, width=0.75\textwidth]{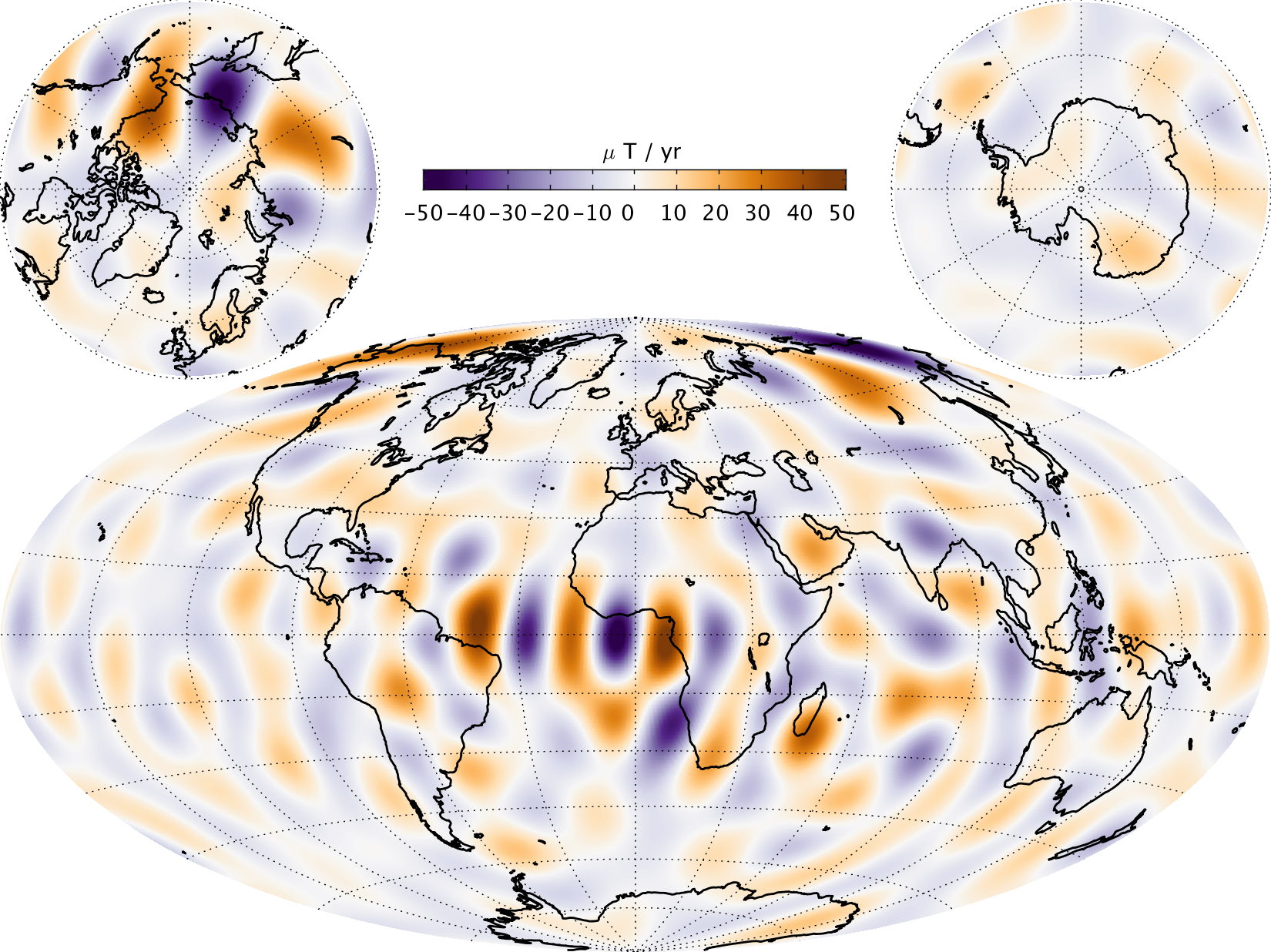}}
\centerline{\includegraphics[angle=0, width=0.75\textwidth]{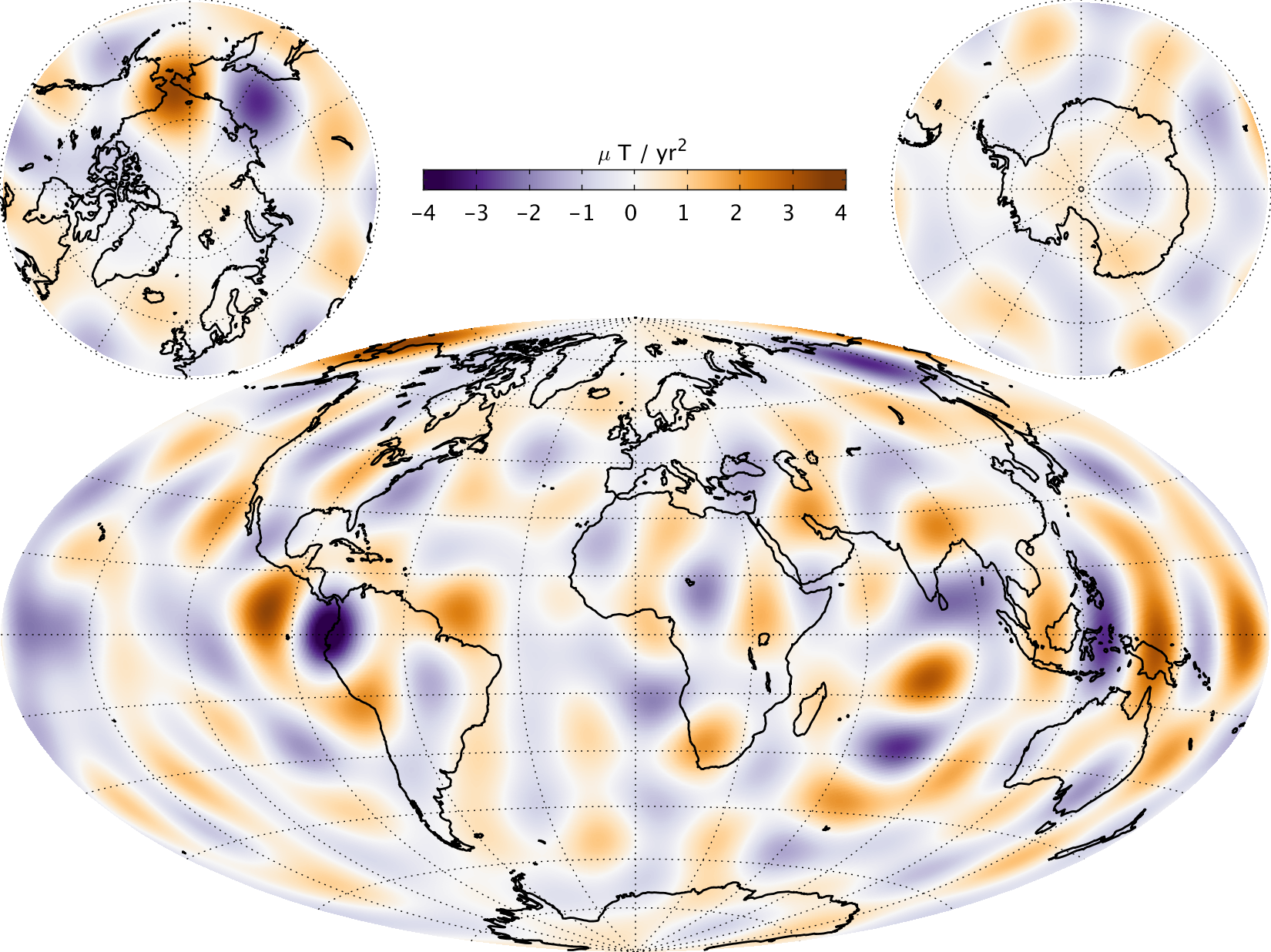}}
\caption{Maps of the radial magnetic field (MF, top row), its first time derivative (SV, middle row) and second time derivative (SA, bottom row) at the core-mantle boundary in 2019.0, Truncation degrees are 13, 17 and 15 respectively for the MF, SV and SA. 
\label{fig:Maps_CMB}}
\end{figure}

Fig.~\ref{fig:Maps} presents maps of the radial component of the main field (MF), its first time derivative (Secular Variation, SV) and its second time derivative (Secular Acceleration, SA) from CHAOS-7 at the Earth's surface, up to SH degree 20, in 2019.0.  Fig.~\ref{fig:Maps_CMB}  presents similar maps but downward-continued to the CMB and truncated at degrees 13, 17 and 15 respectively for the MF, SV and SA.\\

The surface radial magnetic field component has the well-known features of strong high latitude patches and a weaker radial magnetic field in the South Atlantic.  The radial field SV shows the region of largest radial field increase lies in the north-east corner of South America. On the other hand there is a band of decreasing radial field extending south west from Southern Africa.   The radial SA at the surface shows an intense localized dipolar structure in the Pacific (seen here at opposite sides of the map), with negative acceleration in 2019 in the central Pacific, including near Hawaii, and a positive acceleration in 2019 in the western Pacific north-east of Australia.\\

Descending to the CMB involves making the assumption that induction in the mantle plays a minor role  on the timescales of several years and longer that are captured by the time-dependent internal field model.  The truncation degrees (of respectively degree 13 (MF), 17 (SV) and 15 (SA) were chosen to ensure the maps at the CMB in 2019 were stable and well behaved.  The strongest CMB SV features are found in the equatorial region between Africa and South America; they result from the westward motion of intense radial field features this region \citep{Olsen:2014a, Finlay:2016}.  Enhanced CMB SA is seen at eastern longitudes in a band 60 degrees to 90 degrees east.  There are also large scale accelerations in the Pacific region, corresponding to the surface features.  We return to this topic in Section~\ref{sec:rapid_SA_Pac}.\\


\clearpage


\section{Evolution of the South Atlantic Anomaly as seen by \textit{Swarm}}
\label{sec:SAA}
\begin{figure}[!ht]
\centerline{\includegraphics[angle=0, width=1.0\textwidth]{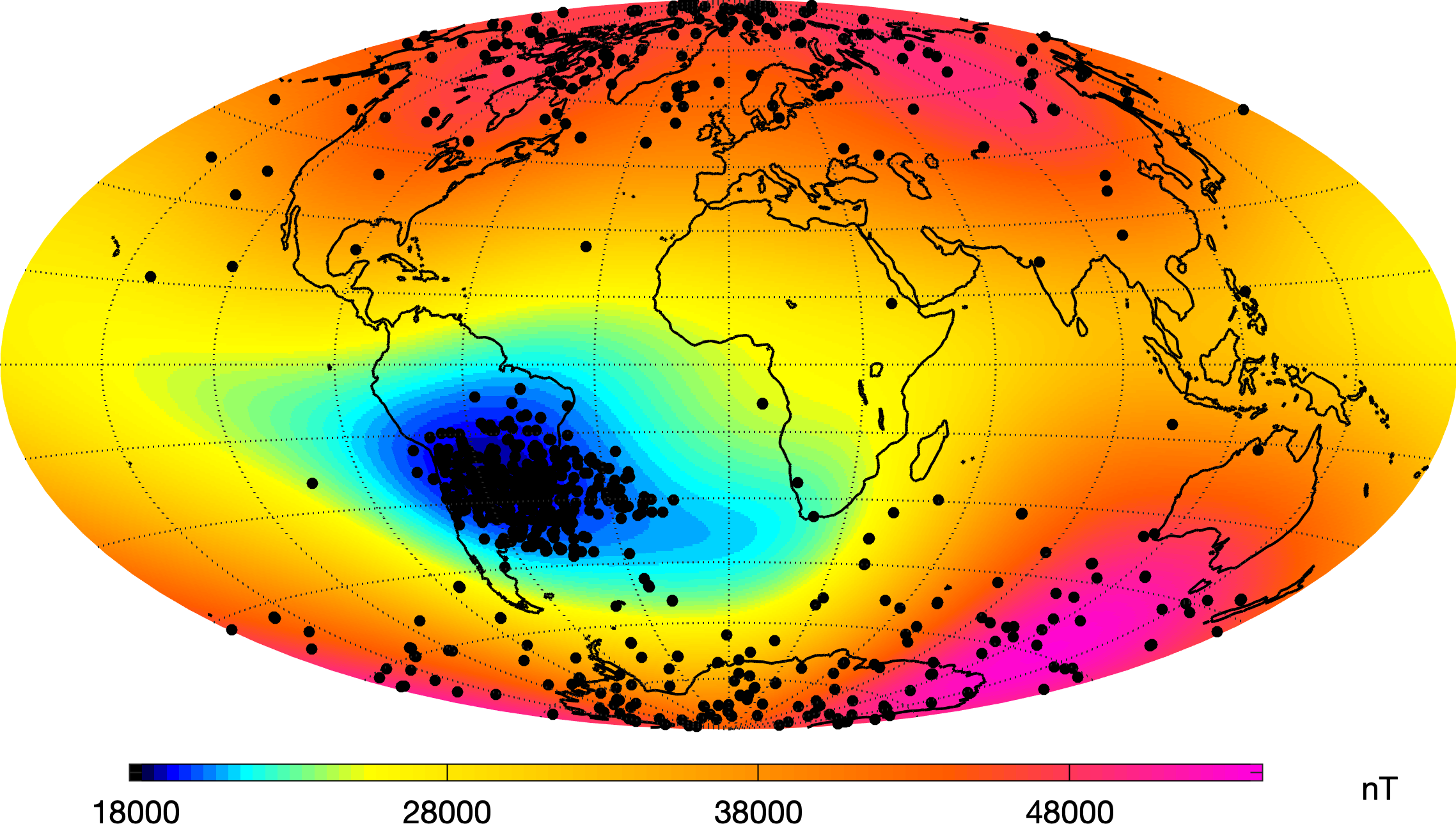} }
\centerline{\includegraphics[angle=0, width=1.0\textwidth]{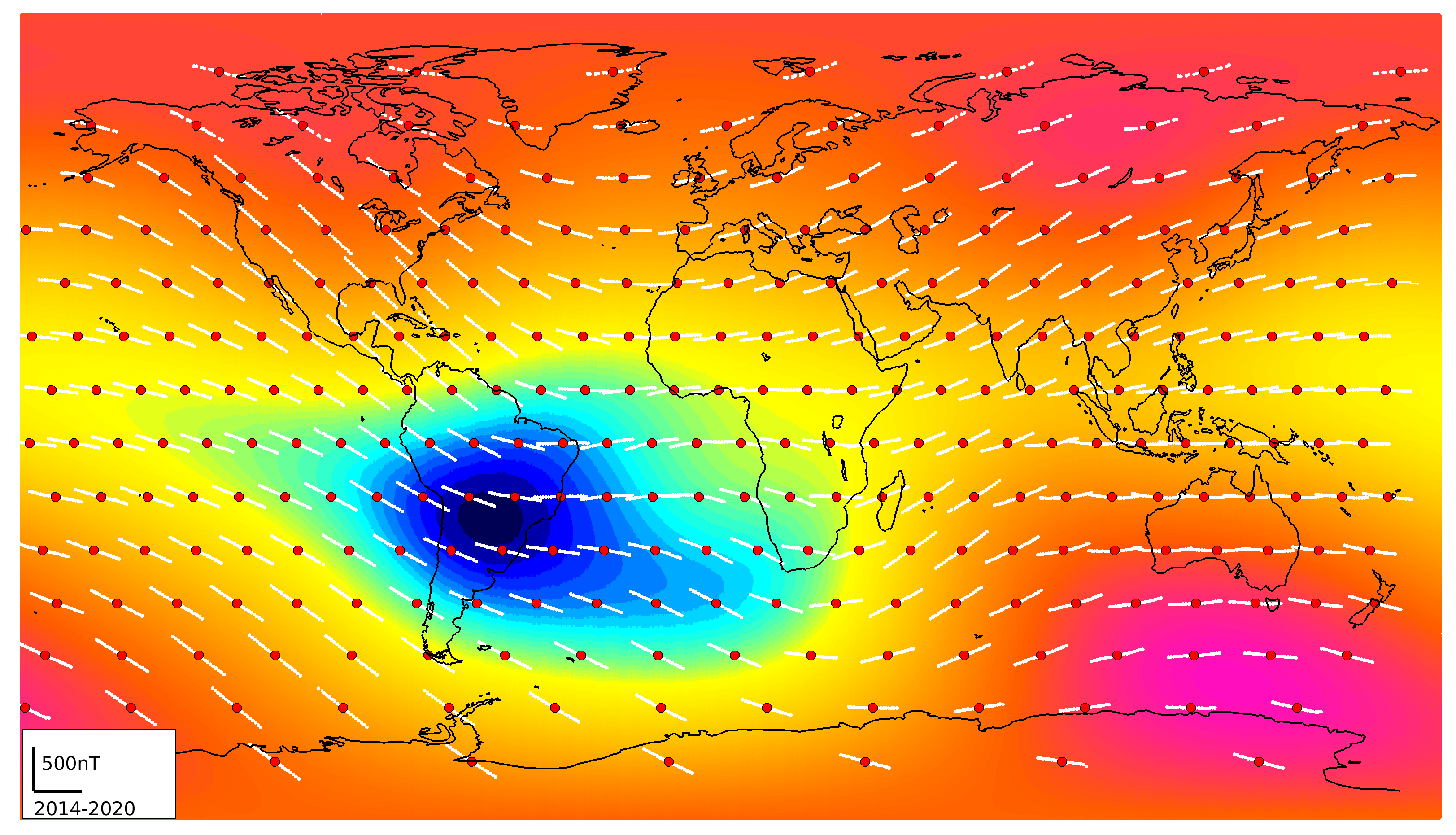} }
\caption{Top: Locations of single event upsets (SEUs) registered onboard the \textit{Swarm} satellites between November 2013 and August 2019 (black dots) plotted on top of the field intensity at 450 km altitude in August 2017 according to the CHAOS-7 model.  Bottom: Field intensity from the CHAOS-7 field model in August 2017 at 450km altitude with local time series of field intensity, constructed using the geomagnetic virtual observatory approach, shown in white for the 300 equal area distributed positions marked by the red dots. The mean value has been removed from each field intensity series and the maximum field change is 492~nT. 
\label{fig:SEU}}
\end{figure}

In this~section we focus on changes in the South Atlantic Anomaly since the launch of the \textit{Swarm} satellites in 2014 and discuss their origin in changes of the CMB radial field. Fig.~\ref{fig:SEU} (top panel) presents contours of the field intensity at 450~km altitude in August 2017 (approximately the middle epoch of the {\it Swarm} data), as given by the internal part of the CHAOS-7 field model, with so-called single event upsets (SEUs), recorded onboard the \textit{Swarm} Alpha, Bravo and Charlie satellites between November 2013 and August 2019, superimposed. These SEUs are routinely recorded as part of bit checking procedures in the onboard electronics and they indicate when the satellite instrumentation has been affected by collisions with high energy charged particles.  The occurrence of SEUs generally increases with latitude towards the polar regions where high energy charge particles are guided along magnetic field lines coupling the magnetosphere and ionosphere.  Nevertheless, the highest concentration of SEUs is clearly observed at mid and low latitudes in the South Atlantic Anomaly weak field region.  This provides a vivid illustration of the impact of the SAA on low-Earth orbit space infrastructure. Considering a sequence of such maps, the weakest field region at satellite altitude, shown by contours of blue colours in Fig.~\ref{fig:SEU}, has over the past six years slowly extended eastwards from South America towards South Africa, at latitudes between 30 and 45 degrees south \cite[see also][]{Rother_etal:2020}.  SEU instances also show a tongue extending eastwards from the center of the anomaly in this region.  

The bottom panel of Fig.~\ref{fig:SEU} shows the field intensity from CHAOS-7, again at 450 km altitude in August 2017, but now overlaid with time series showing the change of the observed field intensity at {\it Swarm} altitude between 2014 to 2019 at a network of 300 geomagnetic virtual observatories (GVOs) 
\citep{Mandea:2006,Olsen:2007c,Barrois:2018}.
 The data presented in these time series are derived from observations within a radius of 700~km of the red target points by fitting a local potential every 4~months. Each time series has had its mean value removed, and the maximum recorded intensity change was 492~nT.   The GVO series show that field intensity at the altitude of the \textit{Swarm} satellites has generally decreased since 2014 over the Americas, with the most rapid decline occurring over North America and over the Pacific to the west of South America. On the other hand the field intensity  has increased over the Indian ocean and Asia.  Of particular interest is what has happened in the South Atlantic Anomaly region.  Here the {\it Swarm} virtual observatory series show that the field intensity has decreased on the western edges of the anomaly, leading to its westwards expansion.  There have been more modest decreases in intensity within the anomaly, and intensity increases at its north-eastern edge in the central Atlantic towards northern Africa.  A striking fall in the field intensity was also seen on Anomaly's south-eastern edge, in the region towards southern Africa around 45$^\circ$S on the Greenwich meridian, leading to an eastward expansion of the anomaly in this direction.  This demonstrates that the development of the South Atlantic Anomaly is more complex than a simple westward motion and expansion of a single anomaly.
 
\begin{figure}[!ht]
\includegraphics[angle=0, width=0.8\textwidth]{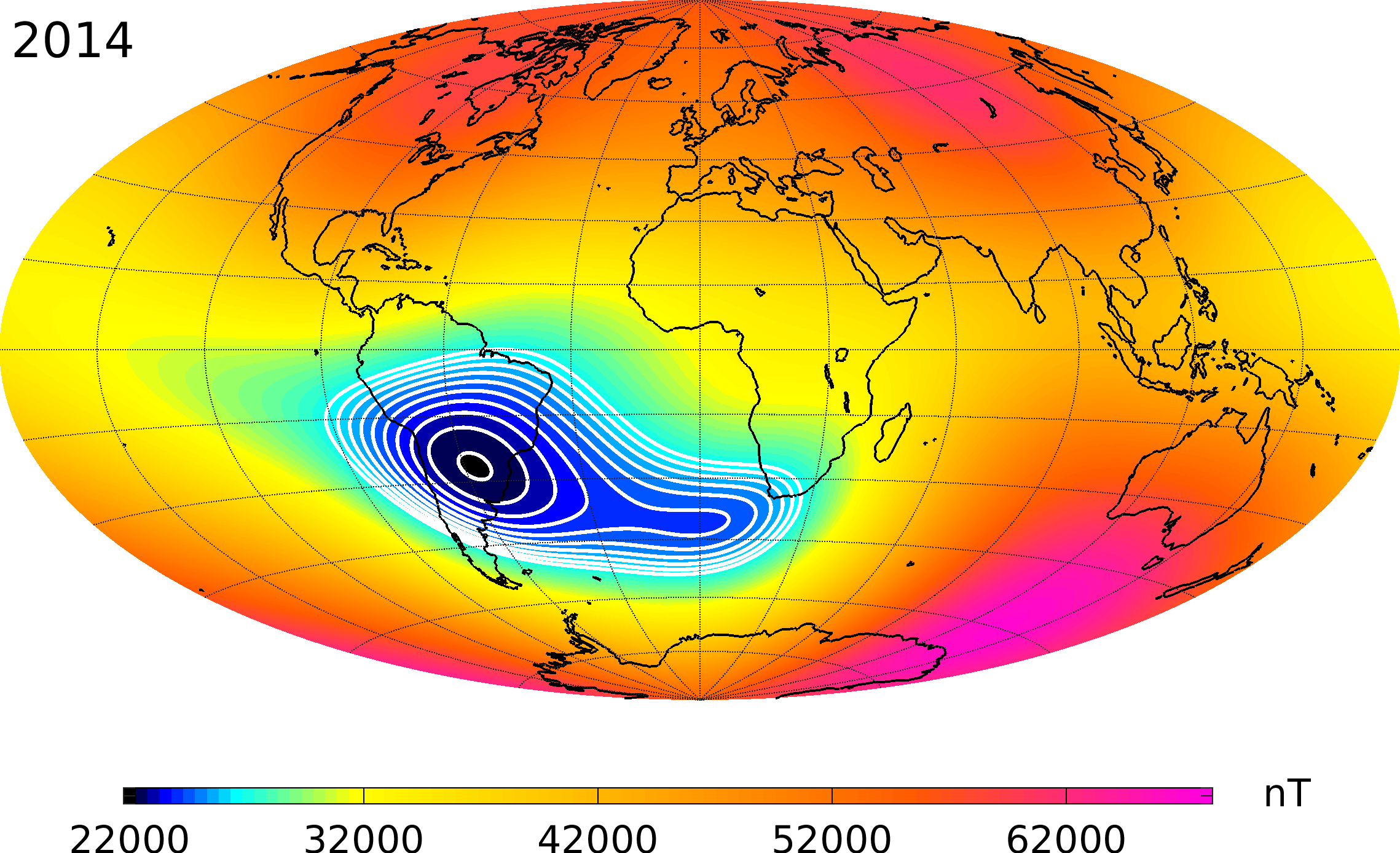}\\
\vspace{0.25cm}

\includegraphics[angle=0, width=0.8\textwidth]{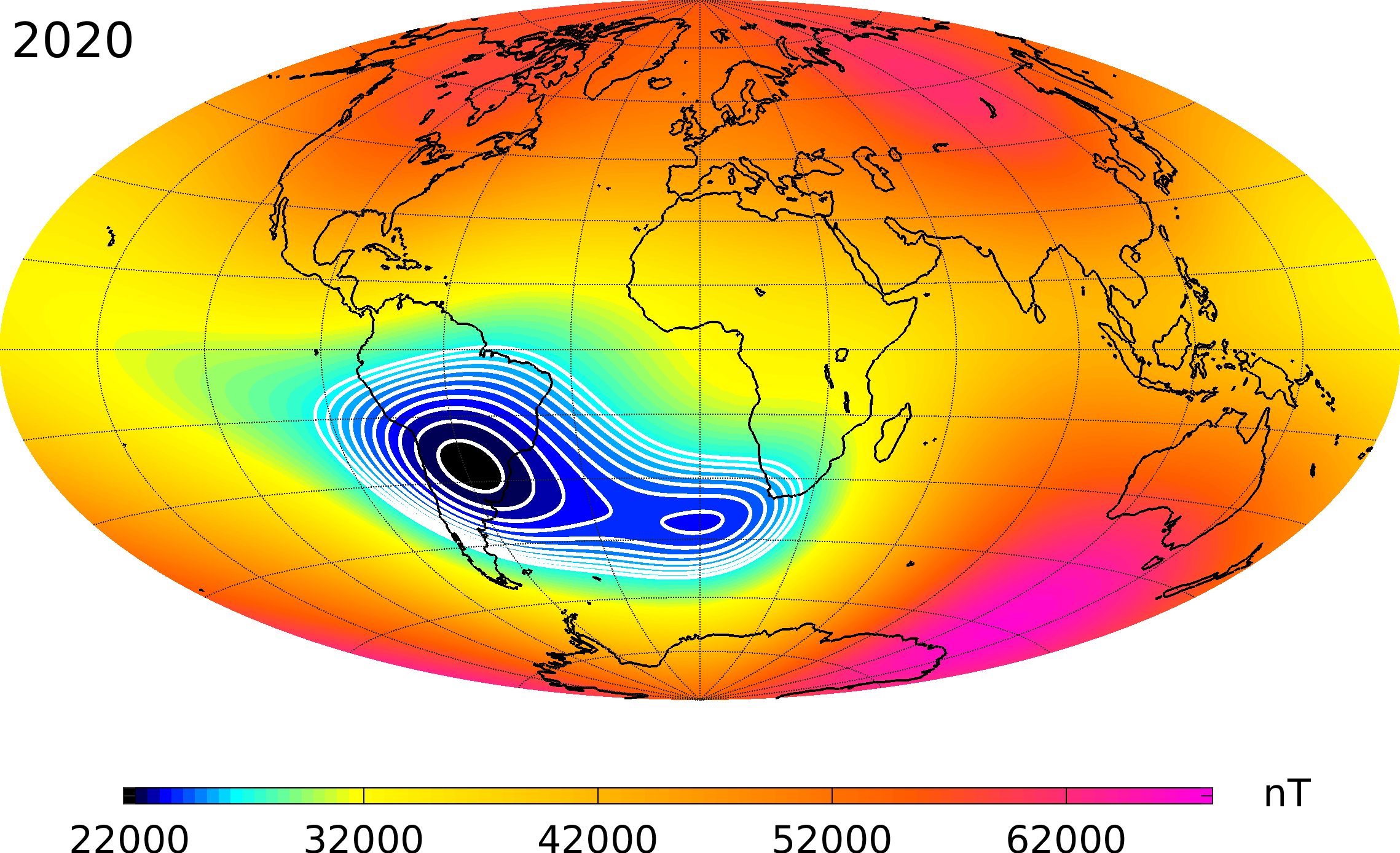}\\

\vspace{0.25cm}
\includegraphics[angle=0, width=0.8\textwidth]{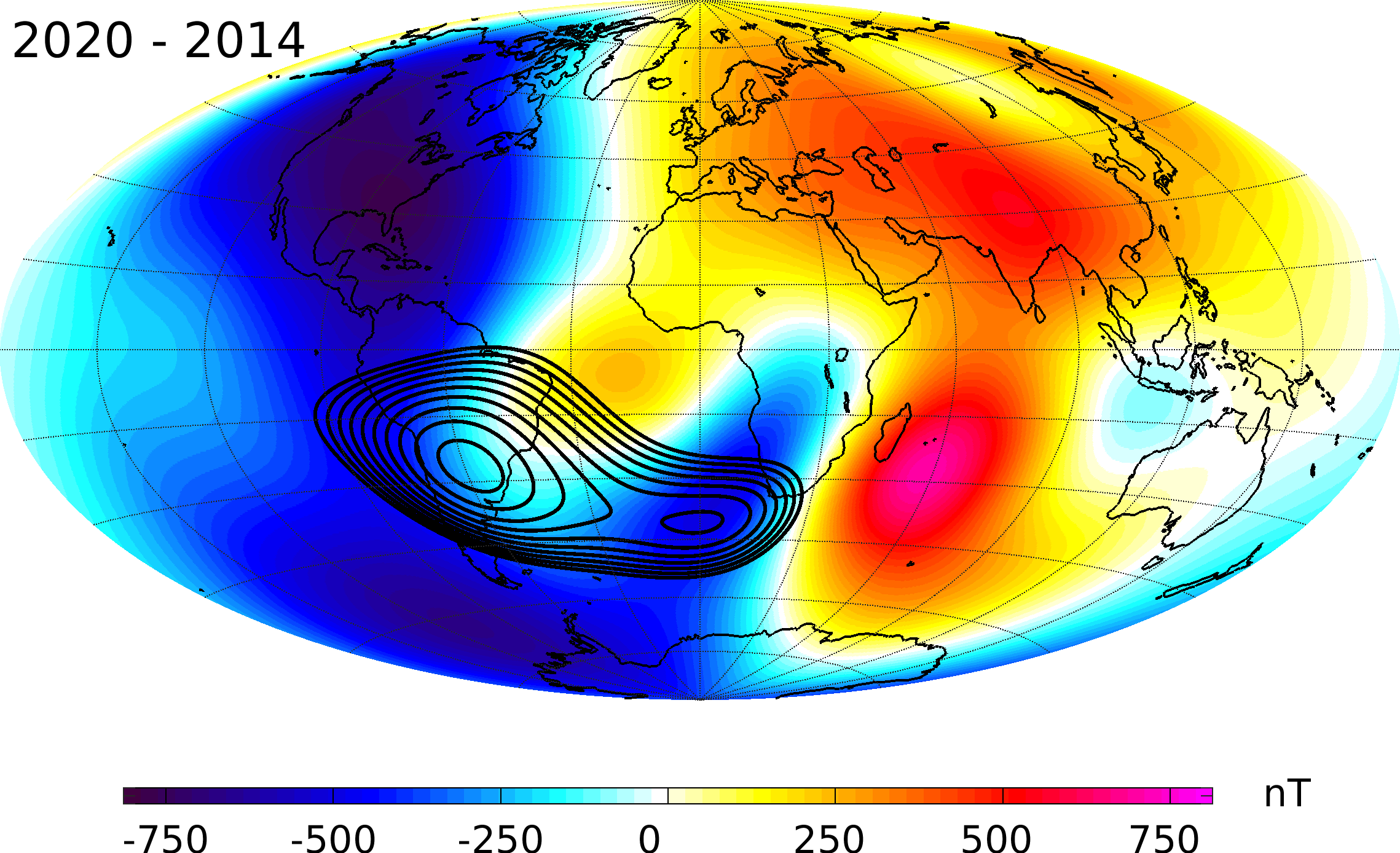}\\

\caption{Field intensity at Earth's surface, highlighting the South Atlantic Anomaly. Top and middle rows show field intensity in~nT in 2014 and 2020, based on the CHAOS-7 internal field up to degree 20. White contour lines are in steps of 500\,nT from 22500\,nT to 27000\,nT.  Bottom row shows the difference in intensity between 2020 and 2014 in~nT, blue colours show a decrease in intensity, red colours an increase, contour lines here again show the field intensity in 2020 in steps of 500\,nT from 22500\,nT to 27000\,nT as black contours.
\label{fig:SAA}
}
\end{figure}

Fig.~\ref{fig:SAA} presents the change in the field intensity at Earth's surface according to CHAOS-7, up to spherical harmonic degree 20, over the past six years when there is excellent data coverage provided by the three {\it Swarm} satellites. The top panel shows the field intensity in 2014, the middle panel the field intensity in 2020 and the bottom panel the difference or accumulated change over the six years.  Contours in steps of 500\,nT between 22500\,nT to 27000\,nT are used to highlight detailed changes in the structure of the South Atlantic Anomaly.  The region of weakest magnetic field over central South America has expanded during these years, and a distinct secondary minimum has developed around 40 degrees South on the Greenwich meridian \citep{Terra-Nova:2019,Rother_etal:2020}. This secondary minimum is seen even if the field is truncated at spherical harmonic degree 9.  The change of intensity displayed in the bottom panel of Fig.~\ref{fig:SAA} illustrates that field intensity decrease is not uniformly distributed across the South Atlantic Anomaly at Earth's surface, but is presently happening fastest at its western and southern edges, as well as in the region south-west of South Africa where the secondary minimum has developed.    The appearance of this distinct secondary minimum of field intensity at Earth's surface indicates that the South Atlantic anomaly must result from the combined action of a number of underlying non-dipolar flux features, and cannot be due to a single flux feature at the CMB.  Due to the attenuation of small scale features with altitude,  the secondary minimum is not yet directly visible at satellite altitude although the changes responsible for its appearance at the Earth's surface are observed in the form of the eastward extension of the weak field region from the central Atlantic toward Africa.

\begin{figure}[!ht]
\includegraphics[angle=0, width=0.45\textwidth]{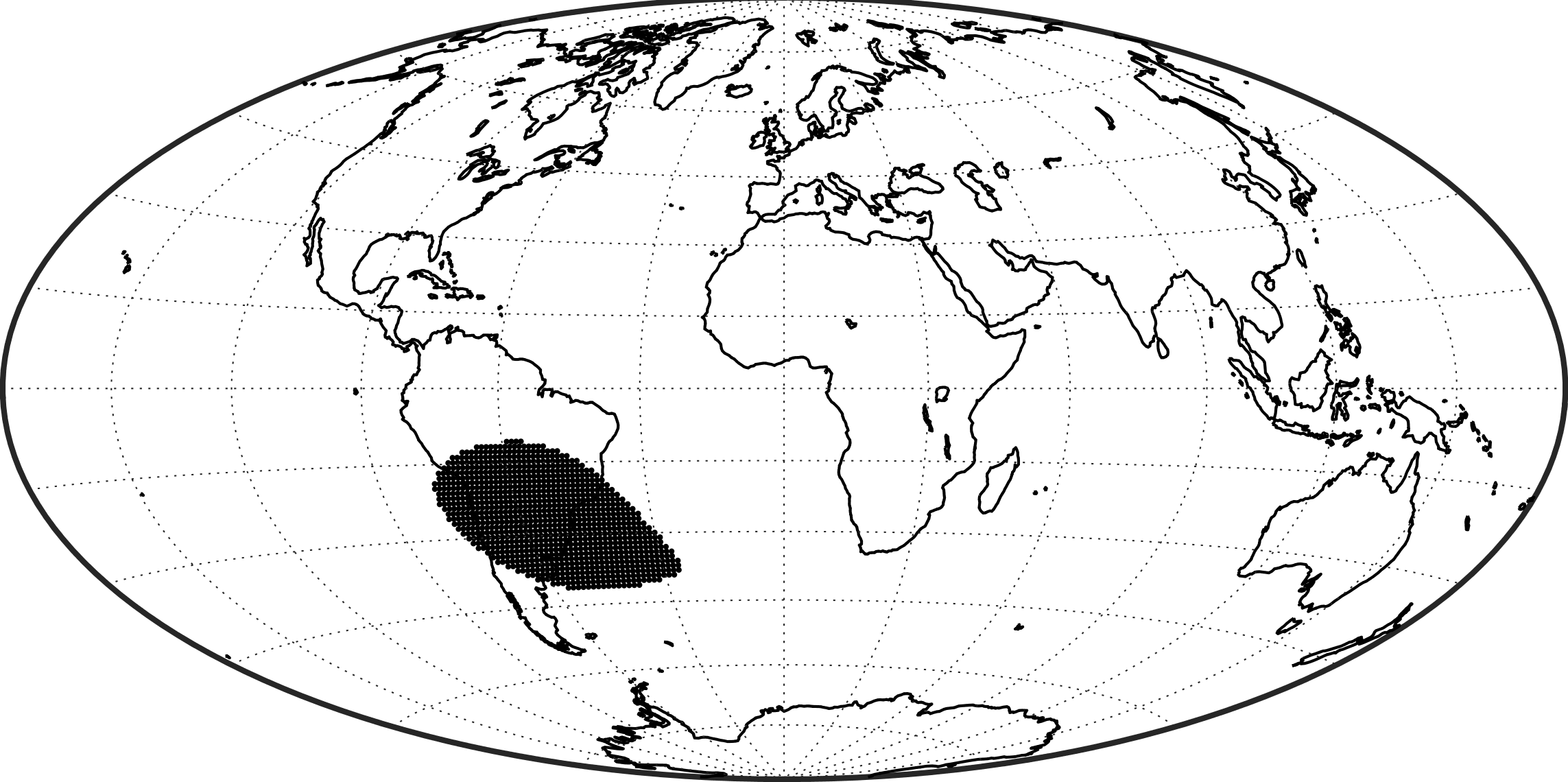}
\hspace{0.1cm}
\includegraphics[angle=0, width=0.45\textwidth]{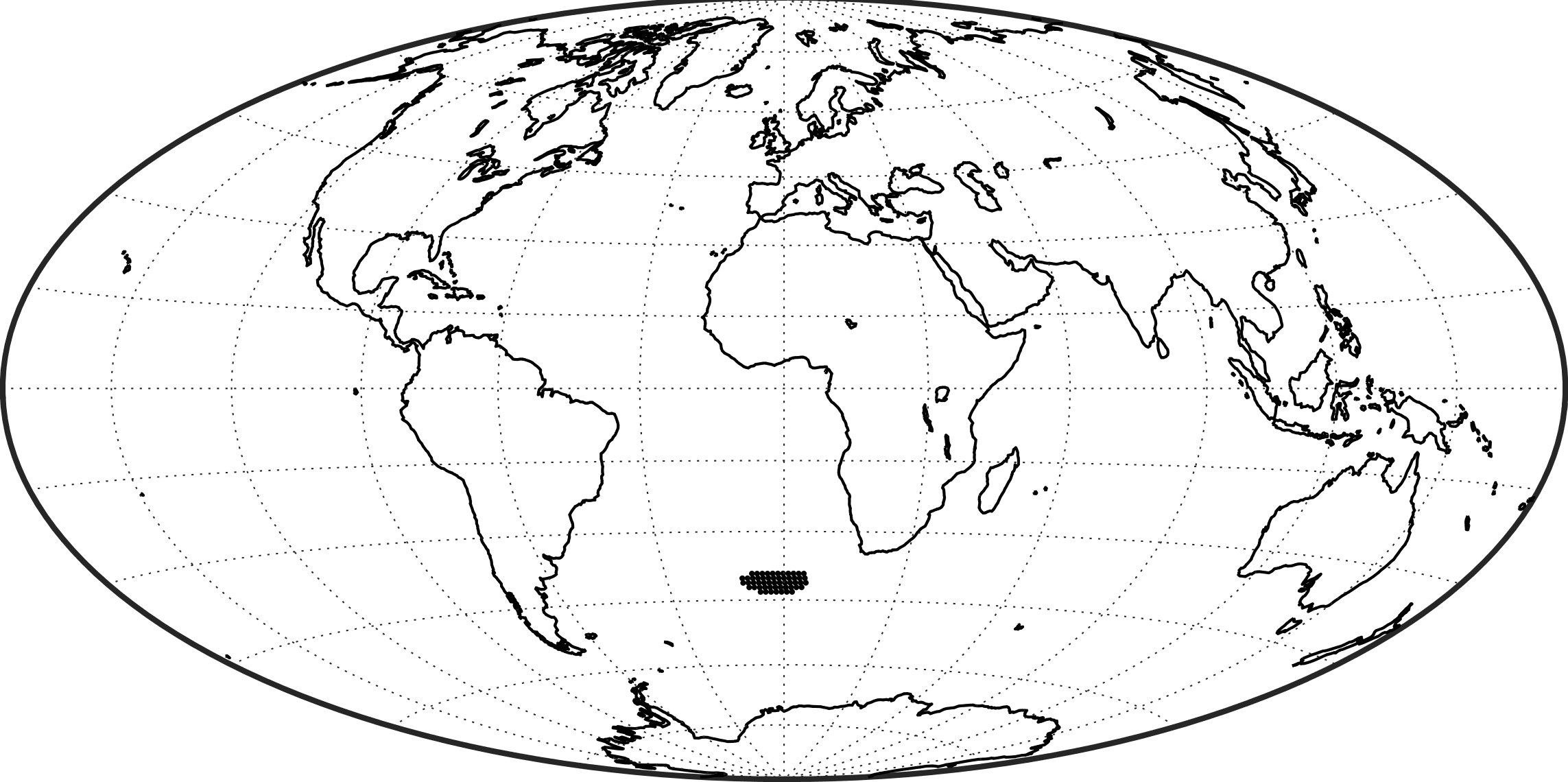}\\
\vspace{0.25cm}

\includegraphics[angle=0, width=0.45\textwidth]{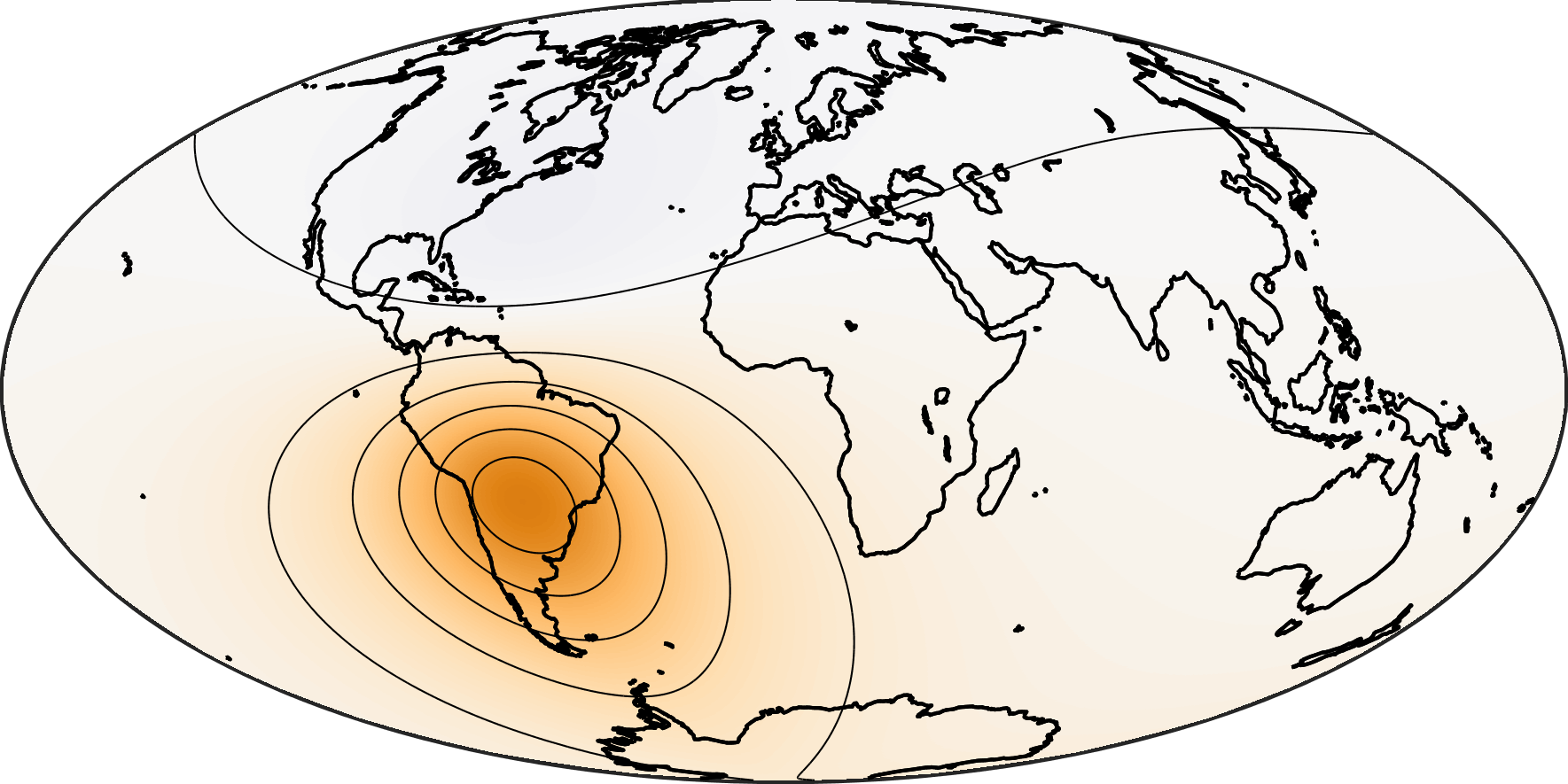}
\hspace{0.1cm}
\includegraphics[angle=0, width=0.45\textwidth]{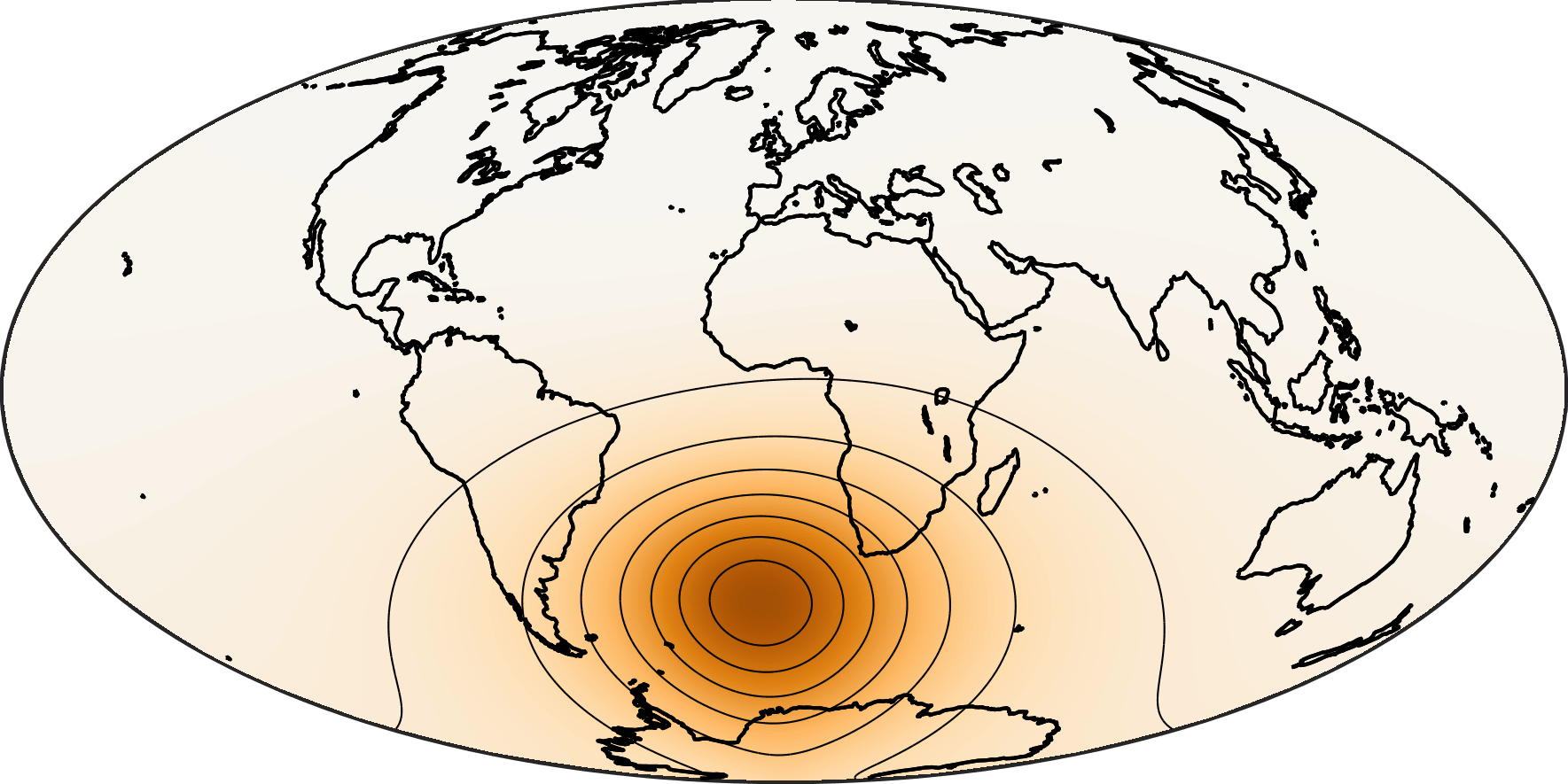}\\
\vspace{0.25cm}

\includegraphics[angle=0, width=0.45\textwidth]{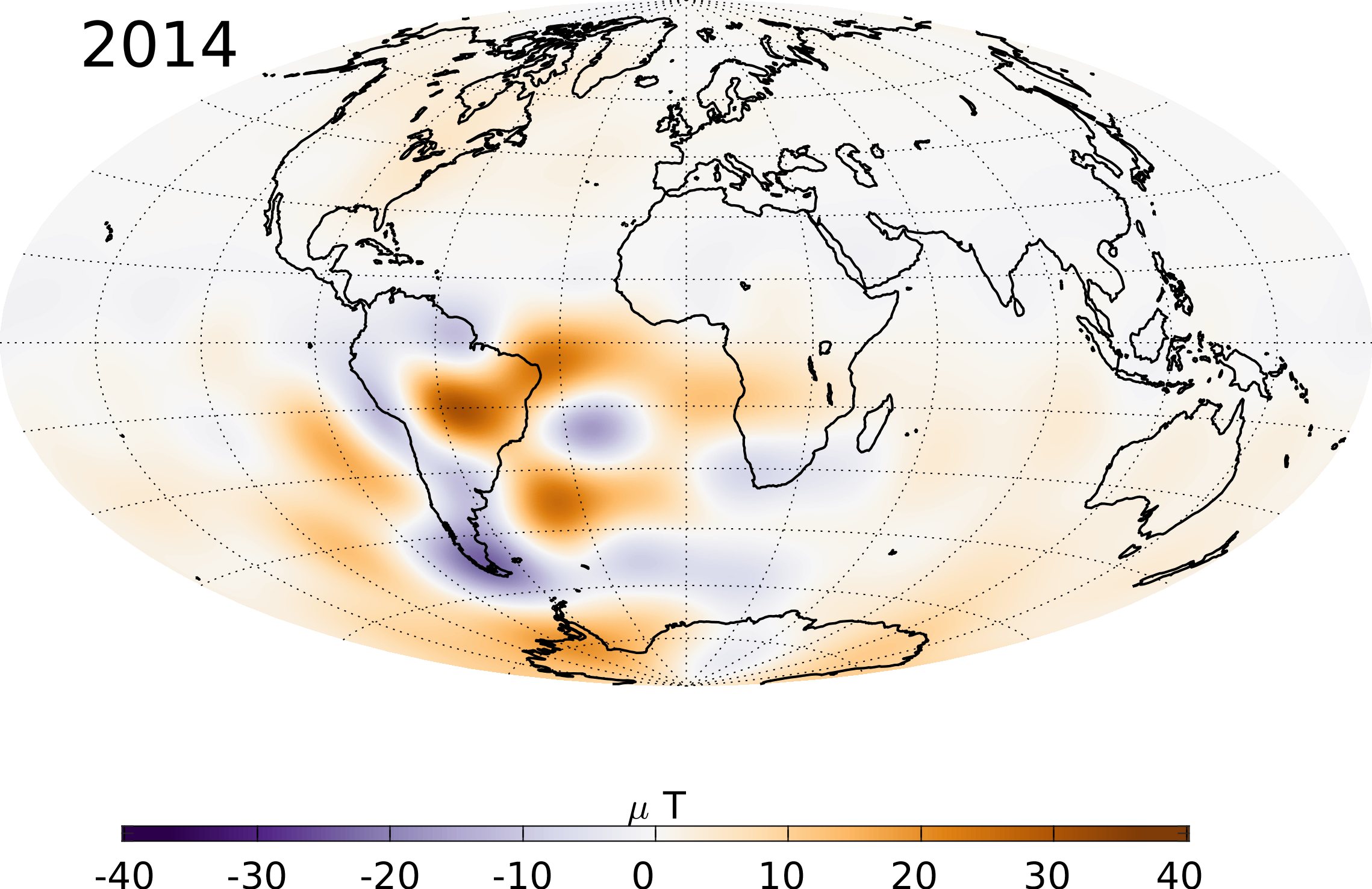}
\hspace{0.1cm}
\includegraphics[angle=0, width=0.45\textwidth]{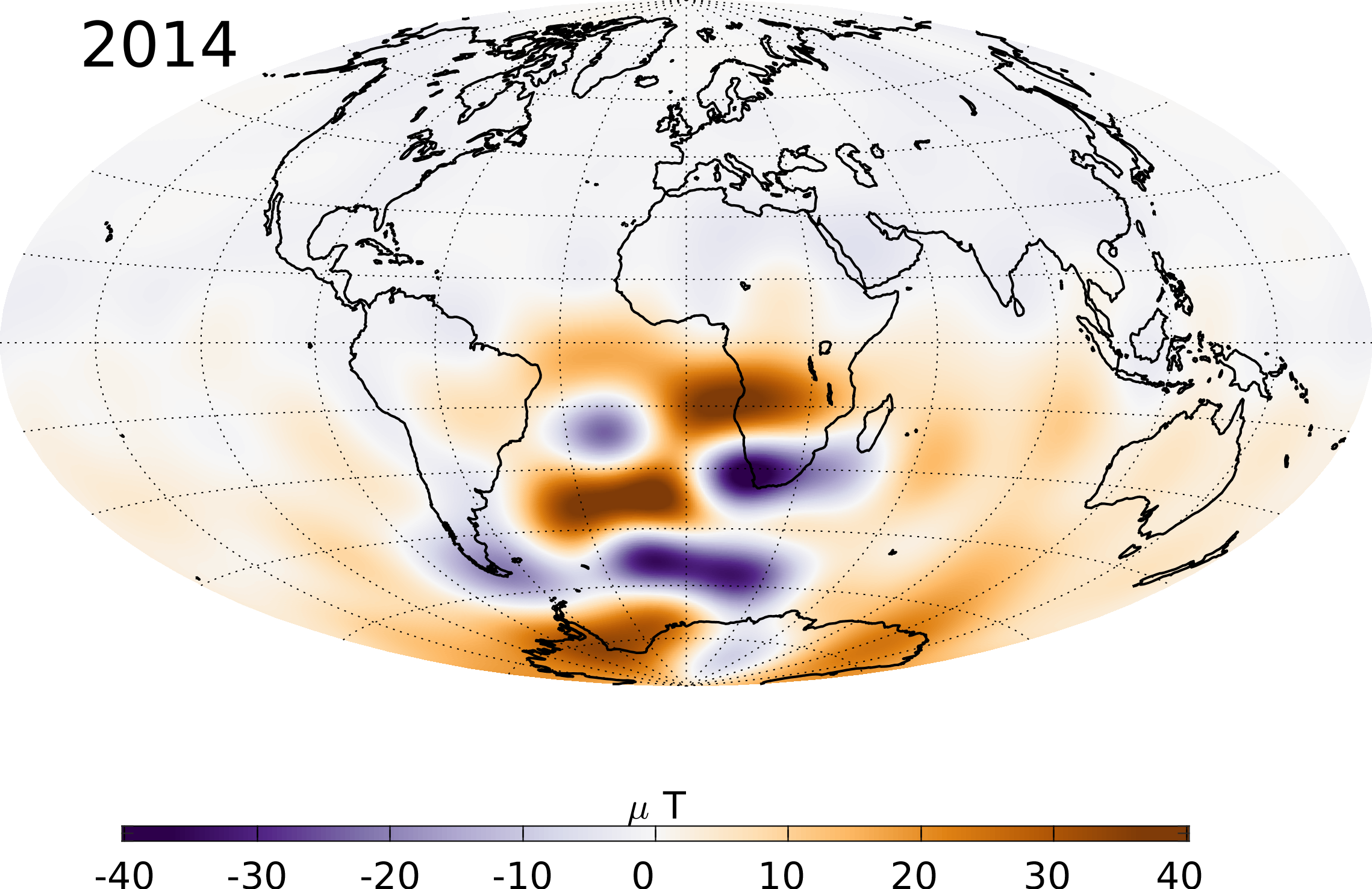}\\
\vspace{0.25cm}

\includegraphics[angle=0, width=0.45\textwidth]{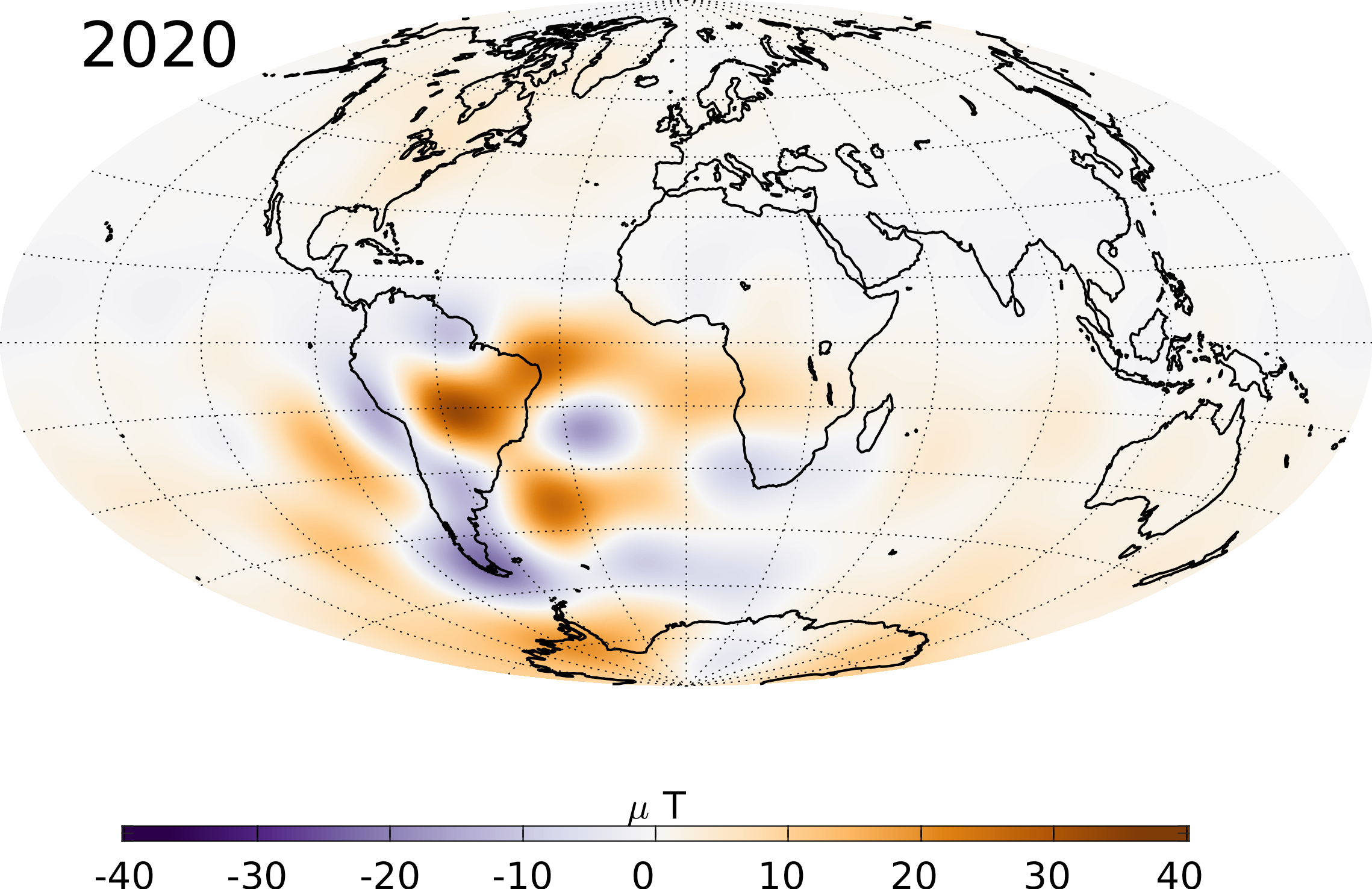}
\hspace{0.1cm}
\includegraphics[angle=0, width=0.45\textwidth]{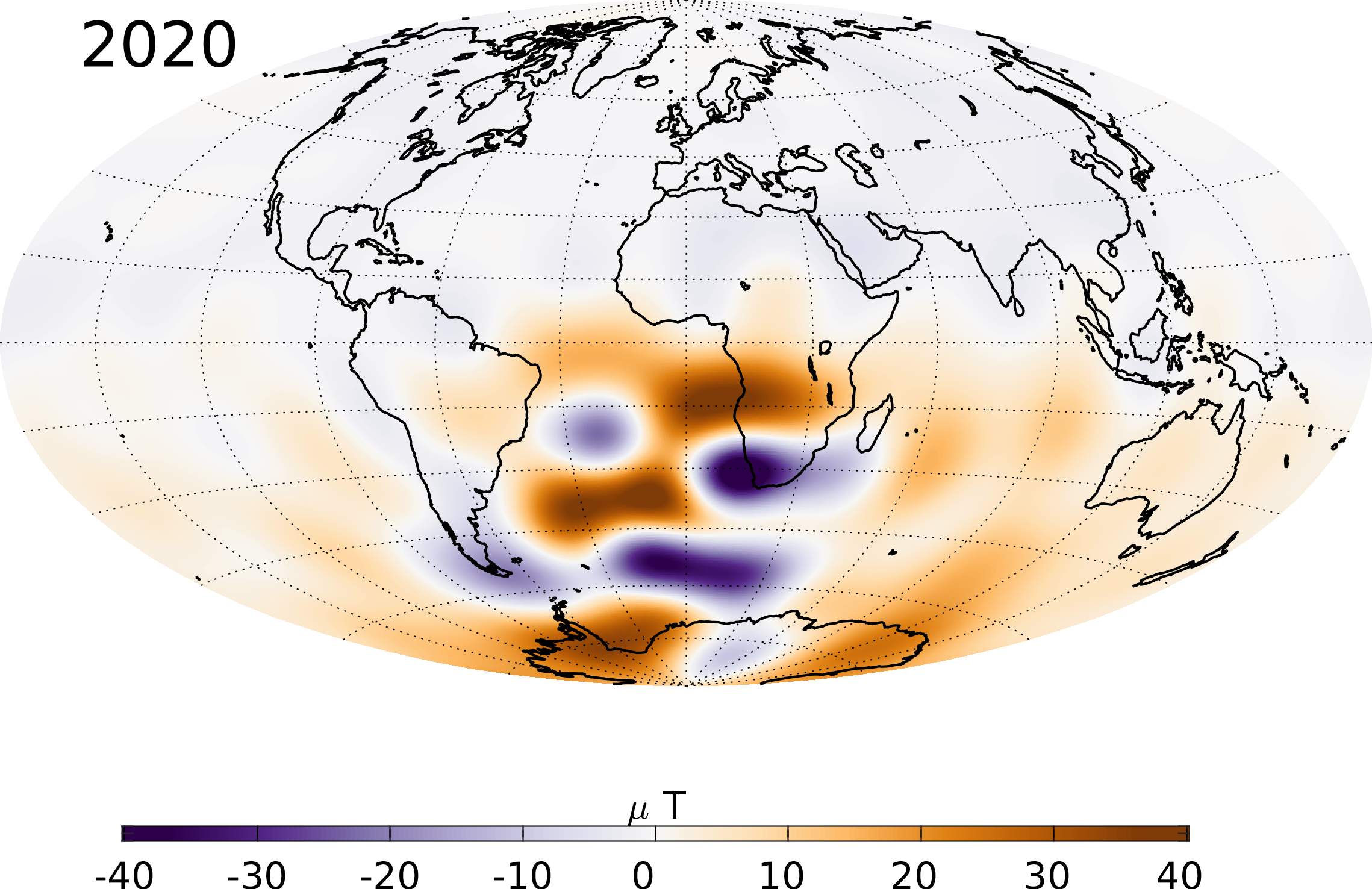}\\
\vspace{0.25cm}


\includegraphics[angle=0, width=0.48\textwidth]{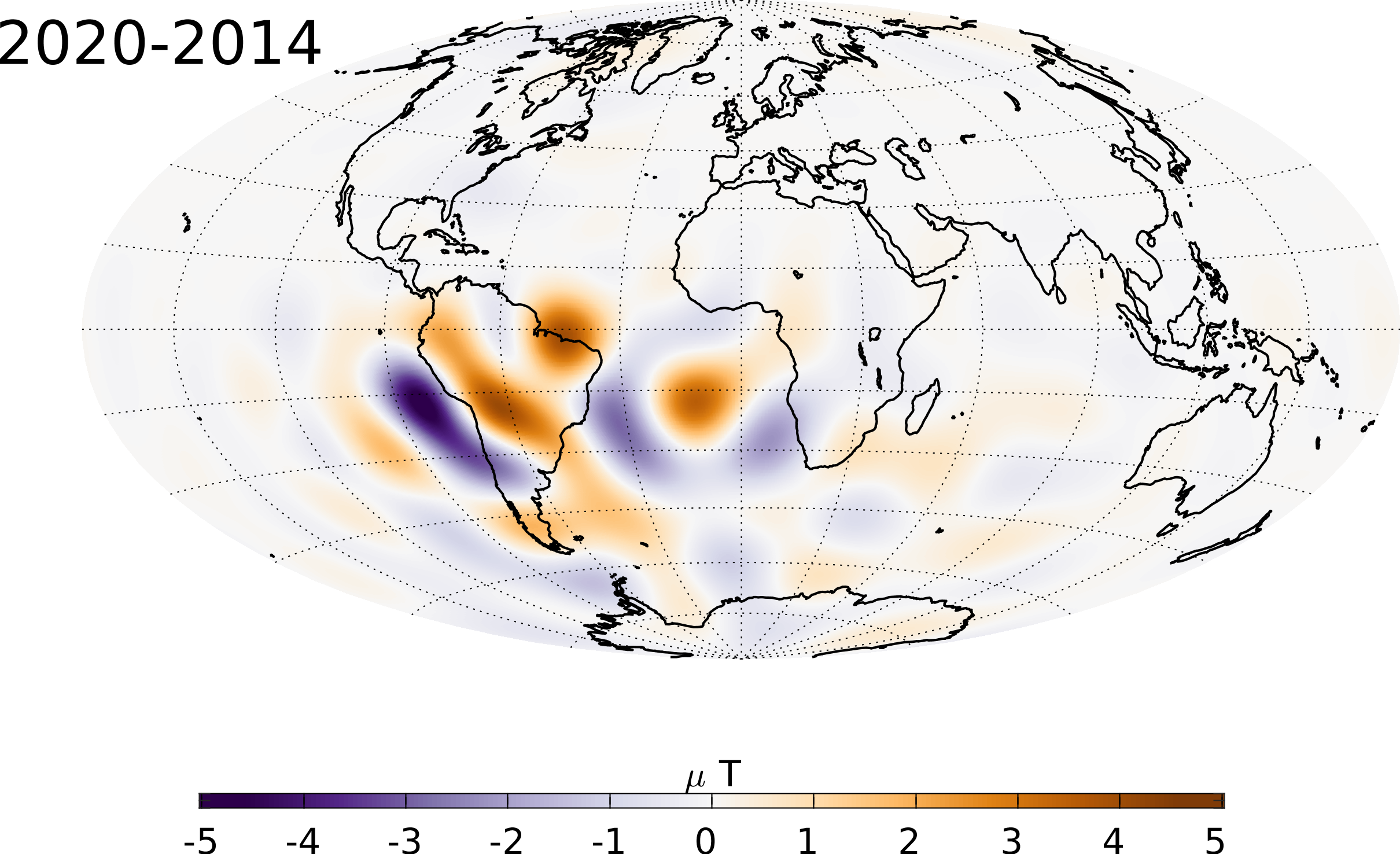}
\hspace{-0.2cm}
\includegraphics[angle=0, width=0.48\textwidth]{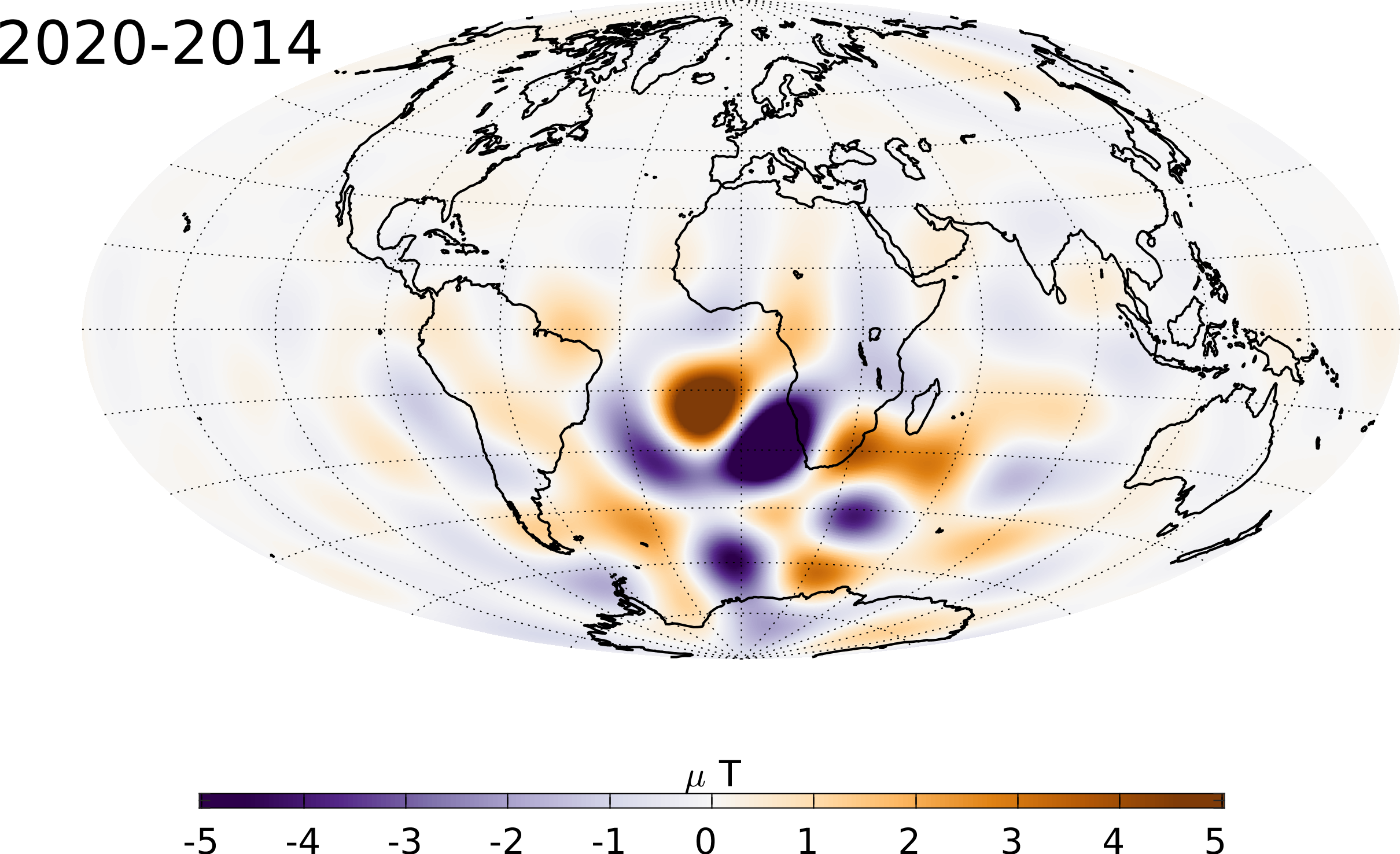}\\

\caption{Core-mantle boundary origin of the South Atlantic Anomaly and its recent changes analysed using Green's functions for the Laplace equation under Neumann boundary conditions.  Top row: black areas show regions of lowest intensity (under 24000~nT) selected for analysis, the main anomaly (left column) and the new secondary minimum (right column).  2nd row: combined sensitivities $G^{\Sigma}_F$ (see eqn (\ref{eq:G_sum_f})) showing the sensitivity of the field intensity, relative to the dipole, to the core-mantle boundary radial field, for each defined region.  3rd row: taking these sensitivities as weights and multiplying by the radial field at the core-mantle boundary.  This shows the parts of the radial field at the core-mantle boundary that in combination are responsible for the main South Atlantic anomaly (left) and the new secondary minimum (right), respectively, in 2020.  Bottom row: change in the sensitivity-weighted radial field at the core-mantle boundary between 2014 and 2020, showing the origin of field intensity changes for the selected regions at Earth's surface.
\label{fig:SAA_Greens}
}
\end{figure}

Fig.~\ref{fig:SAA_Greens} presents an investigation into the origin of these recent changes in the South Atlantic Anomaly.  Our analysis makes use of the Green's function for the Laplace equation under Neumann boundary conditions, which formally links the core-mantle boundary radial field to the observed field \citep{Gubbins_Roberts_1983}.  The top row shows the two regions of weakest intensity at the Earth's surface in 2020 (with intensities below 24000\,nT) considered for the analysis, the second row shows for each region a map of the sensitivity (see below for details) of the average intensity in each region to the underlying radial field at the CMB, multiplying the CMB radial field in 2020 by these weights gives the maps in the third row.  The bottom row shows changes in the CMB radial field with these weights applied between 2014 and 2020.

The sensitivity maps show in the second row of Fig.~\ref{fig:SAA_Greens} were derived as follows.  Considering only sources in the Earth's core, the magnetic field $\mathbf{B}$ measured at a position $\mathbf{r}$ on or above Earth's surface may be written as a weighted integral of the radial magnetic field at the CMB $B_r(\mathbf{\hat{s}})$ \citep{Gubbins_Roberts_1983}
\begin{equation}
	\mathbf{B}(\mathbf{r}) = \int_{\Omega} \mathbf{G}(\mathbf{r},\mathbf{\hat{s}}) B_r(\mathbf{\hat{s}})\, \mathrm{d}S
\end{equation}
where $\mathrm{d}S =\sin\theta_s \mathrm{d}\theta_s \mathrm{d}\phi_s$  is a surface element at the core-mantle boundary $\Omega$  and $\mathbf{G}=\left\{G_r,\,G_{\theta},\,G_{\phi} \right\}$ are the following Green's functions or sensitivities that link $B_r(\mathbf{\hat{s}})$ to the spherical polar vector components of the field $\left\{B_r(\mathbf{r}),\,B_{\theta}(\mathbf{r}),\,B_{\phi}(\mathbf{r})\right\}$
\begin{align}
G_{r}      &= -\frac{\partial N}{\partial r} = \frac{1}{4\pi}\frac{h^2(1-h^2)}{f^3} \label{eq:5.33}  \\
G_{\theta} & = -\frac{1}{r} \frac{\partial N}{\partial \mu} \frac{\partial \mu}{\partial \theta} 
= -\frac{1}{r} \frac{\partial N}{\partial \mu} [\cos\theta \, \sin\theta_s \, \cos(\phi-\phi_s)-\sin\theta \, \cos\theta_s] \label{eq:5.34}  \\
G_{\phi}   &= -\frac{1}{r\sin\theta} \frac{\partial N}{\partial \phi} = \frac{1}{r} \frac{\partial N}{\partial \mu}[\sin\theta_s\, \sin(\phi-\phi_s)] \label{eq:5.35} 
\end{align}
where the derivative with respect to $\mu=\cos \theta$ is 
\begin{equation}
\frac{1}{r} \frac{\partial N}{\partial \mu} =\frac{h}{4\pi} \left[ \frac{1-2h\mu + 3h^2}{f^3} + \frac{\mu}{f(f+h-\mu)} - \frac{1}{1-\mu} \right]
\end{equation}
and $h=c/r$ where $c$ is the radius of the core-mantle boundary.

\clearpage

Here, we are interested not in the vector field components themselves, but in the field intensity $F=\sqrt{B_r^2+B_{\theta}^2+B_{\phi}^2}$, at the Earth's surface.  This is a non-linear function of  $B_r(\mathbf{\hat{s}})$.  However, following \cite{Johnson_Constable_1997} and \cite{Terra-Nova:2017}, we may apply the chain rule and differentiate $F(\mathbf{r})$ with respect to $B_r(\mathbf{\hat{s}})$ we obtain
\begin{equation}
\label{eq:G_f}
\begin{aligned}
	\frac{\partial F(\mathbf{r})}{\partial B_r(\mathbf{\hat{s}})} &= \frac{\partial F(\mathbf{r})}{\partial B_r(\mathbf{r})}\frac{\partial  B_r(\mathbf{r})}{\partial B_r(\mathbf{\hat{s}})} + \frac{\partial F(\mathbf{r})}{\partial  B_{\theta}(\mathbf{r})}\frac{\partial B_{\theta}(\mathbf{r})}{\partial B_r(\mathbf{\hat{s}})} +\frac{\partial F(\mathbf{r})}{\partial B_{\phi}(\mathbf{r})}\frac{\partial B_{\phi}(\mathbf{r})}{\partial B_r(\mathbf{\hat{s}}')} \\
	&= \frac{1}{F(\mathbf{r})} \left[ B_r(\mathbf{r}) G_r(\mathbf{r},\mathbf{\hat{s}}) + B_{\theta}(\mathbf{r}) G_{\theta}(\mathbf{r},\mathbf{\hat{s}}) + B_{\phi}(\mathbf{r})G_{\phi}(\mathbf{r},\mathbf{\hat{s}}) \right]\\
	&= G_F(\mathbf{r},\mathbf{\hat{s}}).
\end{aligned}
\end{equation}
 We can make use of this expression by considering changes of the magnetic field about some defined (known) background reference field $\mathbf{B}(\mathbf{r}, t_0)$; eqn (\ref{eq:G_f}) then defines an appropriate linearized Green's function.   Here, we take the background reference field to be the dipole part of the magnetic field in 2014.0 according to the CHAOS-7 model, $\mathbf{B}(\mathbf{r}, t_0)=\mathbf{B}^\mathrm{dip}(\mathbf{r}, 2014)$, and we consider the sensitivity of departures from this to the CMB radial field.  In particular, we consider 
 \begin{equation}
\begin{aligned}
\label{eq:dFdt}
	 \Delta F (\mathbf{r}, t) = \int_{\Omega} G_F(\mathbf{r},\mathbf{\hat{s}}) \, B_r(\mathbf{\hat{s}}, t) \, \mathrm{d}S.
\end{aligned}
\end{equation}
with $\Delta F(\mathbf{r}, t) = F(\mathbf{r}, t) - F^\mathrm{dip}(\mathbf{r}, 2014)$ and $G_F(\mathbf{r},\mathbf{\hat{s}})$ calculated using eqn (\ref{eq:G_f}) and the reference field $\mathbf{B}(\mathbf{r})= \mathbf{B}^\mathrm{dip}(\mathbf{r}, 2014)$.

Rather than considering sensitivities due to the intensity at a single point (e.g. the position of minimum intensity, as in \cite{Terra-Nova:2017}), we instead consider combined sensitivities for extended low intensity regions, by integrating over regions with intensity below 24000\,nT.  The integral is performed numerically using a dense approximately equal area grid on the Earth's surface, by summing the sensitivities obtained for all grid-points within the region of interest. To aid comparisons we normalized the summed sensitivities by the number of grid-points considered
\begin{equation}
    G_F^{\Sigma} = \frac{\sum_i G_F(\mathbf{r_i},\mathbf{\hat{s}})}{N_i} 
    \label{eq:G_sum_f}
\end{equation}
where $i$ are the indices of grid-points within the chosen region of interest. $G_F^{\Sigma}$ defines the combined sensitivities shown in the second row of Fig~\ref{fig:SAA_Greens}. When considering the region corresponding to the main minimum, there is averaging of spatially-varying sensitivities from a larger number of locations, hence the resulting combined sensitivity is smoothed and of lower peak amplitude, compared to the more focused combined sensitivities obtained for the smaller secondary minimum region.

The third row and fourth rows in Fig.~\ref{fig:SAA_Greens} show how the departure in intensity from the reference dipole field depends on the radial magnetic field at the CMB in 2014 and 2020 respectively.  The integrated values from these maps give, for the region considered, the average departure of the intensity from the intensity of the 2014 dipole.  Weak intensity (compared to the dipole) results from the integral being dominated by negative rather than positive contributions.  The most important features for producing the weak intensity in the region of the secondary minimum are reversed flux features (i) under South Africa and (ii) below the Southern Atlantic between Africa and Antarctica.  Turning to the main minimum, the weak intensity originates from a large reversed flux region that extends beneath the western side of South America, combined with reversed flux regions underneath the central Atlantic east of Brazil and under the Southern Atlantic. 

The bottom panel in Fig.~\ref{fig:SAA_Greens} shows the weighted changes in the CMB radial field responsible for the changes in the average field intensity in the two black regions.  Changes in the CMB radial field at the south-western corner of Africa, associated with the westward movement and development of the reversed flux patch under South Africa are seen to be the dominant influence on the field intensity where the secondary minimum has developed. The change of intensity in the main minimum is dominated by a negative change in the radial field under the region in the Pacific to the west of South America, and also a region of negative field change under the eastern edge of Brazil. Overall, the change to the main minimum appears to be a result of the reversed flux patch east of South America gathering towards the larger reverse flux limb that extends down western South America and is moving slowly westwards.

Based on this analysis using the Green's functions, our interpretation is that the westward motion and deepening of the main minimum of the South Atlantic anomaly is a result of the westward motion of reversed flux patches under South America and the mid-Atlantic, and their gathering under South America due to their different speeds of westward drift. Large intensity variations at Earth's surface caused by the migration of flux patches have also been observed in numerical geodynamo simulations \citep{Davies_Constable_2018}. The growing secondary minimium observed to the south-west of Africa appears to primarily be due to the westward movement of an intense reversed flux feature below South Africa, which is converging towards reversed flux patches under the Southern Atlantic between Africa and Antarctica.




\section{ Field acceleration changes in the Pacific region since 2014}
\label{sec:rapid_SA_Pac}
\begin{figure}[!ht]
\centerline{\includegraphics[angle=0, width=0.75\textwidth]{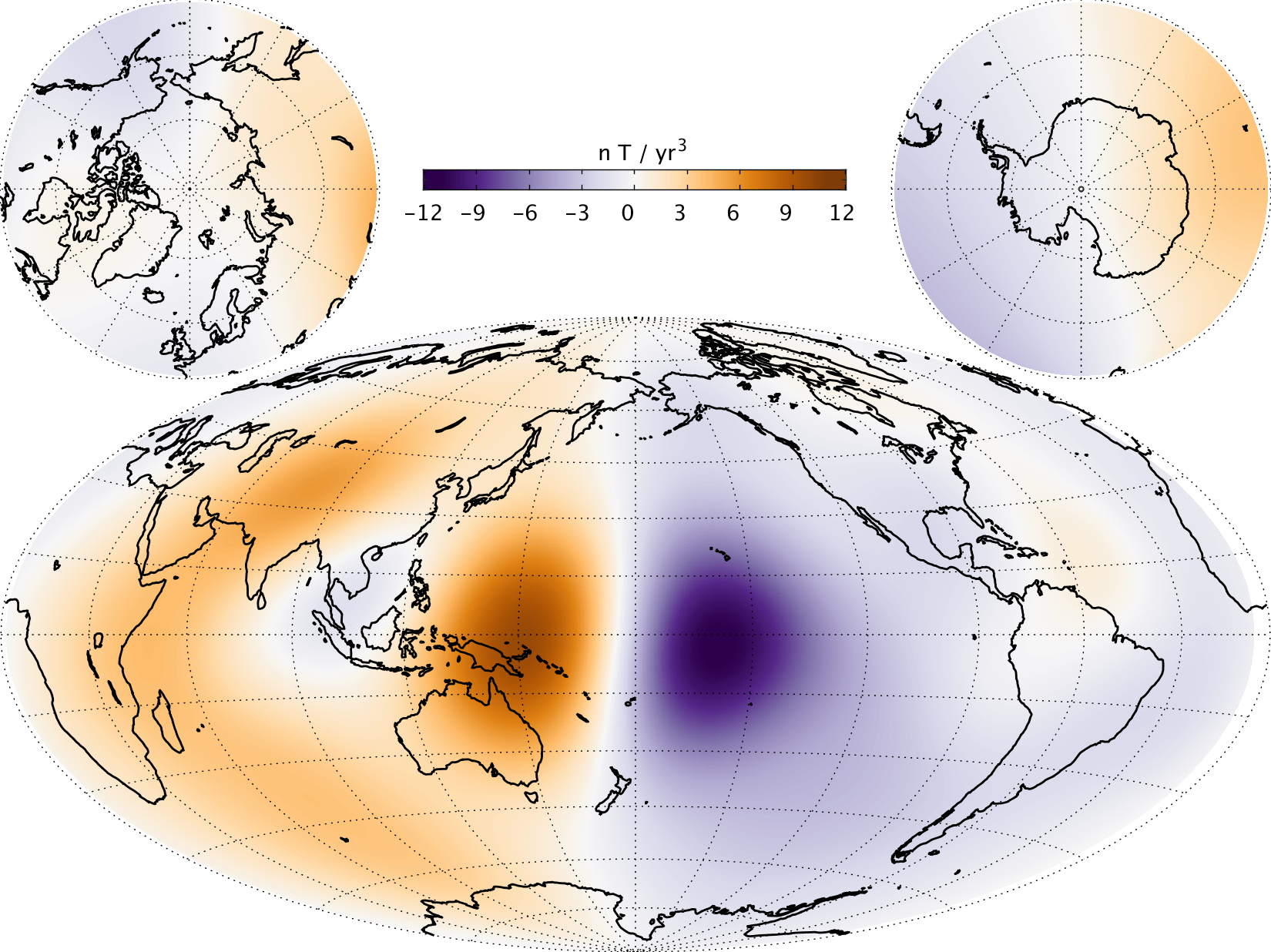}}
\centerline{\includegraphics[angle=0, width=0.75\textwidth]{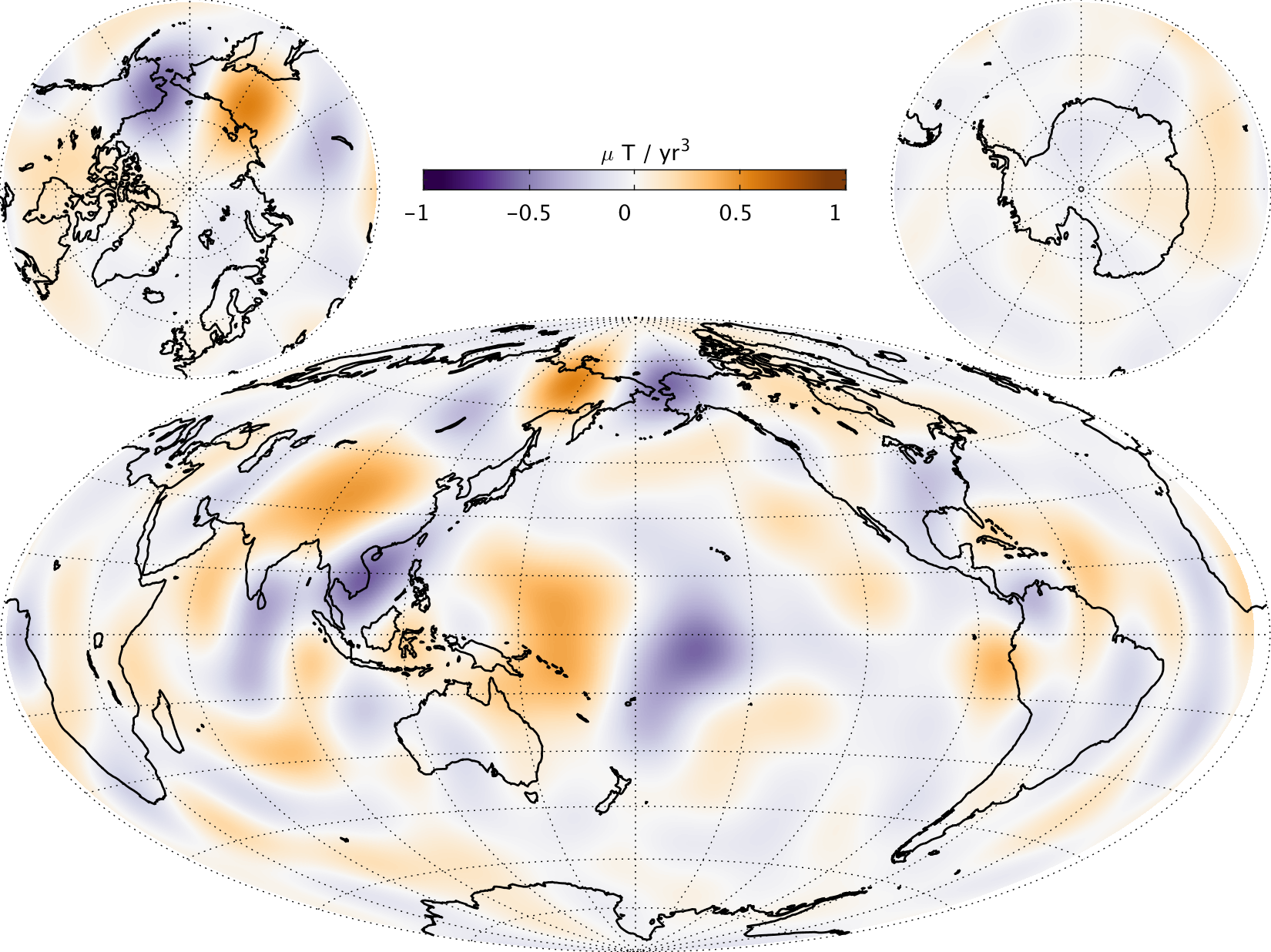}}
\caption{Change in secular acceleration per year, averaged over two consecutive three year time windows during the Swarm era, from 2014 to 2017 and 2017 to 2020. At the Earth's surface (top) and at the CMB (bottom) after truncation at degree $N=15$. Results shown here are taken from the updated CHAOS-7.2 model.
\label{fig:SA change}}
\end{figure}

We finally turn to intriguing, and rapidly evolving, patterns of field acceleration seen since 2014 in the Pacific region.  Fig.~\ref{fig:Maps} showed that in 2019 there was an intense field acceleration in the Pacific region. Fig.~\ref{fig:SA change} has documented the change in the acceleration of the radial component of the field, averaged over consecutive three year windows (2014 to 2017 and 2017 to 2020), at the Earth's surface and at the CMB.\\  

The localized nature of the acceleration change, and its simple structure, consisting of an east-west aligned dipole at Earth's surface are striking.  Downward-continuing the acceleration (and its changes) is difficult due to their power spectra increasing with spherical harmonic degree at the CMB, which means that less well constrained shorter wavelength structures dominate. Furthermore there is uncertainty concerning the electrical conductivity of the lower mantle that is required to be weak in order for the potential field description to hold.  Nevertheless, we find the surface changes reflect coherent localized changes at low latitudes beneath the central Pacific.

\begin{figure}[!ht]
\centerline{\includegraphics[angle=0, width=0.5\textwidth]{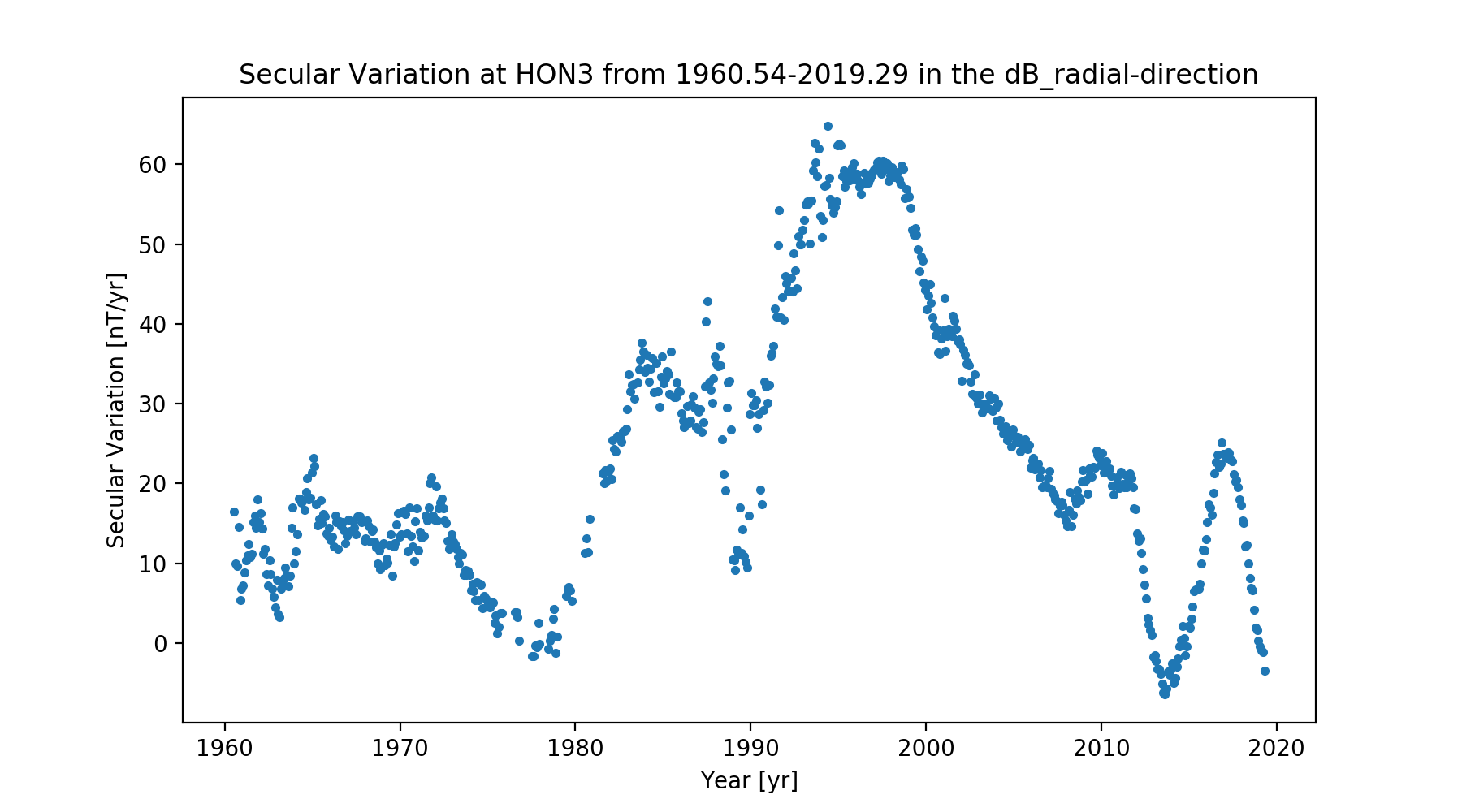}\includegraphics[angle=0, width=0.5\textwidth]{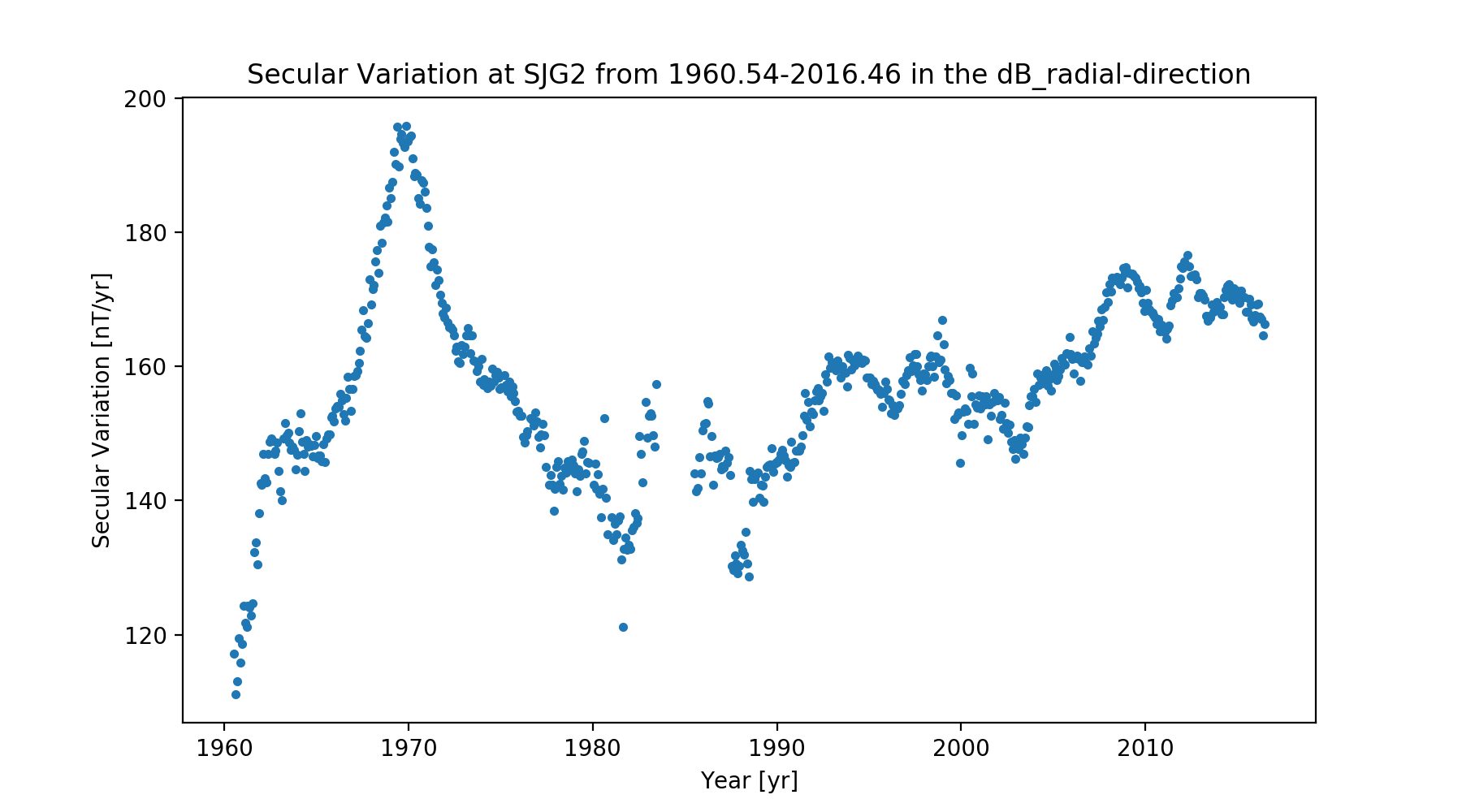}}
\centerline{\includegraphics[angle=0, width=0.5\textwidth]{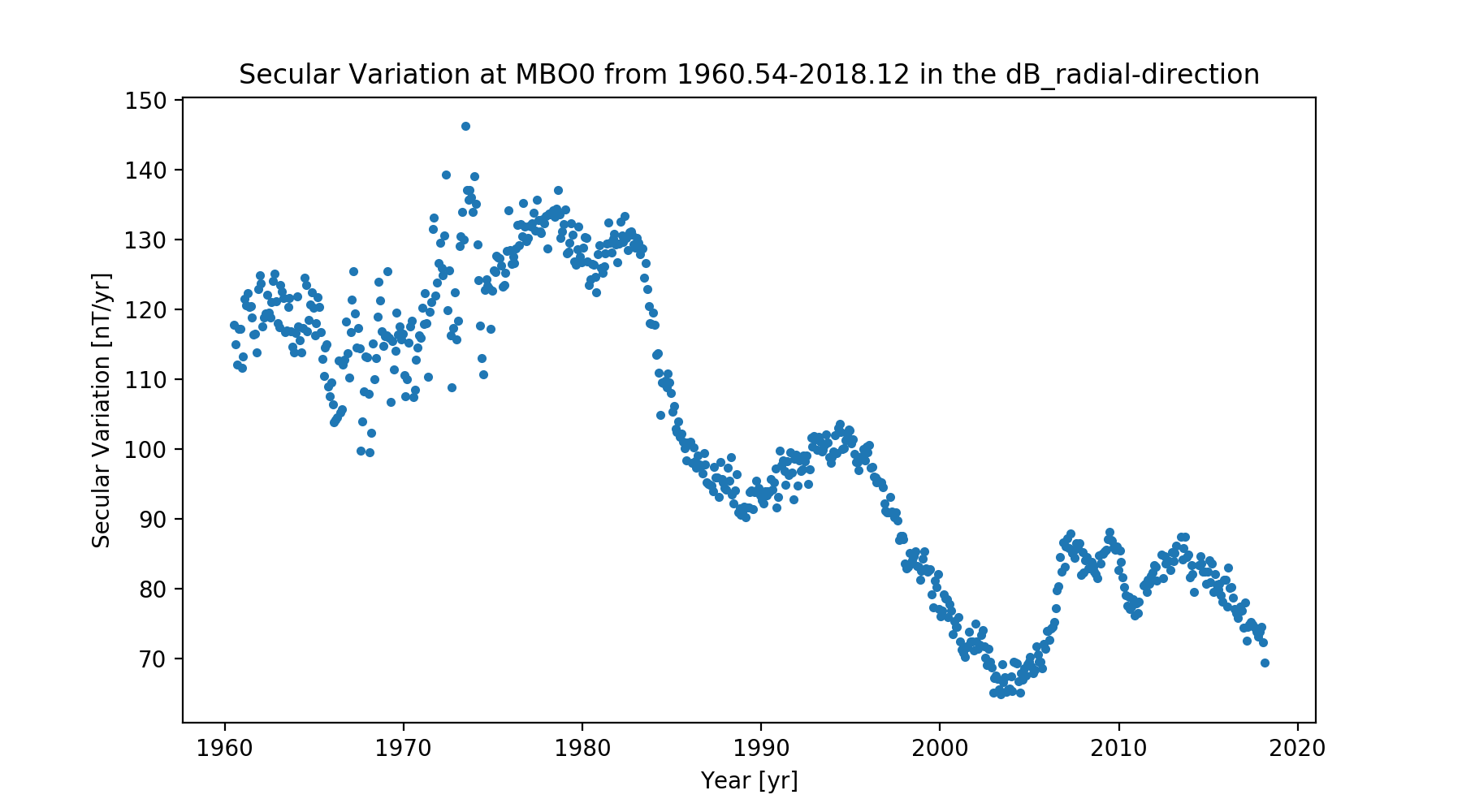}\includegraphics[angle=0, width=0.5\textwidth]{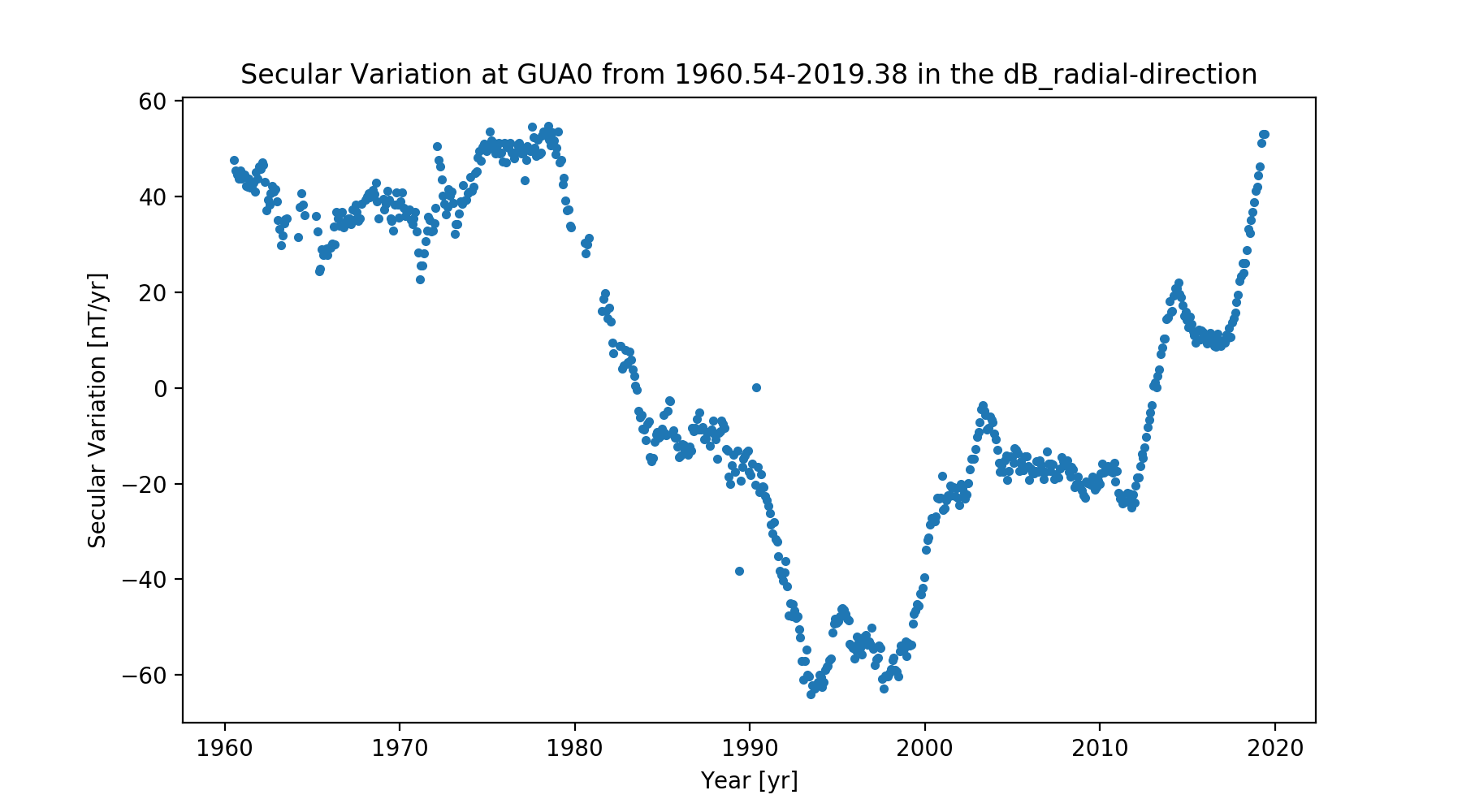}}
\caption{Long observatory series showing annual differences of revised monthly means of the radial field at four low latitudes observatories, Honolulu, Hawaii (top left), San Juan, Puerto Rico (top right), M'Bour, Senegal (bottom left) and Guam, Micronesia (bottom right).
\label{fig:long_obs}}
\end{figure}

An obvious question that then arises is whether this type of localized acceleration change at low latitudes has been seen before, or is it something unusual?  It is well known that similar events have been seen at low latitudes under southern America and in the Atlantic sector in CHAMP data that cover the 2000s \citep{Chulliat:2014, Finlay:2015}, but what about going further back?  To explore this question we present in Fig.~\ref{fig:long_obs} a selection of impressive 60 year long observatory records from mid-to-low latitudes spanning 1960 \--- 2020, showing annual differences of revised monthly means of the radial field component.  These series were computed in the same fashion as the earlier SV series from hourly mean values and using magnetospheric field corrections based on an extended version of the RC index.  The recent episode of acceleration change in the Pacific is clearly visible at the end of the Honolulu record.  Are these variations unusual?  Certainly not in terms of the amplitude of the acceleration change, for example, an even stronger episode of acceleration change was seen at San Juan observatory around 1970; interestingly this event is not obvious in the other four observatories indicating it must also have been a longitudinally-focused event.  Similar events, but of lower acceleration amplitudes were seen at Honolulu in the late 1970s and around 2004 at M'Bour.  The recent changes seen in Guam, which is on the edge of the region of where the rapid acceleration changes have occurred, do not seem extraordinary.  It therefore seems that the recent acceleration changes observed in the Pacific should not be viewed as surprising events, rather they are an integral part of the expected spectrum of rapidly changing SV behaviour that takes place at low latitudes.

By downward-continuing the field to the core-mantle boundary we have assumed a core origin for the changes in the Pacific.  But it is worth pausing to consider whether externally-driven induced currents might instead be responsible. In order not to be seen at all longitudes this would require exotic local conductivity anomalies below each region were such rapid SV events occur.  As shown in Fig.~\ref{fig:Induction}, even if anomalous end-member conductivity profiles are considered, the real part of the $Q$-response for periods of 5 or 6~years is smaller than 0.2, thus to change the radial field acceleration by 10\,nT/yr$^2$ over 3\,years, changes of order 50\,nT/yr$^2$ would be needed in the southward component. There is no evidence for such large changes in the southward component at HON in the past 6\,years, or in SJG in around 1970.  Further study is needed of the relationship between the rapid SV changes occasionally observed at low latitude stations in the radial field and the classical jerk signals most often reported at mid-latitudes in the Eastward component for European observatories.


 \newpage

\section{Conclusions}
\label{sec:Conc}

We have presented the CHAOS-7 geomagnetic field model, the basis for DTU's IGRF-13 candidate field models, and used it to investigate geomagnetic field evolution, focusing on changes occurring over the past six years when  excellent data coverage has been available from the {\it Swarm} trio of satellites.  We find CHAOS-7 adequately represents data from the {\O}rsted, CHAMP, SAC-C, CryoSat-2 and {\it Swarm} satellites over the past 20~years, and is able to follow the trends in secular variation observed at ground observatories.

The DTU candidate model for DGRF~2015 is based directly on values of the internal field from CHAOS-7 in 2015.0,  when data from the {\it Swarm} satellites were available both before and after the target epoch.  The DTU candidate model for IGRF 2020 is based on the internal field from CHAOS-7 in 2019.75 (at the time of the last contributing data) propagated to 2020.0 using the secular variation from CHAOS-7 in 2019.0, when the final constraints from annual differences of ground data were available.  The secular variation in 2019.0 also provides the DTU predictive SV model for the period 2020.0 to 2025.0; we adopted this simple approach since we know of no reliable way to forecast future SV changes for the upcoming five years.

We find that at low Earth orbit the South Atlantic weak field anomaly continues to expand and deepen.  This is directly seen in {\it Swarm} magnetic measurements and in single event electronic upsets recorded on board the satellites.  Mapping the field intensity down to the Earth's surface we find evidence from CHAOS-7 for the development a secondary intensity minimum near 40~degrees South on the Greenwich meridian. This seems to be linked to the westward movement and evolution of a reversed flux feature at the core-mantle boundary under South Africa.  

We find localized changes in radial field acceleration averaged over consecutive three year periods (2014 to 2017 and 2017 to 2020), of amplitude up to 12~nT/yr$^3$ in the central and western Pacific.  The pattern of field change resembles a localized, east-west oriented, dipole spanning 120~degrees in longitude and confined to within 30~degrees of the equator.  Descending to the core-mantle boundary structures more confined in longitude are seen with highest amplitudes close to the equator.  An important task is now to track the development of such features in detail, for example to ascertain whether energy propagates from high latitudes towards lower latitudes where it focuses at the equator.

The CHAOS-7 model, and its updates, which are typically released every 4 to 6~months, depending on \textit{Swarm} data calibration and reprocessing activities, as well associated software, are available from: \\
\url{http://www.spacecenter.dk/files/magnetic-models/CHAOS-7/index.html}.

Users interested in studying the CMB magnetic field are advised to truncate the field itself at SH degree 13, the secular variation, which is less polluted by the lithospheric field, at degree 14 (degree 17 after 2014) and the secular acceleration, which is more challenging to extract, at degree 9 (degree 15 after 2014).  We have recently updated the original CHAOS-7(.1) model to CHAOS-7.2 and further updates will be released in due course.  Henceforth we shall follow the convention for \textit{Swarm} data products that the second version of a given model should be referred to as version 2.  In CHAOS-7.2 we increased the temporal regularization of CryoSat-2 magnetometer sensitivities and increased the maximum degree of the temporal regularization tapering function in eqn (\ref{eqn:taper}) from 11 to 15.  The latter change was designed to mitigate instability observed in the original CHAOS-7 release in the high degree secular variation and acceleration prior to 2005; though unimportant for IGRF-13 this was undesireable.  We recommend that users always use the most recent version of the model available on the webpage or the latest stable release of the python package available at \url{https://pypi.org/project/chaosmagpy/}.\\


\begin{backmatter}

\section*{List of abbreviations}
AT - Along-track,
CHAllenging Minisatellite Payload - CHAMP,
CHAOS - CHAMP, {\O}rsted, and \textit{Swarm} field model,
CM - Comprehensive Model,
CMB - Core-Mantle Boundary,
DGRF - Definitive Geomagnetic Reference Field,
Dst - Disturbance Storm Time,
ECEF - Earth-Centered-Earth-Fixed reference fram
ESA - European Space Agency,
Est - External part of Dst index,
$E_m$ - Merging Electric Field at the Magnetopause,
EW - East-West,
FGM - Fluxgate Magnetometer,
GSM - Geocentric Solar Magnetospheric Coordinate System,
hr - Hour,
IGRF - International Geomagnetic Reference Field,
IMF - Interplanetary Magnetic Field
Iteratively Re-weighted Least-Squares - IRLS,
Kp - K planetary index,
LCS - Lithospheric model from CHAMP and Swarm,
nT - nano-Tesla,
QD - Quasi-Dipole,
RC - Ring-Current index,
rms - Root-Mean Square,
RMM - Revised Monthly Mean,
SA - Secular Acceleration,
sec - seconds,
SH - Spherical Harmonic,
SM - Solar Magnetic coordinate system,
SV - Secular Variation,
yr - year
1D - One-Dimensional,
3D - Three-Dimensional.

\section*{Declarations}

\section*{Availability of datasets and material}
Swarm and Cryosat-2 data are available from \url{https://earth.esa.int/web/guest/swarm/data-access}\\
CHAMP data are available from \url{http://isdc.gfz-potsdam.de} \\
\O rsted and SAC-C data are available from \url{ftp://ftp.spacecenter.dk/data/magnetic-satellites}\\
Ground observatory data are available from \url{ftp://ftp.nerc-murchison.ac.uk/geomag/Swarm/AUX_OBS/hour/}\\
The RC index is available from
\url{http://www.spacecenter.dk/files/magnetic-models/RC/}\\
The CHAOS-7 model and its updates are at \url{http://www.spacecenter.dk/files/magnetic-models/CHAOS-7/}\\
A python package for using the CHAOS model is available at \url{https://pypi.org/project/chaosmagpy/}
\section*{Competing interests}
  The authors declare that they have no competing interests.
  
\section*{Funding}
This study has been partially supported as part of \textit{Swarm} DISC activities, funded by ESA contact no. 4000109587 and also by the Swarm+ 4D Deep Earth: Core project, ESA contract no. 4000127193/19/NL/IA.  CCF and CK were partially funded by the  European Research Council (ERC) under the European Union’s Horizon 2020 research and innovation programme (grant agreement No. 772561).

\section*{Author's contributions}
CCF derived the CHAOS-7 model and led the writing of the manuscript.  CK developed the modelling software for co-estimation of calibration parameters.  NiO developed the CHAOS modelling approach, pre-processed the CryoSat-2 data and participated in the design of the study. LT processed the \textit{Swarm} data. MH derived GVO data and contributed to the analysis of the SAA using Green's functions. AG and AK derived the mantle conductivity models and $Q$-matrices and kernels used to account for induction. All co-authors read and approved the final manuscript.

\section*{Acknowledgements}
We wish  to thank ESA  for the prompt availability of {\it Swarm} L1b data and for providing access to the CryoSat-2 platform magnetometer data and related engineering information.   The support of the CHAMP mission by the German Aerospace Center (DLR) and the Federal Ministry of Education and Research is gratefully acknowledged.  The \O rsted Project was made possible by extensive support from the Danish Government, NASA, ESA, CNES, DARA and the Thomas B. Thriges Foundation.  The staff of the geomagnetic observatories and INTERMAGNET are thanked for supplying high-quality observatory data. Susan Macmillan (BGS) is gratefully acknowledged for collating checked and corrected observatory hourly mean values in the AUX$\_$OBS database. Ignacio Clerigo is thanked for help with the \textit{Swarm} single event upset data. CCF thanks J\"{u}rgen Matzka for helpful discussions related to the SAA and Rasmus M\o ller Blangsb\o ll for assistance in preparing Fig.19.  Two anonymous reviewers and the editor P. Alken are thanked for their comments that helped to improve the quality of the manuscript.

\clearpage

\end{backmatter}
\end{document}